\documentclass[aip,jcp,reprint,onecolumn, citeautoscript]{revtex4-1}

\usepackage{graphicx} 
\usepackage{dcolumn} 
\usepackage{bm}
\usepackage{amssymb} 
\usepackage{hyperref}
\usepackage {natbib}
\usepackage{subfigure}
\usepackage{wrapfig}
\usepackage{placeins}
\usepackage{footnote}
\usepackage{amsmath}

\hypersetup{colorlinks=true, linkcolor=blue, citecolor=blue,filecolor=blue, urlcolor=blue}

\newcommand{\xray}{\mbox{X-ray }}

\newcommand{\blue}{\color{blue} }
\renewcommand{\ae}{\"a}
\renewcommand{\oe}{\"o}
\newcommand{\ue}{\"u}

\graphicspath{{./figures/}}

\begin{document}
\title{Photodissociation of Aligned CH$_\text{3}$I and C$_\text{6}$H$_\text{3}$F$_\text{2}$I Molecules probed with Time-Resolved Coulomb Explosion Imaging by Site-Selective XUV Ionization}

\author{Kasra~Amini}
\altaffiliation{These authors contributed equally to this work.}
\affiliation{The Chemistry Research Laboratory, Department of Chemistry, University of Oxford, Oxford OX1 3TA, United Kingdom}

\author{Evgeny~Savelyev}
\altaffiliation{These authors contributed equally to this work.}
\affiliation{Deutsches Elektronen--Synchrotron DESY, 22607 Hamburg, Germany}
 
\author{Felix~Brau{\ss}e} 
\affiliation{Max Born Institute for Nonlinear Optics and Short Pulse Spectroscopy, 12489 Berlin\mbox{, Germany}}

\author{Nora Berrah} 
\affiliation{Department of Physics, University of Connecticut, Storrs, Connecticut 06269\mbox{,~USA}}

\author{C\'{e}dric~Bomme}
\affiliation{Deutsches Elektronen--Synchrotron DESY, 22607 Hamburg, Germany}

\author{Mark~Brouard} 
\affiliation{The Chemistry Research Laboratory, Department of Chemistry, University of Oxford, Oxford OX1 3TA, United Kingdom}

\author{Michael~Burt} 
\affiliation{The Chemistry Research Laboratory, Department of Chemistry, University of Oxford, Oxford OX1 3TA, United Kingdom}

\author{Lauge~Christensen} 
\affiliation{Department of Chemistry, Aarhus University, 8000 Aarhus C, Denmark}

\author{Stefan~D\ue sterer} 
\affiliation{Deutsches Elektronen--Synchrotron DESY, 22607 Hamburg, Germany}

\author{Benjamin~Erk} 
\affiliation{Deutsches Elektronen--Synchrotron DESY, 22607 Hamburg, Germany}

\author{Hauke~H\oe ppner} 
\affiliation{Deutsches Elektronen--Synchrotron DESY, 22607 Hamburg, Germany}
\affiliation{Helmholtz Zentrum Dresden Rossendorf, Bautzner Landstrasse 400, 01328 Dresden\mbox{, Germany}}

\author{Thomas~Kierspel} 
\affiliation{Center for Free--Electron Laser Science, Deutsches Elektronen--Synchrotron DESY, 22607 Hamburg, Germany}
\affiliation{Center for Ultrafast Imaging, Universit\ae t Hamburg, 22761 Hamburg, Germany}

\author{Faruk~Krecinic} 
\affiliation{Max Born Institute for Nonlinear Optics and Short Pulse Spectroscopy, 12489 Berlin\mbox{, Germany}}

\author{Alexandra~Lauer} 
\affiliation{The Chemistry Research Laboratory, Department of Chemistry, University of Oxford, Oxford OX1 3TA, United Kingdom}

\author{Jason~W.\,L.~Lee} 
\affiliation{The Chemistry Research Laboratory, Department of Chemistry, University of Oxford, Oxford OX1 3TA, United Kingdom}

\author{Maria~M\ue ller} 
\affiliation{Institut f\ue r Optik und Atomare Physik, Technische Universit\ae t Berlin, 10623 Berlin\mbox{, Germany}}

\author{Erland~M\ue ller} 
\affiliation{Deutsches Elektronen--Synchrotron DESY, 22607 Hamburg, Germany}

\author{Terence~Mullins} 
\affiliation{Center for Free--Electron Laser Science, Deutsches Elektronen--Synchrotron DESY, 22607 Hamburg, Germany}

\author{Harald~Redlin} 
\affiliation{Deutsches Elektronen--Synchrotron DESY, 22607 Hamburg, Germany}

\author{Nora~Schirmel} 
\affiliation{Deutsches Elektronen--Synchrotron DESY, 22607 Hamburg, Germany}

\author{Jan~Th{\o}gersen} 
\affiliation{Department of Chemistry, Aarhus University, 8000 Aarhus C, Denmark}

\author{Simone~Techert} 
\affiliation{Deutsches Elektronen--Synchrotron DESY, 22607 Hamburg, Germany}
\affiliation{Max Planck Institute for Biophysical Chemistry, 33077 G\oe ttingen, Germany}
\affiliation{Institute for X-ray Physics, G\oe ttingen University, 33077 G\oe ttingen, Germany}

\author{Sven~Toleikis} 
\affiliation{Deutsches Elektronen--Synchrotron DESY, 22607 Hamburg, Germany}

\author{Rolf~Treusch} 
\affiliation{Deutsches Elektronen--Synchrotron DESY, 22607 Hamburg, Germany}

\author{Sebastian~Trippel} 
\affiliation{Center for Free--Electron Laser Science, Deutsches Elektronen--Synchrotron DESY, 22607 Hamburg, Germany}
\affiliation{Center for Ultrafast Imaging, Universit\ae t Hamburg, 22761 Hamburg, Germany}

\author{Anatoli~Ulmer} 
\affiliation{Institut f\ue r Optik und Atomare Physik, Technische Universit\ae t Berlin, 10623 Berlin\mbox{, Germany}}

\author{Claire~Vallance} 
\affiliation{The Chemistry Research Laboratory, Department of Chemistry, University of Oxford, Oxford OX1 3TA, United Kingdom}

\author{Joss~Wiese} 
\affiliation{Center for Free--Electron Laser Science, Deutsches Elektronen--Synchrotron DESY, 22607 Hamburg, Germany}

\author{Per~Johnsson} 
\affiliation{Department of Physics, Lund University, 22100 Lund, Sweden}

\author{Jochen~K\ue pper} 
\affiliation{Center for Free--Electron Laser Science, Deutsches Elektronen--Synchrotron DESY, 22607 Hamburg, Germany}
\affiliation{Center for Ultrafast Imaging, Universit\ae t Hamburg, 22761 Hamburg, Germany}
\affiliation{Department of Physics, Universit\ae t Hamburg, 22761 Hamburg, Germany}

\author{Artem~Rudenko} 
\affiliation{J. R. Macdonald Laboratory, Department of Physics, Kansas State University, Manhattan, Kansas 66506, USA}

\author{Arnaud~Rouz\'{e}e} 
\affiliation{Max Born Institute for Nonlinear Optics and Short Pulse Spectroscopy, 12489 Berlin\mbox{, Germany}}

\author{Henrik~Stapelfeldt} 
\affiliation{Department of Chemistry, Aarhus University, 8000 Aarhus C, Denmark}

\author{Daniel~Rolles} 
\email[]{rolles@phys.ksu.edu}
\affiliation{J. R. Macdonald Laboratory, Department of Physics, Kansas State University, Manhattan, Kansas 66506, USA}
\affiliation{Deutsches Elektronen--Synchrotron DESY, 22607 Hamburg, Germany}

\author{Rebecca~Boll} 
\email[]{rebecca.boll@desy.de}
\affiliation{Deutsches Elektronen--Synchrotron DESY, 22607 Hamburg, Germany}

\date{\today}

\begin{abstract}
We explore time-resolved Coulomb explosion induced by intense, extreme ultraviolet (XUV) femtosecond pulses from the FLASH free-electron laser as a method to image photo-induced molecular dynamics in two molecules, iodomethane and 2,6-difluoroiodobenzene. At an excitation wavelength of 267\,nm, the dominant reaction pathway in both molecules is neutral dissociation via cleavage of the carbon--iodine bond. This allows investigating the influence of the molecular environment on the absorption of an intense, femtosecond XUV pulse and the subsequent Coulomb explosion process. We find that the XUV probe pulse induces local inner-shell ionization of atomic iodine in dissociating iodomethane, in contrast to non-selective ionization of all photofragments in difluoroiodobenzene. The results reveal evidence of electron transfer from methyl and phenyl moieties to a multiply charged iodine ion. In addition, indications for ultrafast charge rearrangement on the phenyl radical are found, suggesting that time-resolved Coulomb explosion imaging is sensitive to the localization of charge in extended molecules.
\end{abstract}

\maketitle

\section{Introduction}

If several electrons are rapidly removed from a molecule, it fragments into cations by a process termed Coulomb explosion~\cite{vager_coulomb_1989}. Provided the break-up occurs faster than vibrational motion, the momenta of the fragments can be used to determine the structure of gas-phase molecules. Coulomb explosion induced by intense femtosecond~(fs) laser pulses in the visible or the near-infrared region~\cite{posthumus_dynamics_2004} can be used as a time-resolved structural probe of molecular dynamics. A molecular reaction, such as photodissociation, is initiated with a fs pump pulse, and the evolving structure is measured as a function of time using a delayed, intense Coulomb explosion  pulse. Such time-resolved Coulomb explosion imaging~(CEI) has been used to study photoisomerization~\cite{Hishikawa-PhysRevLett.99.258302,ibrahim_tabletop_2014}, photodissociation~\cite{stapelfeldt_wave_1995,legare_imaging_2005,skovsen_imaging_2002}, and torsional motion in an axially chiral molecule~\cite{madsen_manipulating_2009,christensen_dynamic_2014}. An alternative method for inducing Coulomb explosion employs irradiation with extreme ultraviolet (XUV) or X-ray femtosecond pulses, notably from intense free-electron laser sources~\cite{ackermann_operation_2007, shintake_compact_2008, emma_first_2010, ishikawa_compact_2012, allaria_highly_2013}. Several recent experiments have demonstrated the feasibility of studying molecular dynamics such as fragmentation~\cite{picon_hetero-site-specific_2016}, isomerization\cite{jiang_ultrafast_2010,liekhus-schmaltz_ultrafast_2015}, charge transfer~\cite{erk_imaging_2014,boll_charge_2016}, and  interatomic Coulombic decay~\cite{schnorr_time-resolved_2013} in real time.

XUV or X-ray induced Coulomb explosion differs from its strong-field induced equivalent in several respects: While strong-field ionization removes electrons from the molecular valence shell, which is typically highly delocalized, the XUV or X-ray photon energy can be tuned to an inner-shell absorption edge, thus making the photoabsorption site- and element specific. Furthermore, the kinetic energies and angular correlations of the ionic fragments resulting from strong-field induced Coulomb explosion typically strongly depend on the pulse duration~\cite{legare_laser_2005, legare_imaging_2005, legare_laser_2006}, while this dependence can be less pronounced in the case of inner-shell ionization, where the time-scale of the Auger decay is often the most relevant parameter, especially for Coulomb explosion induced by single-photon absorption. 

\begin{figure*} [tb]
\centering
\includegraphics[width = 0.99\textwidth]{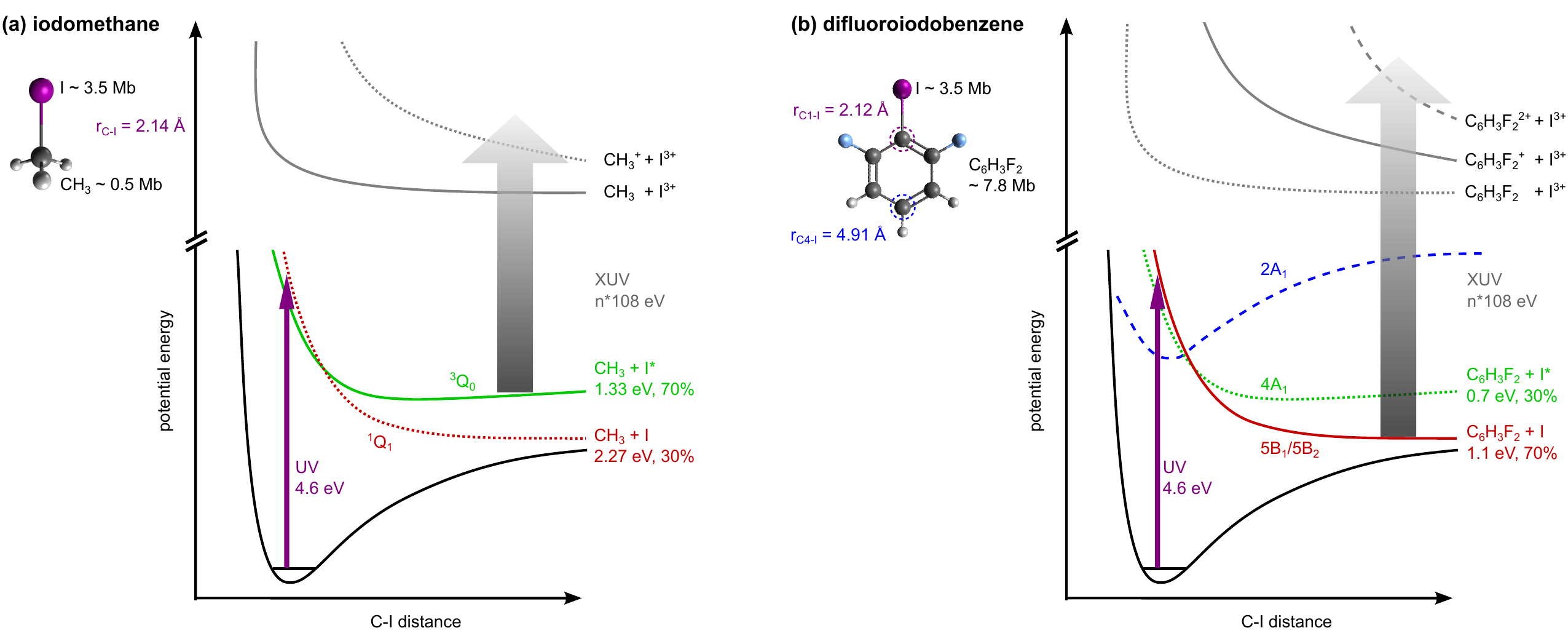}
\caption{Schematic one-dimensional potential energy curves~(PECs) of (a)~iodomethane and (b)~difluoroiodobenzene. The PECs corresponding to the dominant channels in our experiment are shown as solid lines. The dissociative states, $^3$Q$_0$ and $^1$Q$_1$ {\blue in CH$_3$I and 4A$_1$ and 5B$_1$/5B$_2$ in DFIB are shown as red and green lines and} are very similar for both molecules~(the nomenclature follows Ref.~\onlinecite{murdock_uv_2012}). The total kinetic energy release of the resulting products and their asymptotic relative populations after absorption of one 267\,nm photon~(purple arrow) are indicated~\cite{eppink_methyl_1998, murdock_uv_2012, sage_n*_2011}. In DFIB, an additional predissociative state (dashed blue) can also be populated. The XUV probe pulse~(grey arrow) promotes the system to one of many multiply charged potential energy curves through single or multi-photon absorption. For simplicity, only exemplary curves leading to triply charged iodine ions in the final state are shown here. The sum of the atomic cross sections for single-photon ionization at 108\,eV photon energy are indicated next to the sketch of the molecules~\cite{yeh_atomic_1993}. }
\label{pecs}
\end{figure*}

Here, we focus on the role of site-selective ionization in time-resolved Coulomb explosion imaging experiments. To this end, we investigate the ultraviolet~(UV)-induced photoexcitation and subsequent XUV ionization and fragmentation of isolated iodomethane~(CH$_3$I) and 2,6-difluoroiodobenzene~(C$_6$H$_3$F$_2$I, DFIB) molecules, see Fig.~\ref{pecs}. The photochemistry of iodomethane  in the A-band~(210--350\,nm) has been the subject of previous experimental and theoretical studies, see for example Refs.~\onlinecite{eppink_methyl_1998, de_nalda_detailed_2008, rubio-lago_photodissociation_2009} and references therein. In this energy range, the photoexcitation~(purple arrow in Fig.~\ref{pecs}) triggers almost exclusively a resonant one-photon dissociation into two neutrals, by promoting an electron to the $\sigma^*$\,orbital along the C--I bond. This results either in ground state iodine, CH$_3$\,+\,I, with a yield of $\sim$30\,\%, or in spin-orbit excited iodine, CH$_3$\,+\,I$^*$, with a yield of $\sim$70\,\%~\cite{eppink_methyl_1998}, as illustrated in Fig.~\ref{pecs}\,(a). The UV-photochemistry of fluorinated aryl iodides is less well studied, but can be regarded as largely similar to the case of iodobenzene~\cite{murdock_uv_2012}. Two channels analogous to the case of CH$_3$I are accessible through single-photon UV excitation, but leading to opposite yield in the two spin-orbit components~($\sim$70\,\% I, $\sim$30\,\% I$^*$)~\cite{sage_n*_2011}. In addition, bound states involving electron density on the phenyl ring are overlapped with the A-band~(240--320\,nm), and can thus form a predissociative state by mixing with the C--I dissociative state, see the dashed blue line in Fig.~\ref{pecs}\,(b). Creation of ground-state iodine atoms via the predissociative channel is strongly suppressed as compared to the direct dissociation~\cite{murdock_uv_2012}. 

The similarity of the UV-pump step for both molecules gives us the opportunity to study the role of site-selective ionization and the influence of the molecular environment, i.e., methyl versus phenyl moiety, on the XUV-probe step (grey arrow in Fig.~\ref{pecs}). At a certain time delay after the UV excitation, the dissociating molecule is ionized by the FEL probe pulse, leading to a highly excited molecular ion that fragments through Coulomb explosion. This is illustrated in Fig.~\ref{pecs}, showing one-dimensional cuts through the potential energy hypersurfaces of the multiply charged molecular ion that are formed following (multiple) ionization of the excited molecules by the FEL pulse. Note that absorption of more than one XUV photon in the same molecule is possible in our experiment, due to the high peak intensity of the XUV pulse in the focus. The PECs involving three or more charges on the iodine, as illustrated in Fig.~\ref{pecs}, are created by absorption of two or more XUV photons. While the XUV-ionization can be regarded as site-selective in CH$_3$I, as $\sim$90\% of the absorbed photons are absorbed at the iodine atom at 108\,eV photon energy, in DFIB $\sim$70\% of the absorbed photons are absorbed at the difluorobenzene~(DFB) radical~\cite{yeh_atomic_1993}.

\section{Experiment}

The experiment was carried out in the CAMP endstation~\cite{struder_large-format_2010, erk_campflash_2016} at the Free-Electron Laser in Hamburg~(FLASH)~\cite{feldhaus_flashfirst_2010}. The experimental setup as well as the data treatment have been described in detail in Refs.~\onlinecite{savelyev_jitter-correction_2017, amini_alignment_2017}. In brief, CH$_3$I or C$_6$H$_3$F$_2$I molecules were mixed with neon~(20\,bar) at room temperature and supersonically expanded through a pulsed Even-Lavie valve~(opening time 12.5\,$\mu$s) and then passed through two skimmers.  An electrostatic deflector positioned between the skimmers selected the lowest-lying rotational quantum states and also partially separated the molecules from the neon, as the atomic carrier gas is unaffected by the deflector. This increases the maximum degree of molecular alignment that can be achieved~\cite{chang_spatially-controlled_2015, filsinger_quantum-state_2009}, and, moreover, reduces significantly the amount of background ions from the neon carrier gas and thus prevents detector saturation. In the interaction region, the molecular beam was intersected by the free-electron laser~(FEL) and two additional laser beams, which propagated collinearly to the FEL. A near-infrared~(NIR) pulse from an Nd:YAG laser~(1064\,nm, 12\,ns\,(FWHM), 1.2\,J, 50\,$\mu$m focus~(FWHM)) was used to adiabatically align the molecules~\cite{stapelfeldt_colloquium:_2003}, such that their most polarizable axes, the C--I axes, were aligned along the laser polarization direction, parallel to the detector plane. At the peak of the alignment pulse, where the degree of alignment is highest~\cite{stapelfeldt_colloquium:_2003}, the molecules were first photoexcited by an ultraviolet laser pulse~(267\,nm, 150\,fs\,(FWHM), 35\,$\mu$J, 50\,$\mu$m focus\,(FWHM)) and then probed by an intense extreme-ultraviolet FEL pulse~(108\,eV, 120\,fs\,(FWHM), 37\,$\mu$J on average,  20\,$\mu$m focus\,(FWHM)) after a tunable time delay. The polarizations of the UV and the FEL pulse were parallel to each other, in the detection plane. The repetition rate of the experiment was 10\,Hz. The delay between the UV pump pulse and the XUV probe pulse was set using a motorized delay stage in the UV arm. The data was acquired by recording 1000 shots per delay step of 83\,fs in a range of $\pm$\,1\,ps, i.e. 24000 shots for all of the CH$_3$I data. For DFIB, the delay step size was 66\,fs for I$^{2+}$, 20\,fs for I$^{3+}$, and 53\,fs for I$^{4+}$, and the total number of shots contained in the delay scans of the different ion species was 30000, 105000, and 15000, respectively. In the data analysis, these shots are resorted and rebinned according to the information from the beam arrival time monitor of the electron bunch.~\cite{savelyev_jitter-correction_2017}

Ions and electrons resulting from the photoionization were recorded simultaneously with a double-sided velocity-map imaging~(VMI) spectrometer~\cite{struder_large-format_2010, rolles_femtosecond_2014} by multichannel plates coupled to phosphor screens. The corresponding two-dimensional ion momentum distributions were recorded by a commercial {\blue one-Megapixel CCD camera~(Allied Vision Pike F-145B) or a 72\,x\,72 pixels time-stamping camera~(PImMS)~\cite{john_pimms_2012, amini_three-dimensional_2015}. The cameras were mounted outside of vacuum and could be interchanged easily, such that pump-probe scans were recorded for both molecules with both cameras. The CCD camera provides much higher spatial resolution, but it does not have the timing resolution to distinguish between the different ionic species that arrive at the detector with flight-time differences of several hundred nanoseconds after a total flight time of a few microseconds. Therefore, the high voltage on the MCP detector was gated using a fast high-voltage switch such that only a specific ion species was detected at a time, and the pump-probe scans for different ion species were recorded consecutively. The PImMS camera, on the other hand, can record and time-stamp up to four ions hits per pixel with a 12.5\,ns timing precision, which is sufficient to distinguish the different iodine charge states and most other ionic species in this experiment (the corresponding time-of-flight spectra and further details are given in Ref.~\onlinecite{amini_alignment_2017}). With the PImMS camera, the yields and 2D momentum distributions of all ionic species can therefore be recorded within the \emph{same} pump-probe scan, albeit with lower spatial resolution than with the CCD camera.}

{\blue In the subsequent data analysis, the 2D momentum distributions of each ionic species recorded with the CCD or the PImMS camera} were angularly integrated, and the radii were then converted to kinetic energies~(KE) based on ion trajectory simulations carried out using the Simion~8.0 software package, from which an empirical formula was constructed that connects the hit position on the detector with the fragment's kinetic energy. Further discussion of the observed ion kinetic energies in the Coulomb explosion of DFIB is also given in Ref.~\onlinecite{amini_alignment_2017}. {\blue Here, we concentrate on discussing the yields and kinetic energies of the multiply charged iodine fragments from CH$_3$I and DFIB and, in particular, their dependence on the delay between the UV and XUV pulses.} The iodine ion KE distributions were converted to a total kinetic energy release~(TKER) based on the assumption that the cofragment is momentum-matched with the recorded iodine ion.  This assumption is exact for a two-body fragmentation, as is induced by the UV pulse. It is expected to also apply to the majority of the XUV ionization events that occur in already dissociated molecules.

Note that by using strong, adiabatic alignment of the C--I axes parallel to the detector plane, the component of the ion momentum along the spectrometer axis is effectively confined to zero, such that the recorded radial distribution of iodine ions on the detector corresponds to the momentum distribution to a very good approximation, thus making image inversion algorithms that are normally used in VMI spectroscopy unnecessary in this case.\footnote{The finite width of the molecular axis distribution perpendicular to the detector plane leads to an effective smearing out of the extracted kinetic energy spectra towards lower kinetic energies. However, for the data presented here, no significant improvement was found when inverting the ion images, so non-inverted data is shown throughout the manuscript.} The degree of alignment determined from the I$^{3+}$ ion images was $\big\langle$cos$^2 \Theta_{2\textnormal{D}}\big\rangle = 0.92$ for CH$_3$I and 0.94 for DFIB, corresponding to a standard deviation of 17$^\circ$ and 15$^\circ$ with respect to the XUV polarization direction, respectively. 

Adiabatic alignment is an alternative to the retrieval of molecular structure by coincident ion momentum spectroscopy employing delay-line anodes. Our approach allows for experimental conditions with high ion count rates per shot when an MCP/phosphor screen detector is used. Molecular alignment parallel to the detection plane is particularly powerful in combination with a time-stamping camera such as PImMS~\cite{john_pimms_2012, amini_three-dimensional_2015} or TimepixCam~\cite{fisher-levine_timepixcam:_2016, fisher-levine_time-resolved_2017}, or an in-vacuum pixel detector~\cite{long_ion-ion_2017},  facilitating the recording of all ionic species simultaneously. These images provide the opportunity to study angular correlations between different ionic fragments, which, with the help of laser alignment, can be interpreted in a straightforward way, allowing detailed conclusions to be drawn about structure and fragmentation dynamics~\cite{slater_covariance_2014, christensen_dynamic_2014, slater_coulomb-explosion_2015, christensen_using_2015, christensen_deconvoluting_2016, pickering_communication:_2016, hansen_control_2012}.

\section{Results}

\begin{figure*} [tb]
\centering
\includegraphics[width = 0.99\textwidth]{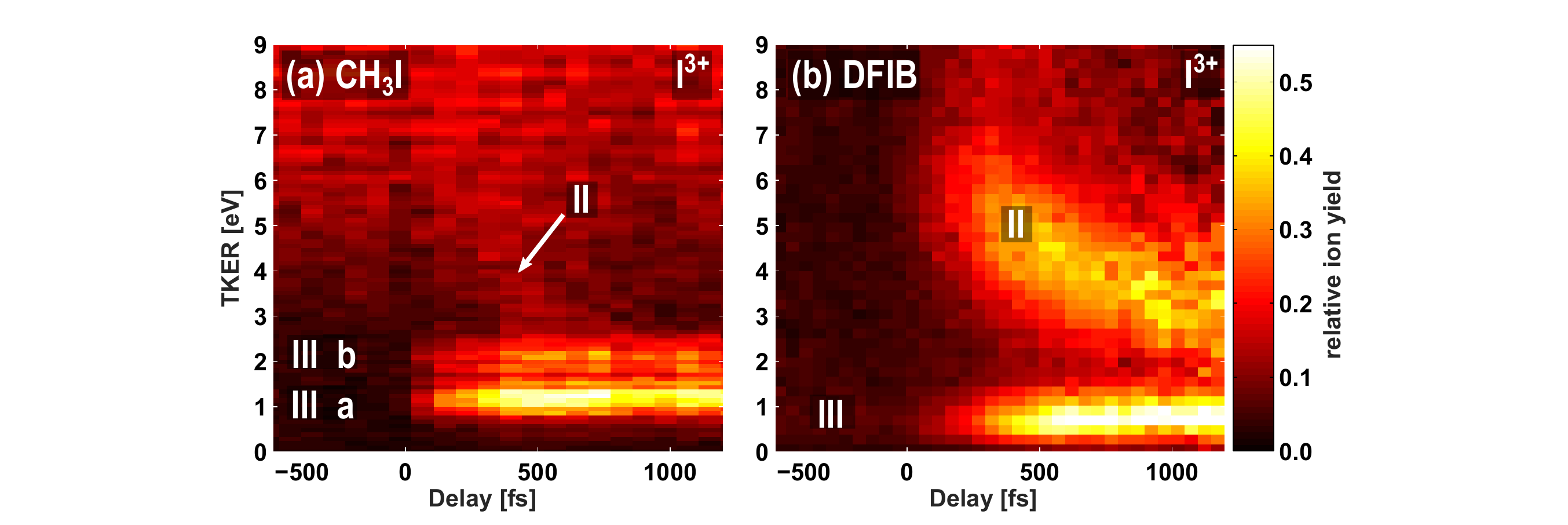}
\caption{Low-kinetic-energy region of the total kinetic energy release of triply charged iodine ions as a function of the delay between the UV-pump and XUV-probe pulses for (a)~iodomethane and (b)~difluoroiodobenzene molecules, {\blue recorded with the CCD camera}. Negative delays correspond to the XUV pulse arriving first, positive delays to the UV pulse arriving first, in accordance with our earlier, related publications~\cite{erk_imaging_2014, boll_charge_2016}. Different fragmentation channels, II and III are indicated and are discussed in the main text.}
\label{I3}
\end{figure*}

Details of the static Coulomb explosion imaging of laser-aligned DFIB molecules, as well as more technical aspects of the laser/FEL pump-probe data analysis have been discussed before~\cite{amini_alignment_2017, savelyev_jitter-correction_2017}. In the present manuscript, we focus on the molecular dynamics due to the UV-dissociation and the subsequent XUV-induced Coulomb explosion, and on the comparison between DFIB and CH$_3$I molecules. 

As described in the introduction, the dominant process upon UV-excitation of both molecules is neutral C--I bond cleavage. In the following, we describe how the XUV-probe signal resulting from the dissociated molecules depends on the molecular environment. At the intensity used in this experiment, {\blue ionization by the UV laser pulse and the Nd:YAG laser pulse alone was minimal and, in particular,  neither laser pulse produced any multiply charged ions}. Figure~\ref{I3}(a) and (b) show the low-energy region of the delay-dependent total kinetic energy releases determined for the two molecules from the triply charged iodine ions. Two dynamic features are visible: a strong feature~(III), which has a TKER independent of the delay; and a second feature~(II), which is strong in DFIB and weak in CH$_3$I and which corresponds to a TKER that varies as a function of the delay. The numbering of the fragmentation channels follows the nomenclature of our earlier publications~\cite{erk_imaging_2014, boll_charge_2016}. Similar features appear also for I$^{2+}$ and  I$^{4+}$, as shown in Fig.~\ref{fig}. Outside the axis range chosen here, an additional broad feature at higher TKER values is present, corresponding to Coulomb explosion of bound molecules by the XUV pulse alone, as discussed in more detail in Ref.~\onlinecite{amini_alignment_2017}.

\subsection{Local XUV ionization at iodine (channel~III)}

\begin{figure*} [bt]
\centering
\includegraphics[width = 0.99\textwidth]{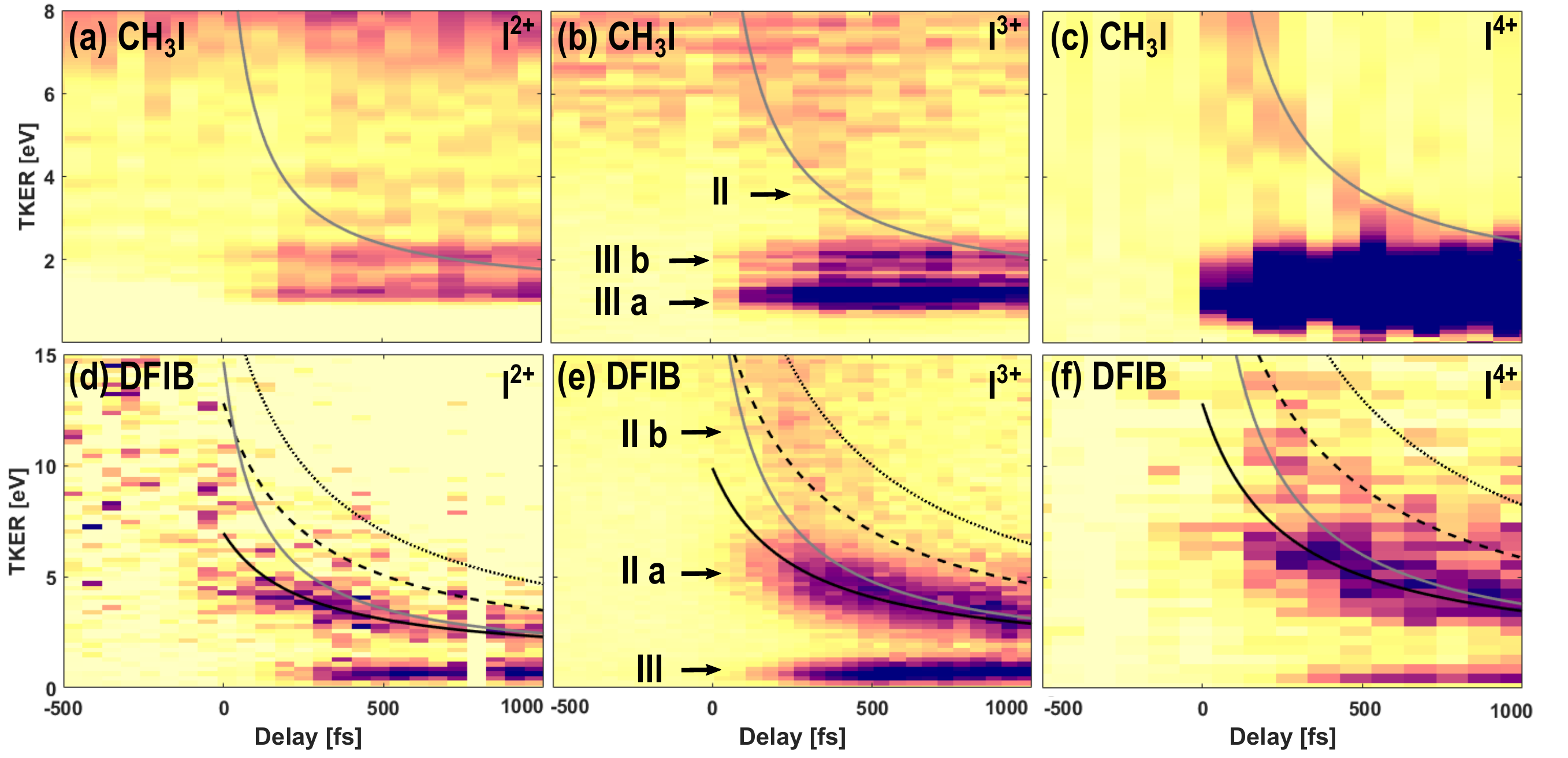}
\caption{Delay-dependent total kinetic energy release of different iodine charge states arising after UV-excitation and subsequent XUV ionization of aligned CH$_{3}$I~(top) and  DFIB~(bottom) molecules, {\blue recorded with the CCD camera}. Results of Coulomb explosion simulations are superimposed as lines. Solid lines correspond to an I$^{n+}$ ion dissociating with a singly charged molecular rest, dashed lines to a doubly and dotted lines to a triply charged partner. For CH$_3$I, a C--I distance of 2.14\,\AA~has been used for the calculation, for DFIB, r$_{\rm{C1-I}}$\,$=$\,2.12\,\AA~(grey) and  r$_{\rm{C4-I}}$\,$=$\,4.91\,\AA~(black) are displayed. The DFIB data were normalized to the sum of the FEL pulse energy in each delay bin, and a jitter-correction was applied~\cite{savelyev_jitter-correction_2017}. For the CH$_3$I data set, the beam arrival time monitor was not operational, and single-shot information was not collected. Therefore, the jitter could not be corrected and the ion yield was normalized to the number of acquisitions. Given the FEL and laser pulse durations in this experiment, the influence of the arrival time jitter between laser and FEL pulses, which is typically less than 200\,fs~(FWHM), should not significantly broaden the effective pump-probe instrument response function. In (d) and (f), the UV late spectrum~(delays\,$\leq$\,$-$270\,fs) has been subtracted from all delay bins, to provide better visibility. In (c), no centroiding could be applied and no normalization was carried out. }
\label{fig}
\end{figure*}

\begin{figure*} [tb]
\centering
\includegraphics[width = 0.99\textwidth]{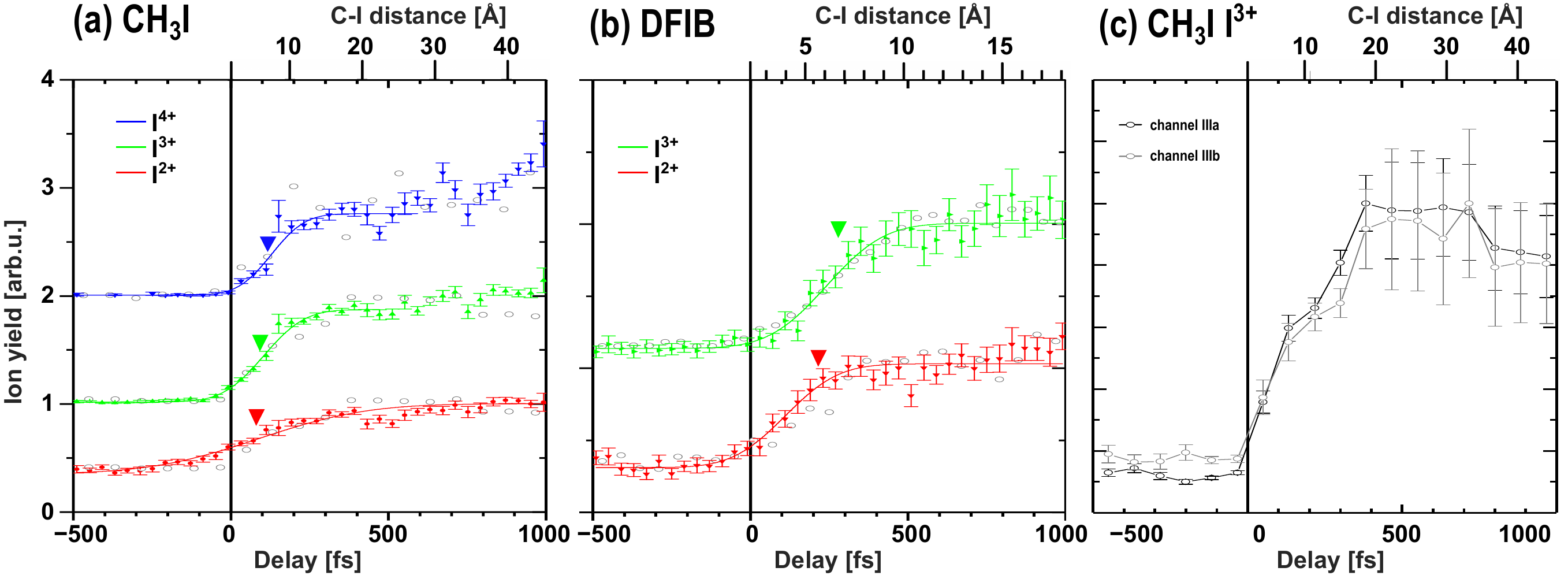}
\caption{Delay-dependent ion yield in channel~III for different iodine charge states from (a)~iodomethane and (b)~difluoroiodobenzene molecules. The data points are obtained by integrating all recorded ions of each species within a TKER range of 0\,$-$2.5\,eV for CH$_3$I and for a TKER of 0\,$-$1.5\,eV for DFIB, {\blue recorded with the CCD camera~(gray circles) or the PImMS camera~(colored symbols). To extract the center of the step functions, a Gaussian cumulative distribution function was fitted to the PImMS data, as shown by the solid lines}. For better visibility, the yields for the different charge states are offset on the vertical axis. Negative delays correspond to the XUV pulse arriving first, positive delays to the UV pulse arriving first. {\blue The large inverted triangles indicate the calculated critical distances for internuclear charge transfer resulting from the over-the-barrier model~(see text). (c) Delay-dependent ion yield of the two components~III\,a~(integrated over the TKER range from 0\,--\,1.5\,eV) and III\,b~(1.5\,--\,2.5\,eV) in the I$^{3+}$ channel in CH$_3$I, recorded with the CCD camera.}}
\label{yields}
\end{figure*}

Channel~III can be assigned to neutral C--I bond cleavage induced by absorption of one UV photon, followed by XUV-ionization of the isolated iodine atom. The cofragment does not interact with the XUV pulse and remains neutral, therefore the resulting TKER in channel~III is determined solely by the translational energy gained during the UV-dissociation when the molecule has dissociated into two independent fragments. {\blue However, closer inspection of the delay-dependencies in the CH$_3$I and C$_6$H$_3$F$_2$I data~(Figs.~\ref{I3} and~\ref{fig}) reveal that these channels are not centered around zero pump-probe delay, but are instead shifted towards positive delays. To show this more clearly, Fig.~\ref{yields} displays the delay-dependent ion yield of channel~III for different iodine charge states. The shift in the onset of this channel for both molecules is attributed to the existence of ultrafast intramolecular electron transfer from the methyl or difluorophenyl radical to the multiply charged iodine ion, which can not happen at large internuclear separations. This process was discussed in detail in Refs.~\onlinecite{erk_imaging_2014, boll_charge_2016} for iodomethane. If the two moieties are at close proximity to each other, which is the case at small delays between the UV and the XUV pulses, the methyl or difluorophenyl radical will not remain neutral in the vicinity of a multiply charged iodine ion but become singly charged via electron transfer. The corresponding fragment pair will gain additional Coulomb energy and thus will not appear in channel~III, but rather at higher TKERs and with an iodine charge state reduced by one.

It was shown in Refs.~\onlinecite{erk_imaging_2014, boll_charge_2016} that for iodine 3d ionization of iodomethane, the critical internuclear distance up to which charge rearrangement is observed, can be well reproduced by a classical over-the-barrier model~\cite{ryufuku_oscillatory_1980, niehaus_classical_1986, schnorr_electron_2014}, which describes the electron transfer as a result of the supression of the potential barrier between the multiply charged iodine ion and the neutral radical at close proximity. The values for the critical internuclear distance resulting from this classical over-the-barrier model applied to the present case are indicated by the large inverted triangles in Fig.~\ref{yields}.

Before discussing these results further, we would like to note that Fig.~\ref{yields} shows the delay-dependent ion yields obtained from multiple pump-probe scans using the CCD (gray circles) and the PImMS camera (colored symbols). In the former case, the data sets were measured consecutively over the course of several days of beamtime. Long-term timing drifts, which could not be measured and corrected for accurately enough, made it impossible to determine one absolute time zero for all delay scans. Therefore, the time zero was determined for each delay scan individually by matching the simulated Coulomb curves for channel~II (see next section) to the experimental data. The accuracy of this method is of the same order as the temporal resolution of the experiment, which was estimated to be 200\,fs~(FHWH)~\cite{savelyev_jitter-correction_2017} and which was limited by the FEL and laser pulse durations. 

In contrast, for the data sets recorded with PImMS camera, the delay-dependence of the different iodine charge states with respect to each other is well defined since all ions were recorded within the same pump-probe run and only the common time zero had to be determined, which was done by fitting channel~II in the I$^{3+}$ ion yield. Therefore, the uncertainty of the absolute time-zero determination does not affect the relative differences observed between different charge states. Furthermore, it is worth pointing out that although time zero has been determined \textit{independently} for the data recorded with the PImMS and the CCD camera, the data sets are in good agreement, suggesting that the method for determining time zero is rather robust.

Unfortunately, in the case of DFIB, the 12.5-ns time resolution of the PImMS camera was not sufficient to distinguish the I$^{4+}$ fragment from the nearby CF$^+$, which has a mass-to-charge ratio that differs by less than one unit and which therefore has a very similar time of flight. Furthermore, the lower spatial resolution of the PImMS camera resulted in a lower kinetic energy resolution, such that channels II and III could not be well separated at large delays, most notably in the case of I$^{4+}$ from CH$_3$I.}

{\blue In order to analyze the relative shift in the onset of channel III for each of the iodine charge states, a Gaussian cumulative distribution function (CDF) was fitted to the PImMS data in Fig.~\ref{yields}. For CH$_3$I, the CDF fits yield center positions of 87$\pm$20\,fs, 104$\pm$7\,fs, and 122$\pm$11\,fs for I$^{2+}$, I$^{3+}$, and I$^{4+}$, respectively. For DFIB, the center positions are 111$\pm$14\,fs and 242$\pm$13\,fs for I$^{2+}$ and I$^{3+}$, respectively. The indicated errors are the statistical errors of the fits. We can thus conclude that the charge transfer process, which was previously observed to occur after iodine 3d ionization of iodomethane, also occurs after iodine 4d ionization of both iodomethane and difluoroiodobenzene, demonstrating that the electron transfer from a neutral molecular fragment to a multiply charged atomic ion as introduced in Refs.~\onlinecite{erk_imaging_2014, boll_charge_2016} is not particular to iodomethane, but also occurs from a phenyl moiety. Furthermore, within the experimental uncertainties and temporal resolution, the classical over-the-barrier-model is consistent with the data. In particular,} the onset of channel~III appears to be shifted to later delays in DFIB as compared to iodomethane. This can, to a large extent, be explained by pure kinematics when taking into account that the UV-excitation of DFIB preferentially results in a two-body dissociation into an iodine atom~(mass 127) and a difluorobenzene radical~(DFB, mass 113), i.\,e., two almost equally heavy partners. By momentum conservation, this leads to a slower dissociation velocity as compared to iodomethane, where the CH$_3$ radical~(mass 15) gains a much higher kinetic energy. Therefore, the internuclear distance in DFIB increases much slower than in CH$_3$I, and the critical distance, at which electron transfer between the two fragments can no longer occur, is reached at larger delays. {\blue The top axes in Fig.~\ref{yields}, which indicate the C--I internuclear distance for a given pump-probe delay and which were used to position the inverted triangles in this plot, were calculated assuming constant velocities of the neutral fragments created by the UV pulse. These velocities were calculated from the corresponding asymptotic TKER values given in the literature~\cite{eppink_methyl_1998, murdock_uv_2012}. The assumption of constant velocities is an approximation that overestimates how quickly the internuclear distances increase since the fragments certainly do not reach their asymptotic energy value instantaneously. A more precise model, which shall not be developed here, would thus require a quantitative knowledge of the dissociative potential curves in the neutral molecule.}

We would also like to point out that without coincident electron spectroscopy and/or detailed quantum chemistry calculations, we can not draw conclusions about the absolute charge transfer probabilities or any details about the underlying electronic processes that lead to the charge transfer in the two molecules. These processes are expected to differ significantly between the methyl and phenyl radicals, as DFIB has many more electrons, which are partly delocalized over the ring. The influence of the molecular environment on the charge rearrangement in iodomethane and iodobenzene molecules has been recently investigated for ultraintense hard X-rays, both experimentally and theoretically~\cite{rudenko_femtosecond_2017}.

Interestingly, in our two earlier \xray experiments on iodomethane, channel~III appeared only for charge states I$^{4+}$ and higher~\cite{erk_imaging_2014, boll_charge_2016}. In the XUV regime however, it can also be observed for I$^{2+}$ and I$^{3+}$ ions, but not for I$^+$~(not shown). This can be understood when considering the fact that in the \xray experiments, the iodine M-shell was accessible for ionization, whereas at 108\,eV, iodine 4d is the deepest energy level that can be ionized by a single photon~\cite{thompson_x-ray_2009}. In the former case, ionization of an isolated iodine atom results predominantly in I$^{4+}$ and I$^{5+}$, as can be concluded by comparison to the M-shell ionization of Xe atoms~\cite{saito_multiple_1992}, which are isoelectronic to I$^-$ and thus have similar cross section and Auger relaxation pathways. I$^{2+}$ and I$^{3+}$ are produced only after ionization of the intact molecule and thus do not exhibit the low-energy channel~III that stems from dissociated molecules. At 108 eV, in contrast, ionization of an isolated iodine atom results predominantly in I$^{2+}$ and I$^{3+}$ (again, we refer to the case of Xe for comparison~\cite{saito_multiple_1992}), while I$^{+}$ is only produced from intact molecules and is thus the only charge state where the low-energy channel~III does not occur. We note that iodine atoms recorded with four or more charges must be created through absorption of at least two XUV photons, thus confirming that multiphoton absorption plays a significant role in this experiment.

{\blue Finally, we would like to investigate the delay-dependence of the two components of channel III, labeled III\,a and III\,b in Figure~\ref{I3}(a), which} correspond to the two different spin-orbit components that contribute to the UV excitation, as illustrated in the PECs in Fig.~\ref{pecs}. Initially, excitation at 267\,nm occurs to the $^3$Q$_0$ state, but a fraction of the dissociating wavepacket is transferred to the $^1$Q$_1$ potential energy surface via nonadiabatic coupling, resulting in a relative population of roughly 70\,\% in the CH$_3$\,+\,I$^*$ channel and 30\,\% in the CH$_3$\,+\,I channel~\cite{eppink_methyl_1998}. According to Ref.~\onlinecite{eppink_methyl_1998}, the total kinetic energy release expected for channels~III\,a and III\,b is 1.33 and 2.27\,eV, respectively. The relative ion yield as well as the TKER values of both channels in the present data set are in accordance with these literature values within the uncertainty of our experimental energy calibration, which is of the order of $\sim$20\,\%.

{\blue As shown in the delay-dependent ion yields in Fig.~\ref{yields}(c),} channels III\,a and III\,b exhibit the same delay-dependence within the uncertainties of the present measurement, which is to be expected, as the non-adiabatic curve crossing between the $^3$Q$_0$ and the $^1$Q$_1$ state takes place within $<$\,50\,fs, i.\,e., at very short internuclear distance~\cite{de_nalda_detailed_2008}. At such early delays, channel~III is not yet observed, since ultrafast charge rearrangement necessarily results in a charged cofragment at such small internuclear distances. Note that the two sub-components were not resolved in our earlier UV-pump, soft X-ray-probe experiment on CH$_3$I~\cite{boll_charge_2016}, utilizing a time-of-flight mass spectrometer with a small aperture and without a position-sensitive detector. 

For DFIB, channel~III comprises two analogous components, with a TKER of 0.7\,eV for the ejection of I$^*$  and 1.1\,eV for the ground state, and an I$^*$ to I ratio of 30/70~(opposite to the case of CH$_3$I)~\cite{murdock_uv_2012}. These components are, however, not resolved in the present data, probably because the DFIB data have been recorded  at a factor of four higher spectrometer voltages as compared to CH$_3$I in order to allow for simultaneous 4$\pi$ electron detection on the opposite side of the VMI. The mean value of 0.7\,eV is in accordance with the published TKER values~\cite{murdock_uv_2012} within the limited energy resolution of this data set.

\subsection{Non-selective XUV ionization of both photofragments (channel~II)}
\label{ch2}

We now turn to the feature exhibiting a rapidly decreasing kinetic energy as a function of the pump-probe delay, labeled channel~II in Figs.~\ref{I3} and~\ref{fig}. The decreasing kinetic energy indicates that this channel is the result of a Coulomb repulsion between two \emph{charged} fragments, whose distance increases as the pump-probe delay increases. Given that the pulse intensity was tuned such that almost no ionization was induced by the UV pulse alone, and noting that channel~II is also present at long pump-probe delays when charge transfer between the two fragments is no longer possible, the origin of this channel must be the absorption of \emph{two} XUV photons, one by each of the fragments.

The observation that channel~II is much stronger for DFIB than for CH$_3$I for all iodine charge states, relative to the yield in channel~III can be understood as follows: considering the sum of the atomic photoabsorption cross sections for each of  the fragments at a photon energy of 108\,eV, $\sim$90\% of the absorbed photons are absorbed by the iodine atom and only $\sim$10\% by the CH$_3$ fragment (for the case of iodomethane), whereas in DFIB, $\sim$70\% of the absorbed photons are absorbed by the difluorobenzene radical and only $\sim$30\% by the iodine~\cite{yeh_atomic_1993}. Following the neutral UV-dissociation, it is thus significantly more probable that an XUV photon is absorbed by the DFB radical than by the isolated methyl group. Therefore, channel~II is much stronger in DFIB than in iodomethane.  We note that channel~II, similar to channel~III discussed above, also seems to start appearing at slightly positive delays, especially for the higher iodine charge states in DFIB. The most likely reason is again an ultrafast charge transfer process, by which an electron from the singly charged methyl fragment is transferred to the highly charged iodine ion when the latter is in very close proximity, thereby increasing the total Coulomb energy of this fragment pair.

In order to further investigate the origin of channel~II and to assign it to a specific Coulomb explosion channel, numerical 2-step Coulomb explosion simulations have been carried out. The fragments resulting from the initial UV-induced dissociation are assumed to travel with a constant velocity (calculated from the corresponding TKER values given in the literature~\cite{eppink_methyl_1998, murdock_uv_2012}) before being ionized by the XUV pulse after a certain time delay $\tau$. The final TKER is thus a function of the delay, $\tau$, and can be calculated as the sum of the TKER of the neutral dissociation, TKER$_{\rm UV}$, and the Coulombic potential energy gained after the two charged fragments, A and B, are created at time $\tau$.
\begin{equation}
\textnormal{TKER}(\tau) = \textnormal{TKER}_{\rm UV} + \frac{k_{\rm e}q_{\rm A}q_{\rm B}}{r_{{\rm AB}}(\tau)}
\label{eq}
\end{equation}
Here, \textit{k$_{{\rm e}}$}, \textit{q}, and \textit{r$_{{\rm AB}}$} are the electrostatic constant, fragment charge, and distance between the charges on fragments A and B. The distance \textit{r$_{{\rm AB}}$} is calculated for each pump-probe delay. The model assumes an instantaneous charging of both fragmentation partners to the final charge states at the given delay time, and a purely Coulombic repulsion between point charges. The TKER$_{\rm UV}$ of the UV-excitation with the higher probability, i.\,e., I$^*$ for CH$_3$I and I for DFIB, has been used for the calculations of the two molecules, respectively. {\blue As noted in section III.A, the assumption that the full TKER$_{\rm UV}$ is added to the Coulomb energy independent of $\tau$ is an approximation, since the fragments do not have this asymptotic energy value for small delays. However, given the difference in magnitude between the Coulomb energy at small delays and the possible change in TKER$_{\rm UV}$, we have neglected the latter for the sake of simplicity}. Further details on the calculations are also given in Refs.~\onlinecite{burt_coulomb_2017, amini_studies_2017}.  

The results are overlaid with the TKER maps in Fig.~\ref{fig}. According to the simulations, the kinetic energy of channel~II in all of the iodine charge states and for both molecules is best matched when assuming a singly charged co-fragment. This is reasonable since absorption of one XUV photon by the CH$_3$ or DFB radical leads, with high probability, to only one charge on this fragment since the photon is absorbed by a valence electron. Core levels of carbon or fluorine are not accessible at 108\,eV photon energy. In Fig.~\ref{fig}(e), a second, weaker feature, labeled~II\,b, is visible in addition to the strong feature,~II\,a. Channel~II\,b has an initially higher, but more rapidly decreasing TKER, which could be attributed to Coulomb explosion with a doubly charged co-fragment. However, the agreement with the corresponding simulated TKER curve is not very good. It might be that there is a mixture of DFB$^{2+}$ and DFB$^{3+}$ fragmentation partners, but it seems that, in particular at delays $>$\,400\,fs, neither of the two simulated curves agree very well with the data. The fact that at least three XUV photoabsorptions are involved in channel~II\,b makes a more quantitative analysis difficult, because there is an unknown delay between the ionization events occurring within the XUV pulse duration.

Finally, for the case of DFIB, we can use the Coulomb explosion simulations to investigate the location of the charge on the co-fragment. For that purpose, we take into consideration two limiting cases for the localization of a point charge on the DFB radical, either at the carbon atom located closest to the iodine~(grey curves in Fig.~\ref{fig}), or at the carbon atom furthest away~(black curves in Fig.~\ref{fig}), see also inset in Fig.~\ref{pecs}(b). In accordance with recent results from a static synchrotron measurement~\cite{ablikim_isomer-dependent_2017}, the time-resolved TKER data agree much better with the simulations using the largest distance between the charges, i.\,e., a situation in which two charges are preferentially located at opposite ends of the molecule when the Coulomb repulsion starts. This suggests that the delocalized charge distribution on the phenyl ring is shifted with respect to the center-of-mass of the DFB radical, due to the dipole moment which is induced by the multiply charged iodine ion that is initially in close vicinity of the DFB radical. It is also consistent with an ultrafast charge migration that is instantaneous within the temporal resolution of our experiment, as is expected for purely electronic intramolecular charge rearrangement~\cite{kuleff_core_2016}. Given the pulse durations of the UV and the XUV pulses in the present experiment, the temporal resolution of the data is not sufficient to draw further conclusions at this point, but experiments with shorter pulse durations may enable studies of such ultrafast charge migration in the near future.

\section{Conclusion and Outlook}

Time-resolved Coulomb explosion imaging of two molecules, iodomethane~(CH$_3$I) and difluoroiodobenzene~(C$_6$H$_3$F$_2$I, DFIB) allowed the influence of the molecular environment on inner-shell ionization to be investigated by an intense, femtosecond XUV pulse at the Free-Electron Laser in Hamburg~(FLASH). UV-excitation at 267\,nm induced a two-body dissociation, resulting in cleavage of the C--I bond into two neutral fragments. At a photon energy of 108\,eV, the XUV probe pulse is absorbed locally at the iodine atom for the case of CH$_3$I, while in DFIB, the photoabsorption is less selective. Both molecules exhibit charge transfer from the multiply charged iodine ion to the methyl and phenyl moieties, respectively, at short internuclear distance. The timescale of this electron rearrangement is slower in DFIB than in iodomethane, because of its slower dissociation velocity.

A non-selective probe pulse ionizing all photofragments can probe the potential energy landscape of a molecule in detail, and enables information about both/all partners of the photoexcitation to be obtained, in particular when used in combination with ion-ion coincidences or covariances. In contrast, site-selective absorption at only one fragment, as is the case in CH$_3$I, leaves the second partner undetected. However, it facilitates, for example, the creation of a localized source of charge in order to study the electron rearrangement~\cite{erk_imaging_2014, boll_charge_2016, erk_ultrafast_2013, erk_inner-shell_2013, rudenko_femtosecond_2017}. In the XUV regime, the internuclear charge transfer process can be probed for smaller internuclear distances as compared to X-ray CEI experiments, therefore these data provide, in principle, better sensitivity to the  orbitals of the intact molecule. 

CEI is suitable for investigating dynamics in halogen-substituted benzene following a simple, two-body dissociation, and we plan to extend this technique to other systems and more complex photochemical reactions in the future. The combination with simultaneous femtosecond-resolved electron spectroscopy is a very promising avenue to gain insight into the electronic dynamics that are interconnected with the nuclear motion~\cite{brausse_time-resolved_2017}, and carrying out complementary pump-probe experiments using either inner-shell or strong-field ionization as a probe can provide valuable additional information on the influence of the probe process~\cite{burt_coulomb_2017, amini_alignment_2017}. Furthermore, we have presented results indicating that with shorter pump and probe pulse durations and, possibly, in combination with coincident or covariant ion detection, time-resolved CEI might be suitable to directly probe charge localization in polyatomic systems.

\section{Acknowledgments}
\begin{footnotesize}
We gratefully acknowledge the work of the scientific and technical teams at FLASH, who have made these experiments possible. We also acknowledge the Max Planck Society for funding the development of the CAMP instrument within the ASG at CFEL. In addition, the installation of CAMP at FLASH was partially funded by BMBF Grant No. 05K10KT2. The support of the UK EPSRC (to M.B. and C.V. via Programme Grant Nos. EP/G00224X/1 and EP/L005913/1), the EU (to M.B. via FP7 EU People ITN Project No. 238671 and to J.K., P.J., H.S., and D.R. via the MEDEA project that has received funding from the European Union's Horizon 2020 research and innovation programme under the Marie Sklodowska-Curie grant agreement No 641789), STFC through PNPAS award and a mini-IPS Grant (No. ST/J002895/1), and a proof of concept grant from ISIS Innovation Ltd. are gratefully acknowledged.  K.A. thanks the EPSRC, Merton College, Oxford University, and RSC for support. A.L. thanks the DFG via Grant No. La 3209/1-1 for support. N.B., A.Ru., and D.R. acknowledge support from the Chemical Sciences, Geosciences, and Biosciences Division, Office of Basic Energy Sciences, Office of Science, U.S. Department of Energy, Grant Nos. DE-FG02-86ER13491 (Kansas group) and DE-SC0012376 (U Conn group). D.R., E.S., R.B., C.B., and B.E. were also supported by the Helmholtz Gemeinschaft through the Helmholtz Young Investigator Program. J.K. and CFEL--CMI were, in addition to DESY, supported by the excellence cluster ``The Hamburg Center for Ultrafast Imaging--Structure, Dynamics, and Control of Matter at the Atomic Scale'' of the Deutsche Forschungsgemeinschaft (CUI, DFG-EXC1074), by the Helmholtz Virtual Institute 419 "Dynamic Pathways in Multidimensional Landscapes", by the Helmholtz Networking and Initiative Funds, and by the European Research Council under the European Union's Seventh Framework Programme (FP7/2007-2013) through the Consolidator Grant COMOTION (ERC-K\ue pper-614507)." S.Te., E.S., and R.B. are grateful for support through the German Science Foundation, project B03/SFB755 ``Nanoscale Photonic Imaging''. S.Te. is also grateful for financial support project through project C02/SFB1073 ``Atomic Control of Energy Conversion". P.J. acknowledges support from the Swedish Research Council and the Swedish Foundation for Strategic Research. A.Ro. is grateful for support through the Deutsche Forschungsgemeinschaft Project No.RO 4577/1-1.
\end{footnotesize}

\FloatBarrier
\bibliographystyle{all_authors_with_title}
\bibliography{FLASH2014} 

\begin{thebibliography}{62}%
\makeatletter
\providecommand \@ifxundefined [1]{%
 \@ifx{#1\undefined}
}%
\providecommand \@ifnum [1]{%
 \ifnum #1\expandafter \@firstoftwo
 \else \expandafter \@secondoftwo
 \fi
}%
\providecommand \@ifx [1]{%
 \ifx #1\expandafter \@firstoftwo
 \else \expandafter \@secondoftwo
 \fi
}%
\providecommand \natexlab [1]{#1}%
\providecommand \enquote  [1]{``#1''}%
\providecommand \bibnamefont  [1]{#1}%
\providecommand \bibfnamefont [1]{#1}%
\providecommand \citenamefont [1]{#1}%
\providecommand \href@noop [0]{\@secondoftwo}%
\providecommand \href [0]{\begingroup \@sanitize@url \@href}%
\providecommand \@href[1]{\@@startlink{#1}\@@href}%
\providecommand \@@href[1]{\endgroup#1\@@endlink}%
\providecommand \@sanitize@url [0]{\catcode `\\12\catcode `\$12\catcode
  `\&12\catcode `\#12\catcode `\^12\catcode `\_12\catcode `\%12\relax}%
\providecommand \@@startlink[1]{}%
\providecommand \@@endlink[0]{}%
\providecommand \url  [0]{\begingroup\@sanitize@url \@url }%
\providecommand \@url [1]{\endgroup\@href {#1}{\urlprefix }}%
\providecommand \urlprefix  [0]{URL }%
\providecommand \Eprint [0]{\href }%
\providecommand \doibase [0]{http://dx.doi.org/}%
\providecommand \selectlanguage [0]{\@gobble}%
\providecommand \bibinfo  [0]{\@secondoftwo}%
\providecommand \bibfield  [0]{\@secondoftwo}%
\providecommand \translation [1]{[#1]}%
\providecommand \BibitemOpen [0]{}%
\providecommand \bibitemStop [0]{}%
\providecommand \bibitemNoStop [0]{.\EOS\space}%
\providecommand \EOS [0]{\spacefactor3000\relax}%
\providecommand \BibitemShut  [1]{\csname bibitem#1\endcsname}%
\let\auto@bib@innerbib\@empty
\bibitem [{\citenamefont {Vager}\ \emph {et~al.}(1989)\citenamefont {Vager},
  \citenamefont {Naaman},\ and\ \citenamefont {Kanter}}]{vager_coulomb_1989}%
  \BibitemOpen
  \bibfield  {author} {\bibinfo {author} {\bibfnamefont {Z.}~\bibnamefont
  {Vager}}, \bibinfo {author} {\bibfnamefont {R.}~\bibnamefont {Naaman}}, \
  and\ \bibinfo {author} {\bibfnamefont {E.~P.}\ \bibnamefont {Kanter}},\
  }\bibfield  {title} {\emph {\bibinfo {title} {Coulomb {Explosion} {Imaging}
  of {Small} {Molecules}}},\ }\href {\doibase 10.1126/science.244.4903.426}
  {\bibfield  {journal} {\bibinfo  {journal} {Science}\ }\textbf {\bibinfo
  {volume} {244}},\ \bibinfo {pages} {426} (\bibinfo {year}
  {1989})}\BibitemShut {NoStop}%
\bibitem [{\citenamefont {Posthumus}(2004)}]{posthumus_dynamics_2004}%
  \BibitemOpen
  \bibfield  {author} {\bibinfo {author} {\bibfnamefont {J.~H.}\ \bibnamefont
  {Posthumus}},\ }\bibfield  {title} {\emph {\bibinfo {title} {The dynamics of
  small molecules in intense laser fields}},\ }\href {\doibase
  10.1088/0034-4885/67/5/R01} {\bibfield  {journal} {\bibinfo  {journal}
  {Reports on Progress in Physics}\ }\textbf {\bibinfo {volume} {67}},\
  \bibinfo {pages} {623} (\bibinfo {year} {2004})}\BibitemShut {NoStop}%
\bibitem [{\citenamefont {Hishikawa}\ \emph {et~al.}(2007)\citenamefont
  {Hishikawa}, \citenamefont {Matsuda}, \citenamefont {Fushitani},\ and\
  \citenamefont {Takahashi}}]{Hishikawa-PhysRevLett.99.258302}%
  \BibitemOpen
  \bibfield  {author} {\bibinfo {author} {\bibfnamefont {A.}~\bibnamefont
  {Hishikawa}}, \bibinfo {author} {\bibfnamefont {A.}~\bibnamefont {Matsuda}},
  \bibinfo {author} {\bibfnamefont {M.}~\bibnamefont {Fushitani}}, \ and\
  \bibinfo {author} {\bibfnamefont {E.~J.}\ \bibnamefont {Takahashi}},\
  }\bibfield  {title} {\emph {\bibinfo {title} {Visualizing recurrently
  migrating hydrogen in acetylene dication by intense ultrashort laser
  pulses}},\ }\href {\doibase 10.1103/PhysRevLett.99.258302} {\bibfield
  {journal} {\bibinfo  {journal} {Physical Review Letters}\ }\textbf {\bibinfo
  {volume} {99}},\ \bibinfo {pages} {258302} (\bibinfo {year}
  {2007})}\BibitemShut {NoStop}%
\bibitem [{\citenamefont {Ibrahim}\ \emph {et~al.}(2014)\citenamefont
  {Ibrahim}, \citenamefont {Wales}, \citenamefont {Beaulieu}, \citenamefont
  {Schmidt}, \citenamefont {Thir{\'e}}, \citenamefont {Fowe}, \citenamefont
  {Bisson}, \citenamefont {Hebeisen}, \citenamefont {Wanie}, \citenamefont
  {Gigu{\'e}re}, \citenamefont {Kieffer}, \citenamefont {Spanner},
  \citenamefont {Bandrauk}, \citenamefont {Sanderson}, \citenamefont
  {Schuurman},\ and\ \citenamefont {L{\'e}gar{\'e}}}]{ibrahim_tabletop_2014}%
  \BibitemOpen
  \bibfield  {author} {\bibinfo {author} {\bibfnamefont {H.}~\bibnamefont
  {Ibrahim}}, \bibinfo {author} {\bibfnamefont {B.}~\bibnamefont {Wales}},
  \bibinfo {author} {\bibfnamefont {S.}~\bibnamefont {Beaulieu}}, \bibinfo
  {author} {\bibfnamefont {B.~E.}\ \bibnamefont {Schmidt}}, \bibinfo {author}
  {\bibfnamefont {N.}~\bibnamefont {Thir{\'e}}}, \bibinfo {author}
  {\bibfnamefont {E.~P.}\ \bibnamefont {Fowe}}, \bibinfo {author}
  {\bibfnamefont {{\'E}.}~\bibnamefont {Bisson}}, \bibinfo {author}
  {\bibfnamefont {C.~T.}\ \bibnamefont {Hebeisen}}, \bibinfo {author}
  {\bibfnamefont {V.}~\bibnamefont {Wanie}}, \bibinfo {author} {\bibfnamefont
  {M.}~\bibnamefont {Gigu{\'e}re}}, \bibinfo {author} {\bibfnamefont {J.-C.}\
  \bibnamefont {Kieffer}}, \bibinfo {author} {\bibfnamefont {M.}~\bibnamefont
  {Spanner}}, \bibinfo {author} {\bibfnamefont {A.~D.}\ \bibnamefont
  {Bandrauk}}, \bibinfo {author} {\bibfnamefont {J.}~\bibnamefont {Sanderson}},
  \bibinfo {author} {\bibfnamefont {M.~S.}\ \bibnamefont {Schuurman}}, \ and\
  \bibinfo {author} {\bibfnamefont {F.}~\bibnamefont {L{\'e}gar{\'e}}},\
  }\bibfield  {title} {\emph {\bibinfo {title} {Tabletop imaging of structural
  evolutions in chemical reactions demonstrated for the acetylene cation}},\
  }\href {\doibase 10.1038/ncomms5422} {\bibfield  {journal} {\bibinfo
  {journal} {Nature Communications}\ }\textbf {\bibinfo {volume} {5}},\
  \bibinfo {pages} {4422} (\bibinfo {year} {2014})}\BibitemShut {NoStop}%
\bibitem [{\citenamefont {Stapelfeldt}\ \emph {et~al.}(1995)\citenamefont
  {Stapelfeldt}, \citenamefont {Constant},\ and\ \citenamefont
  {Corkum}}]{stapelfeldt_wave_1995}%
  \BibitemOpen
  \bibfield  {author} {\bibinfo {author} {\bibfnamefont {H.}~\bibnamefont
  {Stapelfeldt}}, \bibinfo {author} {\bibfnamefont {E.}~\bibnamefont
  {Constant}}, \ and\ \bibinfo {author} {\bibfnamefont {P.~B.}\ \bibnamefont
  {Corkum}},\ }\bibfield  {title} {\emph {\bibinfo {title} {Wave {Packet}
  {Structure} and {Dynamics} {Measured} by {Coulomb} {Explosion}}},\ }\href
  {\doibase 10.1103/PhysRevLett.74.3780} {\bibfield  {journal} {\bibinfo
  {journal} {Physical Review Letters}\ }\textbf {\bibinfo {volume} {74}},\
  \bibinfo {pages} {3780} (\bibinfo {year} {1995})}\BibitemShut {NoStop}%
\bibitem [{\citenamefont {L{\'e}gar{\'e}}\ \emph
  {et~al.}(2005{\natexlab{a}})\citenamefont {L{\'e}gar{\'e}}, \citenamefont
  {Lee}, \citenamefont {Litvinyuk}, \citenamefont {Dooley}, \citenamefont
  {Bandrauk}, \citenamefont {Villeneuve},\ and\ \citenamefont
  {Corkum}}]{legare_imaging_2005}%
  \BibitemOpen
  \bibfield  {author} {\bibinfo {author} {\bibfnamefont {F.}~\bibnamefont
  {L{\'e}gar{\'e}}}, \bibinfo {author} {\bibfnamefont {K.~F.}\ \bibnamefont
  {Lee}}, \bibinfo {author} {\bibfnamefont {I.~V.}\ \bibnamefont {Litvinyuk}},
  \bibinfo {author} {\bibfnamefont {P.~W.}\ \bibnamefont {Dooley}}, \bibinfo
  {author} {\bibfnamefont {A.~D.}\ \bibnamefont {Bandrauk}}, \bibinfo {author}
  {\bibfnamefont {D.~M.}\ \bibnamefont {Villeneuve}}, \ and\ \bibinfo {author}
  {\bibfnamefont {P.~B.}\ \bibnamefont {Corkum}},\ }\bibfield  {title} {\emph
  {\bibinfo {title} {Imaging the time-dependent structure of a molecule as it
  undergoes dynamics}},\ }\href {\doibase 10.1103/PhysRevA.72.052717}
  {\bibfield  {journal} {\bibinfo  {journal} {Physical Review A}\ }\textbf
  {\bibinfo {volume} {72}},\ \bibinfo {pages} {052717} (\bibinfo {year}
  {2005}{\natexlab{a}})}\BibitemShut {NoStop}%
\bibitem [{\citenamefont {Skovsen}\ \emph {et~al.}(2002)\citenamefont
  {Skovsen}, \citenamefont {Machholm}, \citenamefont {Ejdrup}, \citenamefont
  {Th{\o}gersen},\ and\ \citenamefont {Stapelfeldt}}]{skovsen_imaging_2002}%
  \BibitemOpen
  \bibfield  {author} {\bibinfo {author} {\bibfnamefont {E.}~\bibnamefont
  {Skovsen}}, \bibinfo {author} {\bibfnamefont {M.}~\bibnamefont {Machholm}},
  \bibinfo {author} {\bibfnamefont {T.}~\bibnamefont {Ejdrup}}, \bibinfo
  {author} {\bibfnamefont {J.}~\bibnamefont {Th{\o}gersen}}, \ and\ \bibinfo
  {author} {\bibfnamefont {H.}~\bibnamefont {Stapelfeldt}},\ }\bibfield
  {title} {\emph {\bibinfo {title} {Imaging and {Control} of {Interfering}
  {Wave} {Packets} in a {Dissociating} {Molecule}}},\ }\href {\doibase
  10.1103/PhysRevLett.89.133004} {\bibfield  {journal} {\bibinfo  {journal}
  {Physical Review Letters}\ }\textbf {\bibinfo {volume} {89}},\ \bibinfo
  {pages} {133004} (\bibinfo {year} {2002})}\BibitemShut {NoStop}%
\bibitem [{\citenamefont {Madsen}\ \emph {et~al.}(2009)\citenamefont {Madsen},
  \citenamefont {Madsen}, \citenamefont {Viftrup}, \citenamefont {Johansson},
  \citenamefont {Poulsen}, \citenamefont {Holmegaard}, \citenamefont
  {Kumarappan}, \citenamefont {J{\o}rgensen},\ and\ \citenamefont
  {Stapelfeldt}}]{madsen_manipulating_2009}%
  \BibitemOpen
  \bibfield  {author} {\bibinfo {author} {\bibfnamefont {C.~B.}\ \bibnamefont
  {Madsen}}, \bibinfo {author} {\bibfnamefont {L.~B.}\ \bibnamefont {Madsen}},
  \bibinfo {author} {\bibfnamefont {S.~S.}\ \bibnamefont {Viftrup}}, \bibinfo
  {author} {\bibfnamefont {M.~P.}\ \bibnamefont {Johansson}}, \bibinfo {author}
  {\bibfnamefont {T.~B.}\ \bibnamefont {Poulsen}}, \bibinfo {author}
  {\bibfnamefont {L.}~\bibnamefont {Holmegaard}}, \bibinfo {author}
  {\bibfnamefont {V.}~\bibnamefont {Kumarappan}}, \bibinfo {author}
  {\bibfnamefont {K.~A.}\ \bibnamefont {J{\o}rgensen}}, \ and\ \bibinfo
  {author} {\bibfnamefont {H.}~\bibnamefont {Stapelfeldt}},\ }\bibfield
  {title} {\emph {\bibinfo {title} {Manipulating the {Torsion} of {Molecules}
  by {Strong} {Laser} {Pulses}}},\ }\href {\doibase
  10.1103/PhysRevLett.102.073007} {\bibfield  {journal} {\bibinfo  {journal}
  {Physical Review Letters}\ }\textbf {\bibinfo {volume} {102}},\ \bibinfo
  {pages} {073007} (\bibinfo {year} {2009})}\BibitemShut {NoStop}%
\bibitem [{\citenamefont {Christensen}\ \emph {et~al.}(2014)\citenamefont
  {Christensen}, \citenamefont {Nielsen}, \citenamefont {Brandt}, \citenamefont
  {Madsen}, \citenamefont {Madsen}, \citenamefont {Slater}, \citenamefont
  {Lauer}, \citenamefont {Brouard}, \citenamefont {Johansson}, \citenamefont
  {Shepperson},\ and\ \citenamefont {Stapelfeldt}}]{christensen_dynamic_2014}%
  \BibitemOpen
  \bibfield  {author} {\bibinfo {author} {\bibfnamefont {L.}~\bibnamefont
  {Christensen}}, \bibinfo {author} {\bibfnamefont {J.~H.}\ \bibnamefont
  {Nielsen}}, \bibinfo {author} {\bibfnamefont {C.~B.}\ \bibnamefont {Brandt}},
  \bibinfo {author} {\bibfnamefont {C.~B.}\ \bibnamefont {Madsen}}, \bibinfo
  {author} {\bibfnamefont {L.~B.}\ \bibnamefont {Madsen}}, \bibinfo {author}
  {\bibfnamefont {C.~S.}\ \bibnamefont {Slater}}, \bibinfo {author}
  {\bibfnamefont {A.}~\bibnamefont {Lauer}}, \bibinfo {author} {\bibfnamefont
  {M.}~\bibnamefont {Brouard}}, \bibinfo {author} {\bibfnamefont {M.~P.}\
  \bibnamefont {Johansson}}, \bibinfo {author} {\bibfnamefont {B.}~\bibnamefont
  {Shepperson}}, \ and\ \bibinfo {author} {\bibfnamefont {H.}~\bibnamefont
  {Stapelfeldt}},\ }\bibfield  {title} {\emph {\bibinfo {title} {Dynamic
  {Stark} {Control} of {Torsional} {Motion} by a {Pair} of {Laser} {Pulses}}},\
  }\href {\doibase 10.1103/PhysRevLett.113.073005} {\bibfield  {journal}
  {\bibinfo  {journal} {Physical Review Letters}\ }\textbf {\bibinfo {volume}
  {113}},\ \bibinfo {pages} {073005} (\bibinfo {year} {2014})}\BibitemShut
  {NoStop}%
\bibitem [{\citenamefont {Ackermann}\ \emph {et~al.}(2007)\citenamefont
  {Ackermann}, \citenamefont {Asova}, \citenamefont {Ayvazyan}, \citenamefont
  {Azima}, \citenamefont {Baboi}, \citenamefont {B{\"a}hr}, \citenamefont
  {Balandin}, \citenamefont {Beutner}, \citenamefont {Brandt}, \citenamefont
  {Bolzmann}, \citenamefont {Brinkmann}, \citenamefont {Brovko}, \citenamefont
  {Castellano}, \citenamefont {Castro}, \citenamefont {Catani}, \citenamefont
  {Chiadroni}, \citenamefont {Choroba}, \citenamefont {Cianchi}, \citenamefont
  {Costello}, \citenamefont {Cubaynes}, \citenamefont {Dardis}, \citenamefont
  {Decking}, \citenamefont {Delsim-Hashemi}, \citenamefont {Delserieys},
  \citenamefont {Di~Pirro}, \citenamefont {Dohlus}, \citenamefont
  {D{\"u}sterer}, \citenamefont {Eckhardt}, \citenamefont {Edwards},
  \citenamefont {Faatz}, \citenamefont {Feldhaus}, \citenamefont
  {Fl{\"o}ttmann}, \citenamefont {Frisch}, \citenamefont {Fr{\"o}hlich},
  \citenamefont {Garvey}, \citenamefont {Gensch}, \citenamefont {Gerth},
  \citenamefont {G{\"o}rler}, \citenamefont {Golubeva}, \citenamefont
  {Grabosch}, \citenamefont {Grecki}, \citenamefont {Grimm}, \citenamefont
  {Hacker}, \citenamefont {Hahn}, \citenamefont {Han}, \citenamefont
  {Honkavaara}, \citenamefont {Hott}, \citenamefont {H{\"u}ning}, \citenamefont
  {Ivanisenko}, \citenamefont {Jaeschke}, \citenamefont {Jalmuzna},
  \citenamefont {Jezynski}, \citenamefont {Kammering}, \citenamefont {Katalev},
  \citenamefont {Kavanagh}, \citenamefont {Kennedy}, \citenamefont
  {Khodyachykh}, \citenamefont {Klose}, \citenamefont {Kocharyan},
  \citenamefont {K{\"o}rfer}, \citenamefont {Kollewe}, \citenamefont {Koprek},
  \citenamefont {Korepanov}, \citenamefont {Kostin}, \citenamefont
  {Krassilnikov}, \citenamefont {Kube}, \citenamefont {Kuhlmann}, \citenamefont
  {Lewis}, \citenamefont {Lilje}, \citenamefont {Limberg}, \citenamefont
  {Lipka}, \citenamefont {L{\"o}hl}, \citenamefont {Luna}, \citenamefont
  {Luong}, \citenamefont {Martins}, \citenamefont {Meyer}, \citenamefont
  {Michelato}, \citenamefont {Miltchev}, \citenamefont {M{\"o}ller},
  \citenamefont {Monaco}, \citenamefont {M{\"u}ller}, \citenamefont
  {Napieralski}, \citenamefont {Napoly}, \citenamefont {Nicolosi},
  \citenamefont {N{\"o}lle}, \citenamefont {Nu{\~n}ez}, \citenamefont {Oppelt},
  \citenamefont {Pagani}, \citenamefont {Paparella}, \citenamefont {Pchalek},
  \citenamefont {Pedregosa-Gutierrez}, \citenamefont {Petersen}, \citenamefont
  {Petrosyan}, \citenamefont {Petrosyan}, \citenamefont {Petrosyan},
  \citenamefont {Pfl{\"u}ger}, \citenamefont {Pl{\"o}njes}, \citenamefont
  {Poletto}, \citenamefont {Pozniak}, \citenamefont {Prat}, \citenamefont
  {Proch}, \citenamefont {Pucyk}, \citenamefont {Radcliffe}, \citenamefont
  {Redlin}, \citenamefont {Rehlich}, \citenamefont {Richter}, \citenamefont
  {Roehrs}, \citenamefont {Roensch}, \citenamefont {Romaniuk}, \citenamefont
  {Ross}, \citenamefont {Rossbach}, \citenamefont {Rybnikov}, \citenamefont
  {Sachwitz}, \citenamefont {Saldin}, \citenamefont {Sandner}, \citenamefont
  {Schlarb}, \citenamefont {Schmidt}, \citenamefont {Schmitz}, \citenamefont
  {Schm{\"u}ser}, \citenamefont {Schneider}, \citenamefont {Schneidmiller},
  \citenamefont {Schnepp}, \citenamefont {Schreiber}, \citenamefont {Seidel},
  \citenamefont {Sertore}, \citenamefont {Shabunov}, \citenamefont {Simon},
  \citenamefont {Simrock}, \citenamefont {Sombrowski}, \citenamefont {Sorokin},
  \citenamefont {Spanknebel}, \citenamefont {Spesyvtsev}, \citenamefont
  {Staykov}, \citenamefont {Steffen}, \citenamefont {Stephan}, \citenamefont
  {Stulle}, \citenamefont {Thom}, \citenamefont {Tiedtke}, \citenamefont
  {Tischer}, \citenamefont {Toleikis}, \citenamefont {Treusch}, \citenamefont
  {Trines}, \citenamefont {Tsakov}, \citenamefont {Vogel}, \citenamefont
  {Weiland}, \citenamefont {Weise}, \citenamefont {Wellh{\"o}fer},
  \citenamefont {Wendt}, \citenamefont {Will}, \citenamefont {Winter},
  \citenamefont {Wittenburg}, \citenamefont {Wurth}, \citenamefont {Yeates},
  \citenamefont {Yurkov}, \citenamefont {Zagorodnov},\ and\ \citenamefont
  {Zapfe}}]{ackermann_operation_2007}%
  \BibitemOpen
  \bibfield  {author} {\bibinfo {author} {\bibfnamefont {W.}~\bibnamefont
  {Ackermann}}, \bibinfo {author} {\bibfnamefont {G.}~\bibnamefont {Asova}},
  \bibinfo {author} {\bibfnamefont {V.}~\bibnamefont {Ayvazyan}}, \bibinfo
  {author} {\bibfnamefont {A.}~\bibnamefont {Azima}}, \bibinfo {author}
  {\bibfnamefont {N.}~\bibnamefont {Baboi}}, \bibinfo {author} {\bibfnamefont
  {J.}~\bibnamefont {B{\"a}hr}}, \bibinfo {author} {\bibfnamefont
  {V.}~\bibnamefont {Balandin}}, \bibinfo {author} {\bibfnamefont
  {B.}~\bibnamefont {Beutner}}, \bibinfo {author} {\bibfnamefont
  {A.}~\bibnamefont {Brandt}}, \bibinfo {author} {\bibfnamefont
  {A.}~\bibnamefont {Bolzmann}}, \bibinfo {author} {\bibfnamefont
  {R.}~\bibnamefont {Brinkmann}}, \bibinfo {author} {\bibfnamefont {O.~I.}\
  \bibnamefont {Brovko}}, \bibinfo {author} {\bibfnamefont {M.}~\bibnamefont
  {Castellano}}, \bibinfo {author} {\bibfnamefont {P.}~\bibnamefont {Castro}},
  \bibinfo {author} {\bibfnamefont {L.}~\bibnamefont {Catani}}, \bibinfo
  {author} {\bibfnamefont {E.}~\bibnamefont {Chiadroni}}, \bibinfo {author}
  {\bibfnamefont {S.}~\bibnamefont {Choroba}}, \bibinfo {author} {\bibfnamefont
  {A.}~\bibnamefont {Cianchi}}, \bibinfo {author} {\bibfnamefont {J.~T.}\
  \bibnamefont {Costello}}, \bibinfo {author} {\bibfnamefont {D.}~\bibnamefont
  {Cubaynes}}, \bibinfo {author} {\bibfnamefont {J.}~\bibnamefont {Dardis}},
  \bibinfo {author} {\bibfnamefont {W.}~\bibnamefont {Decking}}, \bibinfo
  {author} {\bibfnamefont {H.}~\bibnamefont {Delsim-Hashemi}}, \bibinfo
  {author} {\bibfnamefont {A.}~\bibnamefont {Delserieys}}, \bibinfo {author}
  {\bibfnamefont {G.}~\bibnamefont {Di~Pirro}}, \bibinfo {author}
  {\bibfnamefont {M.}~\bibnamefont {Dohlus}}, \bibinfo {author} {\bibfnamefont
  {S.}~\bibnamefont {D{\"u}sterer}}, \bibinfo {author} {\bibfnamefont
  {A.}~\bibnamefont {Eckhardt}}, \bibinfo {author} {\bibfnamefont {H.~T.}\
  \bibnamefont {Edwards}}, \bibinfo {author} {\bibfnamefont {B.}~\bibnamefont
  {Faatz}}, \bibinfo {author} {\bibfnamefont {J.}~\bibnamefont {Feldhaus}},
  \bibinfo {author} {\bibfnamefont {K.}~\bibnamefont {Fl{\"o}ttmann}}, \bibinfo
  {author} {\bibfnamefont {J.}~\bibnamefont {Frisch}}, \bibinfo {author}
  {\bibfnamefont {L.}~\bibnamefont {Fr{\"o}hlich}}, \bibinfo {author}
  {\bibfnamefont {T.}~\bibnamefont {Garvey}}, \bibinfo {author} {\bibfnamefont
  {U.}~\bibnamefont {Gensch}}, \bibinfo {author} {\bibfnamefont
  {C.}~\bibnamefont {Gerth}}, \bibinfo {author} {\bibfnamefont
  {M.}~\bibnamefont {G{\"o}rler}}, \bibinfo {author} {\bibfnamefont
  {N.}~\bibnamefont {Golubeva}}, \bibinfo {author} {\bibfnamefont {H.-J.}\
  \bibnamefont {Grabosch}}, \bibinfo {author} {\bibfnamefont {M.}~\bibnamefont
  {Grecki}}, \bibinfo {author} {\bibfnamefont {O.}~\bibnamefont {Grimm}},
  \bibinfo {author} {\bibfnamefont {K.}~\bibnamefont {Hacker}}, \bibinfo
  {author} {\bibfnamefont {U.}~\bibnamefont {Hahn}}, \bibinfo {author}
  {\bibfnamefont {J.~H.}\ \bibnamefont {Han}}, \bibinfo {author} {\bibfnamefont
  {K.}~\bibnamefont {Honkavaara}}, \bibinfo {author} {\bibfnamefont
  {T.}~\bibnamefont {Hott}}, \bibinfo {author} {\bibfnamefont {M.}~\bibnamefont
  {H{\"u}ning}}, \bibinfo {author} {\bibfnamefont {Y.}~\bibnamefont
  {Ivanisenko}}, \bibinfo {author} {\bibfnamefont {E.}~\bibnamefont
  {Jaeschke}}, \bibinfo {author} {\bibfnamefont {W.}~\bibnamefont {Jalmuzna}},
  \bibinfo {author} {\bibfnamefont {T.}~\bibnamefont {Jezynski}}, \bibinfo
  {author} {\bibfnamefont {R.}~\bibnamefont {Kammering}}, \bibinfo {author}
  {\bibfnamefont {V.}~\bibnamefont {Katalev}}, \bibinfo {author} {\bibfnamefont
  {K.}~\bibnamefont {Kavanagh}}, \bibinfo {author} {\bibfnamefont {E.~T.}\
  \bibnamefont {Kennedy}}, \bibinfo {author} {\bibfnamefont {S.}~\bibnamefont
  {Khodyachykh}}, \bibinfo {author} {\bibfnamefont {K.}~\bibnamefont {Klose}},
  \bibinfo {author} {\bibfnamefont {V.}~\bibnamefont {Kocharyan}}, \bibinfo
  {author} {\bibfnamefont {M.}~\bibnamefont {K{\"o}rfer}}, \bibinfo {author}
  {\bibfnamefont {M.}~\bibnamefont {Kollewe}}, \bibinfo {author} {\bibfnamefont
  {W.}~\bibnamefont {Koprek}}, \bibinfo {author} {\bibfnamefont
  {S.}~\bibnamefont {Korepanov}}, \bibinfo {author} {\bibfnamefont
  {D.}~\bibnamefont {Kostin}}, \bibinfo {author} {\bibfnamefont
  {M.}~\bibnamefont {Krassilnikov}}, \bibinfo {author} {\bibfnamefont
  {G.}~\bibnamefont {Kube}}, \bibinfo {author} {\bibfnamefont {M.}~\bibnamefont
  {Kuhlmann}}, \bibinfo {author} {\bibfnamefont {C.~L.~S.}\ \bibnamefont
  {Lewis}}, \bibinfo {author} {\bibfnamefont {L.}~\bibnamefont {Lilje}},
  \bibinfo {author} {\bibfnamefont {T.}~\bibnamefont {Limberg}}, \bibinfo
  {author} {\bibfnamefont {D.}~\bibnamefont {Lipka}}, \bibinfo {author}
  {\bibfnamefont {F.}~\bibnamefont {L{\"o}hl}}, \bibinfo {author}
  {\bibfnamefont {H.}~\bibnamefont {Luna}}, \bibinfo {author} {\bibfnamefont
  {M.}~\bibnamefont {Luong}}, \bibinfo {author} {\bibfnamefont
  {M.}~\bibnamefont {Martins}}, \bibinfo {author} {\bibfnamefont
  {M.}~\bibnamefont {Meyer}}, \bibinfo {author} {\bibfnamefont
  {P.}~\bibnamefont {Michelato}}, \bibinfo {author} {\bibfnamefont
  {V.}~\bibnamefont {Miltchev}}, \bibinfo {author} {\bibfnamefont {W.~D.}\
  \bibnamefont {M{\"o}ller}}, \bibinfo {author} {\bibfnamefont
  {L.}~\bibnamefont {Monaco}}, \bibinfo {author} {\bibfnamefont {W.~F.~O.}\
  \bibnamefont {M{\"u}ller}}, \bibinfo {author} {\bibfnamefont
  {O.}~\bibnamefont {Napieralski}}, \bibinfo {author} {\bibfnamefont
  {O.}~\bibnamefont {Napoly}}, \bibinfo {author} {\bibfnamefont
  {P.}~\bibnamefont {Nicolosi}}, \bibinfo {author} {\bibfnamefont
  {D.}~\bibnamefont {N{\"o}lle}}, \bibinfo {author} {\bibfnamefont
  {T.}~\bibnamefont {Nu{\~n}ez}}, \bibinfo {author} {\bibfnamefont
  {A.}~\bibnamefont {Oppelt}}, \bibinfo {author} {\bibfnamefont
  {C.}~\bibnamefont {Pagani}}, \bibinfo {author} {\bibfnamefont
  {R.}~\bibnamefont {Paparella}}, \bibinfo {author} {\bibfnamefont
  {N.}~\bibnamefont {Pchalek}}, \bibinfo {author} {\bibfnamefont
  {J.}~\bibnamefont {Pedregosa-Gutierrez}}, \bibinfo {author} {\bibfnamefont
  {B.}~\bibnamefont {Petersen}}, \bibinfo {author} {\bibfnamefont
  {B.}~\bibnamefont {Petrosyan}}, \bibinfo {author} {\bibfnamefont
  {G.}~\bibnamefont {Petrosyan}}, \bibinfo {author} {\bibfnamefont
  {L.}~\bibnamefont {Petrosyan}}, \bibinfo {author} {\bibfnamefont
  {J.}~\bibnamefont {Pfl{\"u}ger}}, \bibinfo {author} {\bibfnamefont
  {E.}~\bibnamefont {Pl{\"o}njes}}, \bibinfo {author} {\bibfnamefont
  {L.}~\bibnamefont {Poletto}}, \bibinfo {author} {\bibfnamefont
  {K.}~\bibnamefont {Pozniak}}, \bibinfo {author} {\bibfnamefont
  {E.}~\bibnamefont {Prat}}, \bibinfo {author} {\bibfnamefont {D.}~\bibnamefont
  {Proch}}, \bibinfo {author} {\bibfnamefont {P.}~\bibnamefont {Pucyk}},
  \bibinfo {author} {\bibfnamefont {P.}~\bibnamefont {Radcliffe}}, \bibinfo
  {author} {\bibfnamefont {H.}~\bibnamefont {Redlin}}, \bibinfo {author}
  {\bibfnamefont {K.}~\bibnamefont {Rehlich}}, \bibinfo {author} {\bibfnamefont
  {M.}~\bibnamefont {Richter}}, \bibinfo {author} {\bibfnamefont
  {M.}~\bibnamefont {Roehrs}}, \bibinfo {author} {\bibfnamefont
  {J.}~\bibnamefont {Roensch}}, \bibinfo {author} {\bibfnamefont
  {R.}~\bibnamefont {Romaniuk}}, \bibinfo {author} {\bibfnamefont
  {M.}~\bibnamefont {Ross}}, \bibinfo {author} {\bibfnamefont {J.}~\bibnamefont
  {Rossbach}}, \bibinfo {author} {\bibfnamefont {V.}~\bibnamefont {Rybnikov}},
  \bibinfo {author} {\bibfnamefont {M.}~\bibnamefont {Sachwitz}}, \bibinfo
  {author} {\bibfnamefont {E.~L.}\ \bibnamefont {Saldin}}, \bibinfo {author}
  {\bibfnamefont {W.}~\bibnamefont {Sandner}}, \bibinfo {author} {\bibfnamefont
  {H.}~\bibnamefont {Schlarb}}, \bibinfo {author} {\bibfnamefont
  {B.}~\bibnamefont {Schmidt}}, \bibinfo {author} {\bibfnamefont
  {M.}~\bibnamefont {Schmitz}}, \bibinfo {author} {\bibfnamefont
  {P.}~\bibnamefont {Schm{\"u}ser}}, \bibinfo {author} {\bibfnamefont {J.~R.}\
  \bibnamefont {Schneider}}, \bibinfo {author} {\bibfnamefont {E.~A.}\
  \bibnamefont {Schneidmiller}}, \bibinfo {author} {\bibfnamefont
  {S.}~\bibnamefont {Schnepp}}, \bibinfo {author} {\bibfnamefont
  {S.}~\bibnamefont {Schreiber}}, \bibinfo {author} {\bibfnamefont
  {M.}~\bibnamefont {Seidel}}, \bibinfo {author} {\bibfnamefont
  {D.}~\bibnamefont {Sertore}}, \bibinfo {author} {\bibfnamefont {A.~V.}\
  \bibnamefont {Shabunov}}, \bibinfo {author} {\bibfnamefont {C.}~\bibnamefont
  {Simon}}, \bibinfo {author} {\bibfnamefont {S.}~\bibnamefont {Simrock}},
  \bibinfo {author} {\bibfnamefont {E.}~\bibnamefont {Sombrowski}}, \bibinfo
  {author} {\bibfnamefont {A.~A.}\ \bibnamefont {Sorokin}}, \bibinfo {author}
  {\bibfnamefont {P.}~\bibnamefont {Spanknebel}}, \bibinfo {author}
  {\bibfnamefont {R.}~\bibnamefont {Spesyvtsev}}, \bibinfo {author}
  {\bibfnamefont {L.}~\bibnamefont {Staykov}}, \bibinfo {author} {\bibfnamefont
  {B.}~\bibnamefont {Steffen}}, \bibinfo {author} {\bibfnamefont
  {F.}~\bibnamefont {Stephan}}, \bibinfo {author} {\bibfnamefont
  {F.}~\bibnamefont {Stulle}}, \bibinfo {author} {\bibfnamefont
  {H.}~\bibnamefont {Thom}}, \bibinfo {author} {\bibfnamefont {K.}~\bibnamefont
  {Tiedtke}}, \bibinfo {author} {\bibfnamefont {M.}~\bibnamefont {Tischer}},
  \bibinfo {author} {\bibfnamefont {S.}~\bibnamefont {Toleikis}}, \bibinfo
  {author} {\bibfnamefont {R.}~\bibnamefont {Treusch}}, \bibinfo {author}
  {\bibfnamefont {D.}~\bibnamefont {Trines}}, \bibinfo {author} {\bibfnamefont
  {I.}~\bibnamefont {Tsakov}}, \bibinfo {author} {\bibfnamefont
  {E.}~\bibnamefont {Vogel}}, \bibinfo {author} {\bibfnamefont
  {T.}~\bibnamefont {Weiland}}, \bibinfo {author} {\bibfnamefont
  {H.}~\bibnamefont {Weise}}, \bibinfo {author} {\bibfnamefont
  {M.}~\bibnamefont {Wellh{\"o}fer}}, \bibinfo {author} {\bibfnamefont
  {M.}~\bibnamefont {Wendt}}, \bibinfo {author} {\bibfnamefont
  {I.}~\bibnamefont {Will}}, \bibinfo {author} {\bibfnamefont {A.}~\bibnamefont
  {Winter}}, \bibinfo {author} {\bibfnamefont {K.}~\bibnamefont {Wittenburg}},
  \bibinfo {author} {\bibfnamefont {W.}~\bibnamefont {Wurth}}, \bibinfo
  {author} {\bibfnamefont {P.}~\bibnamefont {Yeates}}, \bibinfo {author}
  {\bibfnamefont {M.~V.}\ \bibnamefont {Yurkov}}, \bibinfo {author}
  {\bibfnamefont {I.}~\bibnamefont {Zagorodnov}}, \ and\ \bibinfo {author}
  {\bibfnamefont {K.}~\bibnamefont {Zapfe}},\ }\bibfield  {title} {\emph
  {\bibinfo {title} {Operation of a free-electron laser from the extreme
  ultraviolet to the water window}},\ }\href {\doibase 10.1038/nphoton.2007.76}
  {\bibfield  {journal} {\bibinfo  {journal} {Nature Photonics}\ }\textbf
  {\bibinfo {volume} {1}},\ \bibinfo {pages} {336} (\bibinfo {year}
  {2007})}\BibitemShut {NoStop}%
\bibitem [{\citenamefont {Shintake}\ \emph {et~al.}(2008)\citenamefont
  {Shintake}, \citenamefont {Tanaka}, \citenamefont {Hara}, \citenamefont
  {Tanaka}, \citenamefont {Togawa}, \citenamefont {Yabashi}, \citenamefont
  {Otake}, \citenamefont {Asano}, \citenamefont {Bizen}, \citenamefont {Fukui},
  \citenamefont {Goto}, \citenamefont {Higashiya}, \citenamefont {Hirono},
  \citenamefont {Hosoda}, \citenamefont {Inagaki}, \citenamefont {Inoue},
  \citenamefont {Ishii}, \citenamefont {Kim}, \citenamefont {Kimura},
  \citenamefont {Kitamura}, \citenamefont {Kobayashi}, \citenamefont {Maesaka},
  \citenamefont {Masuda}, \citenamefont {Matsui}, \citenamefont {Matsushita},
  \citenamefont {Mar{\'e}chal}, \citenamefont {Nagasono}, \citenamefont
  {Ohashi}, \citenamefont {Ohata}, \citenamefont {Ohshima}, \citenamefont
  {Onoe}, \citenamefont {Shirasawa}, \citenamefont {Takagi}, \citenamefont
  {Takahashi}, \citenamefont {Takeuchi}, \citenamefont {Tamasaku},
  \citenamefont {Tanaka}, \citenamefont {Tanaka}, \citenamefont {Tanikawa},
  \citenamefont {Togashi}, \citenamefont {Wu}, \citenamefont {Yamashita},
  \citenamefont {Yanagida}, \citenamefont {Zhang}, \citenamefont {Kitamura},\
  and\ \citenamefont {Ishikawa}}]{shintake_compact_2008}%
  \BibitemOpen
  \bibfield  {author} {\bibinfo {author} {\bibfnamefont {T.}~\bibnamefont
  {Shintake}}, \bibinfo {author} {\bibfnamefont {H.}~\bibnamefont {Tanaka}},
  \bibinfo {author} {\bibfnamefont {T.}~\bibnamefont {Hara}}, \bibinfo {author}
  {\bibfnamefont {T.}~\bibnamefont {Tanaka}}, \bibinfo {author} {\bibfnamefont
  {K.}~\bibnamefont {Togawa}}, \bibinfo {author} {\bibfnamefont
  {M.}~\bibnamefont {Yabashi}}, \bibinfo {author} {\bibfnamefont
  {Y.}~\bibnamefont {Otake}}, \bibinfo {author} {\bibfnamefont
  {Y.}~\bibnamefont {Asano}}, \bibinfo {author} {\bibfnamefont
  {T.}~\bibnamefont {Bizen}}, \bibinfo {author} {\bibfnamefont
  {T.}~\bibnamefont {Fukui}}, \bibinfo {author} {\bibfnamefont
  {S.}~\bibnamefont {Goto}}, \bibinfo {author} {\bibfnamefont {A.}~\bibnamefont
  {Higashiya}}, \bibinfo {author} {\bibfnamefont {T.}~\bibnamefont {Hirono}},
  \bibinfo {author} {\bibfnamefont {N.}~\bibnamefont {Hosoda}}, \bibinfo
  {author} {\bibfnamefont {T.}~\bibnamefont {Inagaki}}, \bibinfo {author}
  {\bibfnamefont {S.}~\bibnamefont {Inoue}}, \bibinfo {author} {\bibfnamefont
  {M.}~\bibnamefont {Ishii}}, \bibinfo {author} {\bibfnamefont
  {Y.}~\bibnamefont {Kim}}, \bibinfo {author} {\bibfnamefont {H.}~\bibnamefont
  {Kimura}}, \bibinfo {author} {\bibfnamefont {M.}~\bibnamefont {Kitamura}},
  \bibinfo {author} {\bibfnamefont {T.}~\bibnamefont {Kobayashi}}, \bibinfo
  {author} {\bibfnamefont {H.}~\bibnamefont {Maesaka}}, \bibinfo {author}
  {\bibfnamefont {T.}~\bibnamefont {Masuda}}, \bibinfo {author} {\bibfnamefont
  {S.}~\bibnamefont {Matsui}}, \bibinfo {author} {\bibfnamefont
  {T.}~\bibnamefont {Matsushita}}, \bibinfo {author} {\bibfnamefont
  {X.}~\bibnamefont {Mar{\'e}chal}}, \bibinfo {author} {\bibfnamefont
  {M.}~\bibnamefont {Nagasono}}, \bibinfo {author} {\bibfnamefont
  {H.}~\bibnamefont {Ohashi}}, \bibinfo {author} {\bibfnamefont
  {T.}~\bibnamefont {Ohata}}, \bibinfo {author} {\bibfnamefont
  {T.}~\bibnamefont {Ohshima}}, \bibinfo {author} {\bibfnamefont
  {K.}~\bibnamefont {Onoe}}, \bibinfo {author} {\bibfnamefont {K.}~\bibnamefont
  {Shirasawa}}, \bibinfo {author} {\bibfnamefont {T.}~\bibnamefont {Takagi}},
  \bibinfo {author} {\bibfnamefont {S.}~\bibnamefont {Takahashi}}, \bibinfo
  {author} {\bibfnamefont {M.}~\bibnamefont {Takeuchi}}, \bibinfo {author}
  {\bibfnamefont {K.}~\bibnamefont {Tamasaku}}, \bibinfo {author}
  {\bibfnamefont {R.}~\bibnamefont {Tanaka}}, \bibinfo {author} {\bibfnamefont
  {Y.}~\bibnamefont {Tanaka}}, \bibinfo {author} {\bibfnamefont
  {T.}~\bibnamefont {Tanikawa}}, \bibinfo {author} {\bibfnamefont
  {T.}~\bibnamefont {Togashi}}, \bibinfo {author} {\bibfnamefont
  {S.}~\bibnamefont {Wu}}, \bibinfo {author} {\bibfnamefont {A.}~\bibnamefont
  {Yamashita}}, \bibinfo {author} {\bibfnamefont {K.}~\bibnamefont {Yanagida}},
  \bibinfo {author} {\bibfnamefont {C.}~\bibnamefont {Zhang}}, \bibinfo
  {author} {\bibfnamefont {H.}~\bibnamefont {Kitamura}}, \ and\ \bibinfo
  {author} {\bibfnamefont {T.}~\bibnamefont {Ishikawa}},\ }\bibfield  {title}
  {\emph {\bibinfo {title} {A compact free-electron laser for generating
  coherent radiation in the extreme ultraviolet region}},\ }\href {\doibase
  10.1038/nphoton.2008.134} {\bibfield  {journal} {\bibinfo  {journal} {Nature
  Photonics}\ }\textbf {\bibinfo {volume} {2}},\ \bibinfo {pages} {555}
  (\bibinfo {year} {2008})}\BibitemShut {NoStop}%
\bibitem [{\citenamefont {Emma}\ \emph {et~al.}(2010)\citenamefont {Emma},
  \citenamefont {Akre}, \citenamefont {Arthur}, \citenamefont {Bionta},
  \citenamefont {Bostedt}, \citenamefont {Bozek}, \citenamefont {Brachmann},
  \citenamefont {Bucksbaum}, \citenamefont {Coffee}, \citenamefont {Decker},
  \citenamefont {Ding}, \citenamefont {Dowell}, \citenamefont {Edstrom},
  \citenamefont {Fisher}, \citenamefont {Frisch}, \citenamefont {Gilevich},
  \citenamefont {Hastings}, \citenamefont {Hays}, \citenamefont {Hering},
  \citenamefont {Huang}, \citenamefont {Iverson}, \citenamefont {Loos},
  \citenamefont {Messerschmidt}, \citenamefont {Miahnahri}, \citenamefont
  {Moeller}, \citenamefont {Nuhn}, \citenamefont {Pile}, \citenamefont
  {Ratner}, \citenamefont {Rzepiela}, \citenamefont {Schultz}, \citenamefont
  {Smith}, \citenamefont {Stefan}, \citenamefont {Tompkins}, \citenamefont
  {Turner}, \citenamefont {Welch}, \citenamefont {White}, \citenamefont {Wu},
  \citenamefont {Yocky},\ and\ \citenamefont {Galayda}}]{emma_first_2010}%
  \BibitemOpen
  \bibfield  {author} {\bibinfo {author} {\bibfnamefont {P.}~\bibnamefont
  {Emma}}, \bibinfo {author} {\bibfnamefont {R.}~\bibnamefont {Akre}}, \bibinfo
  {author} {\bibfnamefont {J.}~\bibnamefont {Arthur}}, \bibinfo {author}
  {\bibfnamefont {R.}~\bibnamefont {Bionta}}, \bibinfo {author} {\bibfnamefont
  {C.}~\bibnamefont {Bostedt}}, \bibinfo {author} {\bibfnamefont
  {J.}~\bibnamefont {Bozek}}, \bibinfo {author} {\bibfnamefont
  {A.}~\bibnamefont {Brachmann}}, \bibinfo {author} {\bibfnamefont
  {P.}~\bibnamefont {Bucksbaum}}, \bibinfo {author} {\bibfnamefont
  {R.}~\bibnamefont {Coffee}}, \bibinfo {author} {\bibfnamefont {F.-J.}\
  \bibnamefont {Decker}}, \bibinfo {author} {\bibfnamefont {Y.}~\bibnamefont
  {Ding}}, \bibinfo {author} {\bibfnamefont {D.}~\bibnamefont {Dowell}},
  \bibinfo {author} {\bibfnamefont {S.}~\bibnamefont {Edstrom}}, \bibinfo
  {author} {\bibfnamefont {A.}~\bibnamefont {Fisher}}, \bibinfo {author}
  {\bibfnamefont {J.}~\bibnamefont {Frisch}}, \bibinfo {author} {\bibfnamefont
  {S.}~\bibnamefont {Gilevich}}, \bibinfo {author} {\bibfnamefont
  {J.}~\bibnamefont {Hastings}}, \bibinfo {author} {\bibfnamefont
  {G.}~\bibnamefont {Hays}}, \bibinfo {author} {\bibfnamefont {P.}~\bibnamefont
  {Hering}}, \bibinfo {author} {\bibfnamefont {Z.}~\bibnamefont {Huang}},
  \bibinfo {author} {\bibfnamefont {R.}~\bibnamefont {Iverson}}, \bibinfo
  {author} {\bibfnamefont {H.}~\bibnamefont {Loos}}, \bibinfo {author}
  {\bibfnamefont {M.}~\bibnamefont {Messerschmidt}}, \bibinfo {author}
  {\bibfnamefont {A.}~\bibnamefont {Miahnahri}}, \bibinfo {author}
  {\bibfnamefont {S.}~\bibnamefont {Moeller}}, \bibinfo {author} {\bibfnamefont
  {H.-D.}\ \bibnamefont {Nuhn}}, \bibinfo {author} {\bibfnamefont
  {G.}~\bibnamefont {Pile}}, \bibinfo {author} {\bibfnamefont {D.}~\bibnamefont
  {Ratner}}, \bibinfo {author} {\bibfnamefont {J.}~\bibnamefont {Rzepiela}},
  \bibinfo {author} {\bibfnamefont {D.}~\bibnamefont {Schultz}}, \bibinfo
  {author} {\bibfnamefont {T.}~\bibnamefont {Smith}}, \bibinfo {author}
  {\bibfnamefont {P.}~\bibnamefont {Stefan}}, \bibinfo {author} {\bibfnamefont
  {H.}~\bibnamefont {Tompkins}}, \bibinfo {author} {\bibfnamefont
  {J.}~\bibnamefont {Turner}}, \bibinfo {author} {\bibfnamefont
  {J.}~\bibnamefont {Welch}}, \bibinfo {author} {\bibfnamefont
  {W.}~\bibnamefont {White}}, \bibinfo {author} {\bibfnamefont
  {J.}~\bibnamefont {Wu}}, \bibinfo {author} {\bibfnamefont {G.}~\bibnamefont
  {Yocky}}, \ and\ \bibinfo {author} {\bibfnamefont {J.}~\bibnamefont
  {Galayda}},\ }\bibfield  {title} {\emph {\bibinfo {title} {First lasing and
  operation of an {\r a}ngstrom-wavelength free-electron laser}},\ }\href
  {\doibase 10.1038/nphoton.2010.176} {\bibfield  {journal} {\bibinfo
  {journal} {Nature Photonics}\ }\textbf {\bibinfo {volume} {4}},\ \bibinfo
  {pages} {641} (\bibinfo {year} {2010})}\BibitemShut {NoStop}%
\bibitem [{\citenamefont {Ishikawa}\ \emph {et~al.}(2012)\citenamefont
  {Ishikawa}, \citenamefont {Aoyagi}, \citenamefont {Asaka}, \citenamefont
  {Asano}, \citenamefont {Azumi}, \citenamefont {Bizen}, \citenamefont {Ego},
  \citenamefont {Fukami}, \citenamefont {Fukui}, \citenamefont {Furukawa},
  \citenamefont {Goto}, \citenamefont {Hanaki}, \citenamefont {Hara},
  \citenamefont {Hasegawa}, \citenamefont {Hatsui}, \citenamefont {Higashiya},
  \citenamefont {Hirono}, \citenamefont {Hosoda}, \citenamefont {Ishii},
  \citenamefont {Inagaki}, \citenamefont {Inubushi}, \citenamefont {Itoga},
  \citenamefont {Joti}, \citenamefont {Kago}, \citenamefont {Kameshima},
  \citenamefont {Kimura}, \citenamefont {Kirihara}, \citenamefont {Kiyomichi},
  \citenamefont {Kobayashi}, \citenamefont {Kondo}, \citenamefont {Kudo},
  \citenamefont {Maesaka}, \citenamefont {Mar{\'e}chal}, \citenamefont
  {Masuda}, \citenamefont {Matsubara}, \citenamefont {Matsumoto}, \citenamefont
  {Matsushita}, \citenamefont {Matsui}, \citenamefont {Nagasono}, \citenamefont
  {Nariyama}, \citenamefont {Ohashi}, \citenamefont {Ohata}, \citenamefont
  {Ohshima}, \citenamefont {Ono}, \citenamefont {Otake}, \citenamefont {Saji},
  \citenamefont {Sakurai}, \citenamefont {Sato}, \citenamefont {Sawada},
  \citenamefont {Seike}, \citenamefont {Shirasawa}, \citenamefont {Sugimoto},
  \citenamefont {Suzuki}, \citenamefont {Takahashi}, \citenamefont {Takebe},
  \citenamefont {Takeshita}, \citenamefont {Tamasaku}, \citenamefont {Tanaka},
  \citenamefont {Tanaka}, \citenamefont {Tanaka}, \citenamefont {Togashi},
  \citenamefont {Togawa}, \citenamefont {Tokuhisa}, \citenamefont {Tomizawa},
  \citenamefont {Tono}, \citenamefont {Wu}, \citenamefont {Yabashi},
  \citenamefont {Yamaga}, \citenamefont {Yamashita}, \citenamefont {Yanagida},
  \citenamefont {Zhang}, \citenamefont {Shintake}, \citenamefont {Kitamura},\
  and\ \citenamefont {Kumagai}}]{ishikawa_compact_2012}%
  \BibitemOpen
  \bibfield  {author} {\bibinfo {author} {\bibfnamefont {T.}~\bibnamefont
  {Ishikawa}}, \bibinfo {author} {\bibfnamefont {H.}~\bibnamefont {Aoyagi}},
  \bibinfo {author} {\bibfnamefont {T.}~\bibnamefont {Asaka}}, \bibinfo
  {author} {\bibfnamefont {Y.}~\bibnamefont {Asano}}, \bibinfo {author}
  {\bibfnamefont {N.}~\bibnamefont {Azumi}}, \bibinfo {author} {\bibfnamefont
  {T.}~\bibnamefont {Bizen}}, \bibinfo {author} {\bibfnamefont
  {H.}~\bibnamefont {Ego}}, \bibinfo {author} {\bibfnamefont {K.}~\bibnamefont
  {Fukami}}, \bibinfo {author} {\bibfnamefont {T.}~\bibnamefont {Fukui}},
  \bibinfo {author} {\bibfnamefont {Y.}~\bibnamefont {Furukawa}}, \bibinfo
  {author} {\bibfnamefont {S.}~\bibnamefont {Goto}}, \bibinfo {author}
  {\bibfnamefont {H.}~\bibnamefont {Hanaki}}, \bibinfo {author} {\bibfnamefont
  {T.}~\bibnamefont {Hara}}, \bibinfo {author} {\bibfnamefont {T.}~\bibnamefont
  {Hasegawa}}, \bibinfo {author} {\bibfnamefont {T.}~\bibnamefont {Hatsui}},
  \bibinfo {author} {\bibfnamefont {A.}~\bibnamefont {Higashiya}}, \bibinfo
  {author} {\bibfnamefont {T.}~\bibnamefont {Hirono}}, \bibinfo {author}
  {\bibfnamefont {N.}~\bibnamefont {Hosoda}}, \bibinfo {author} {\bibfnamefont
  {M.}~\bibnamefont {Ishii}}, \bibinfo {author} {\bibfnamefont
  {T.}~\bibnamefont {Inagaki}}, \bibinfo {author} {\bibfnamefont
  {Y.}~\bibnamefont {Inubushi}}, \bibinfo {author} {\bibfnamefont
  {T.}~\bibnamefont {Itoga}}, \bibinfo {author} {\bibfnamefont
  {Y.}~\bibnamefont {Joti}}, \bibinfo {author} {\bibfnamefont {M.}~\bibnamefont
  {Kago}}, \bibinfo {author} {\bibfnamefont {T.}~\bibnamefont {Kameshima}},
  \bibinfo {author} {\bibfnamefont {H.}~\bibnamefont {Kimura}}, \bibinfo
  {author} {\bibfnamefont {Y.}~\bibnamefont {Kirihara}}, \bibinfo {author}
  {\bibfnamefont {A.}~\bibnamefont {Kiyomichi}}, \bibinfo {author}
  {\bibfnamefont {T.}~\bibnamefont {Kobayashi}}, \bibinfo {author}
  {\bibfnamefont {C.}~\bibnamefont {Kondo}}, \bibinfo {author} {\bibfnamefont
  {T.}~\bibnamefont {Kudo}}, \bibinfo {author} {\bibfnamefont {H.}~\bibnamefont
  {Maesaka}}, \bibinfo {author} {\bibfnamefont {X.~M.}\ \bibnamefont
  {Mar{\'e}chal}}, \bibinfo {author} {\bibfnamefont {T.}~\bibnamefont
  {Masuda}}, \bibinfo {author} {\bibfnamefont {S.}~\bibnamefont {Matsubara}},
  \bibinfo {author} {\bibfnamefont {T.}~\bibnamefont {Matsumoto}}, \bibinfo
  {author} {\bibfnamefont {T.}~\bibnamefont {Matsushita}}, \bibinfo {author}
  {\bibfnamefont {S.}~\bibnamefont {Matsui}}, \bibinfo {author} {\bibfnamefont
  {M.}~\bibnamefont {Nagasono}}, \bibinfo {author} {\bibfnamefont
  {N.}~\bibnamefont {Nariyama}}, \bibinfo {author} {\bibfnamefont
  {H.}~\bibnamefont {Ohashi}}, \bibinfo {author} {\bibfnamefont
  {T.}~\bibnamefont {Ohata}}, \bibinfo {author} {\bibfnamefont
  {T.}~\bibnamefont {Ohshima}}, \bibinfo {author} {\bibfnamefont
  {S.}~\bibnamefont {Ono}}, \bibinfo {author} {\bibfnamefont {Y.}~\bibnamefont
  {Otake}}, \bibinfo {author} {\bibfnamefont {C.}~\bibnamefont {Saji}},
  \bibinfo {author} {\bibfnamefont {T.}~\bibnamefont {Sakurai}}, \bibinfo
  {author} {\bibfnamefont {T.}~\bibnamefont {Sato}}, \bibinfo {author}
  {\bibfnamefont {K.}~\bibnamefont {Sawada}}, \bibinfo {author} {\bibfnamefont
  {T.}~\bibnamefont {Seike}}, \bibinfo {author} {\bibfnamefont
  {K.}~\bibnamefont {Shirasawa}}, \bibinfo {author} {\bibfnamefont
  {T.}~\bibnamefont {Sugimoto}}, \bibinfo {author} {\bibfnamefont
  {S.}~\bibnamefont {Suzuki}}, \bibinfo {author} {\bibfnamefont
  {S.}~\bibnamefont {Takahashi}}, \bibinfo {author} {\bibfnamefont
  {H.}~\bibnamefont {Takebe}}, \bibinfo {author} {\bibfnamefont
  {K.}~\bibnamefont {Takeshita}}, \bibinfo {author} {\bibfnamefont
  {K.}~\bibnamefont {Tamasaku}}, \bibinfo {author} {\bibfnamefont
  {H.}~\bibnamefont {Tanaka}}, \bibinfo {author} {\bibfnamefont
  {R.}~\bibnamefont {Tanaka}}, \bibinfo {author} {\bibfnamefont
  {T.}~\bibnamefont {Tanaka}}, \bibinfo {author} {\bibfnamefont
  {T.}~\bibnamefont {Togashi}}, \bibinfo {author} {\bibfnamefont
  {K.}~\bibnamefont {Togawa}}, \bibinfo {author} {\bibfnamefont
  {A.}~\bibnamefont {Tokuhisa}}, \bibinfo {author} {\bibfnamefont
  {H.}~\bibnamefont {Tomizawa}}, \bibinfo {author} {\bibfnamefont
  {K.}~\bibnamefont {Tono}}, \bibinfo {author} {\bibfnamefont {S.}~\bibnamefont
  {Wu}}, \bibinfo {author} {\bibfnamefont {M.}~\bibnamefont {Yabashi}},
  \bibinfo {author} {\bibfnamefont {M.}~\bibnamefont {Yamaga}}, \bibinfo
  {author} {\bibfnamefont {A.}~\bibnamefont {Yamashita}}, \bibinfo {author}
  {\bibfnamefont {K.}~\bibnamefont {Yanagida}}, \bibinfo {author}
  {\bibfnamefont {C.}~\bibnamefont {Zhang}}, \bibinfo {author} {\bibfnamefont
  {T.}~\bibnamefont {Shintake}}, \bibinfo {author} {\bibfnamefont
  {H.}~\bibnamefont {Kitamura}}, \ and\ \bibinfo {author} {\bibfnamefont
  {N.}~\bibnamefont {Kumagai}},\ }\bibfield  {title} {\emph {\bibinfo {title}
  {A compact {X}-ray free-electron laser emitting in the sub-{\r a}ngstr{\"o}m
  region}},\ }\href {\doibase 10.1038/nphoton.2012.141} {\bibfield  {journal}
  {\bibinfo  {journal} {Nature Photonics}\ }\textbf {\bibinfo {volume} {6}},\
  \bibinfo {pages} {540} (\bibinfo {year} {2012})}\BibitemShut {NoStop}%
\bibitem [{\citenamefont {Allaria}\ \emph {et~al.}(2013)\citenamefont
  {Allaria}, \citenamefont {Appio}, \citenamefont {Badano}, \citenamefont
  {Barletta}, \citenamefont {Bassanese}, \citenamefont {Biedron}, \citenamefont
  {Borga}, \citenamefont {Busetto}, \citenamefont {Castronovo}, \citenamefont
  {Cinquegrana}, \citenamefont {Cleva}, \citenamefont {Cocco}, \citenamefont
  {Cornacchia}, \citenamefont {Craievich}, \citenamefont {Cudin}, \citenamefont
  {D'Auria}, \citenamefont {Dal~Forno}, \citenamefont {Danailov}, \citenamefont
  {De~Monte}, \citenamefont {De~Ninno}, \citenamefont {Delgiusto},
  \citenamefont {Demidovich}, \citenamefont {Di~Mitri}, \citenamefont
  {Diviacco}, \citenamefont {Fabris}, \citenamefont {Fabris}, \citenamefont
  {Fawley}, \citenamefont {Ferianis}, \citenamefont {Ferrari}, \citenamefont
  {Ferry}, \citenamefont {Froehlich}, \citenamefont {Furlan}, \citenamefont
  {Gaio}, \citenamefont {Gelmetti}, \citenamefont {Giannessi}, \citenamefont
  {Giannini}, \citenamefont {Gobessi}, \citenamefont {Ivanov}, \citenamefont
  {Karantzoulis}, \citenamefont {Lonza}, \citenamefont {Lutman}, \citenamefont
  {Mahieu}, \citenamefont {Milloch}, \citenamefont {Milton}, \citenamefont
  {Musardo}, \citenamefont {Nikolov}, \citenamefont {Noe}, \citenamefont
  {Parmigiani}, \citenamefont {Penco}, \citenamefont {Petronio}, \citenamefont
  {Pivetta}, \citenamefont {Predonzani}, \citenamefont {Rossi}, \citenamefont
  {Rumiz}, \citenamefont {Salom}, \citenamefont {Scafuri}, \citenamefont
  {Serpico}, \citenamefont {Sigalotti}, \citenamefont {Spampinati},
  \citenamefont {Spezzani}, \citenamefont {Svandrlik}, \citenamefont {Svetina},
  \citenamefont {Tazzari}, \citenamefont {Trovo}, \citenamefont {Umer},
  \citenamefont {Vascotto}, \citenamefont {Veronese}, \citenamefont
  {Visintini}, \citenamefont {Zaccaria}, \citenamefont {Zangrando},\ and\
  \citenamefont {Zangrando}}]{allaria_highly_2013}%
  \BibitemOpen
  \bibfield  {author} {\bibinfo {author} {\bibfnamefont {E.}~\bibnamefont
  {Allaria}}, \bibinfo {author} {\bibfnamefont {R.}~\bibnamefont {Appio}},
  \bibinfo {author} {\bibfnamefont {L.}~\bibnamefont {Badano}}, \bibinfo
  {author} {\bibfnamefont {W.}~\bibnamefont {Barletta}}, \bibinfo {author}
  {\bibfnamefont {S.}~\bibnamefont {Bassanese}}, \bibinfo {author}
  {\bibfnamefont {S.}~\bibnamefont {Biedron}}, \bibinfo {author} {\bibfnamefont
  {A.}~\bibnamefont {Borga}}, \bibinfo {author} {\bibfnamefont
  {E.}~\bibnamefont {Busetto}}, \bibinfo {author} {\bibfnamefont
  {D.}~\bibnamefont {Castronovo}}, \bibinfo {author} {\bibfnamefont
  {P.}~\bibnamefont {Cinquegrana}}, \bibinfo {author} {\bibfnamefont
  {S.}~\bibnamefont {Cleva}}, \bibinfo {author} {\bibfnamefont
  {D.}~\bibnamefont {Cocco}}, \bibinfo {author} {\bibfnamefont
  {M.}~\bibnamefont {Cornacchia}}, \bibinfo {author} {\bibfnamefont
  {P.}~\bibnamefont {Craievich}}, \bibinfo {author} {\bibfnamefont
  {I.}~\bibnamefont {Cudin}}, \bibinfo {author} {\bibfnamefont
  {G.}~\bibnamefont {D'Auria}}, \bibinfo {author} {\bibfnamefont
  {M.}~\bibnamefont {Dal~Forno}}, \bibinfo {author} {\bibfnamefont
  {M.}~\bibnamefont {Danailov}}, \bibinfo {author} {\bibfnamefont
  {R.}~\bibnamefont {De~Monte}}, \bibinfo {author} {\bibfnamefont
  {G.}~\bibnamefont {De~Ninno}}, \bibinfo {author} {\bibfnamefont
  {P.}~\bibnamefont {Delgiusto}}, \bibinfo {author} {\bibfnamefont
  {A.}~\bibnamefont {Demidovich}}, \bibinfo {author} {\bibfnamefont
  {S.}~\bibnamefont {Di~Mitri}}, \bibinfo {author} {\bibfnamefont
  {B.}~\bibnamefont {Diviacco}}, \bibinfo {author} {\bibfnamefont
  {A.}~\bibnamefont {Fabris}}, \bibinfo {author} {\bibfnamefont
  {R.}~\bibnamefont {Fabris}}, \bibinfo {author} {\bibfnamefont
  {W.}~\bibnamefont {Fawley}}, \bibinfo {author} {\bibfnamefont
  {M.}~\bibnamefont {Ferianis}}, \bibinfo {author} {\bibfnamefont
  {E.}~\bibnamefont {Ferrari}}, \bibinfo {author} {\bibfnamefont
  {S.}~\bibnamefont {Ferry}}, \bibinfo {author} {\bibfnamefont
  {L.}~\bibnamefont {Froehlich}}, \bibinfo {author} {\bibfnamefont
  {P.}~\bibnamefont {Furlan}}, \bibinfo {author} {\bibfnamefont
  {G.}~\bibnamefont {Gaio}}, \bibinfo {author} {\bibfnamefont {F.}~\bibnamefont
  {Gelmetti}}, \bibinfo {author} {\bibfnamefont {L.}~\bibnamefont {Giannessi}},
  \bibinfo {author} {\bibfnamefont {M.}~\bibnamefont {Giannini}}, \bibinfo
  {author} {\bibfnamefont {R.}~\bibnamefont {Gobessi}}, \bibinfo {author}
  {\bibfnamefont {R.}~\bibnamefont {Ivanov}}, \bibinfo {author} {\bibfnamefont
  {E.}~\bibnamefont {Karantzoulis}}, \bibinfo {author} {\bibfnamefont
  {M.}~\bibnamefont {Lonza}}, \bibinfo {author} {\bibfnamefont
  {A.}~\bibnamefont {Lutman}}, \bibinfo {author} {\bibfnamefont
  {B.}~\bibnamefont {Mahieu}}, \bibinfo {author} {\bibfnamefont
  {M.}~\bibnamefont {Milloch}}, \bibinfo {author} {\bibfnamefont
  {S.}~\bibnamefont {Milton}}, \bibinfo {author} {\bibfnamefont
  {M.}~\bibnamefont {Musardo}}, \bibinfo {author} {\bibfnamefont
  {I.}~\bibnamefont {Nikolov}}, \bibinfo {author} {\bibfnamefont
  {S.}~\bibnamefont {Noe}}, \bibinfo {author} {\bibfnamefont {F.}~\bibnamefont
  {Parmigiani}}, \bibinfo {author} {\bibfnamefont {G.}~\bibnamefont {Penco}},
  \bibinfo {author} {\bibfnamefont {M.}~\bibnamefont {Petronio}}, \bibinfo
  {author} {\bibfnamefont {L.}~\bibnamefont {Pivetta}}, \bibinfo {author}
  {\bibfnamefont {M.}~\bibnamefont {Predonzani}}, \bibinfo {author}
  {\bibfnamefont {F.}~\bibnamefont {Rossi}}, \bibinfo {author} {\bibfnamefont
  {L.}~\bibnamefont {Rumiz}}, \bibinfo {author} {\bibfnamefont
  {A.}~\bibnamefont {Salom}}, \bibinfo {author} {\bibfnamefont
  {C.}~\bibnamefont {Scafuri}}, \bibinfo {author} {\bibfnamefont
  {C.}~\bibnamefont {Serpico}}, \bibinfo {author} {\bibfnamefont
  {P.}~\bibnamefont {Sigalotti}}, \bibinfo {author} {\bibfnamefont
  {S.}~\bibnamefont {Spampinati}}, \bibinfo {author} {\bibfnamefont
  {C.}~\bibnamefont {Spezzani}}, \bibinfo {author} {\bibfnamefont
  {M.}~\bibnamefont {Svandrlik}}, \bibinfo {author} {\bibfnamefont
  {C.}~\bibnamefont {Svetina}}, \bibinfo {author} {\bibfnamefont
  {S.}~\bibnamefont {Tazzari}}, \bibinfo {author} {\bibfnamefont
  {M.}~\bibnamefont {Trovo}}, \bibinfo {author} {\bibfnamefont
  {R.}~\bibnamefont {Umer}}, \bibinfo {author} {\bibfnamefont {A.}~\bibnamefont
  {Vascotto}}, \bibinfo {author} {\bibfnamefont {M.}~\bibnamefont {Veronese}},
  \bibinfo {author} {\bibfnamefont {R.}~\bibnamefont {Visintini}}, \bibinfo
  {author} {\bibfnamefont {M.}~\bibnamefont {Zaccaria}}, \bibinfo {author}
  {\bibfnamefont {D.}~\bibnamefont {Zangrando}}, \ and\ \bibinfo {author}
  {\bibfnamefont {M.}~\bibnamefont {Zangrando}},\ }\bibfield  {title} {\emph
  {\bibinfo {title} {Highly coherent and stable pulses from the {FERMI} seeded
  free-electron laser in the extreme ultraviolet}},\ }\href {\doibase
  10.1038/nphoton.2012.233} {\bibfield  {journal} {\bibinfo  {journal} {Nature
  Photonics}\ }\textbf {\bibinfo {volume} {6}},\ \bibinfo {pages} {699}
  (\bibinfo {year} {2013})}\BibitemShut {NoStop}%
\bibitem [{\citenamefont {Pic{\'o}n}\ \emph {et~al.}(2016)\citenamefont
  {Pic{\'o}n}, \citenamefont {Lehmann}, \citenamefont {Bostedt}, \citenamefont
  {Rudenko}, \citenamefont {Marinelli}, \citenamefont {Osipov}, \citenamefont
  {Rolles}, \citenamefont {Berrah}, \citenamefont {Bomme}, \citenamefont
  {Bucher}, \citenamefont {Doumy}, \citenamefont {Erk}, \citenamefont
  {Ferguson}, \citenamefont {Gorkhover}, \citenamefont {Ho}, \citenamefont
  {Kanter}, \citenamefont {Kr{\"a}ssig}, \citenamefont {Krzywinski},
  \citenamefont {Lutman}, \citenamefont {March}, \citenamefont {Moonshiram},
  \citenamefont {Ray}, \citenamefont {Young}, \citenamefont {Pratt},\ and\
  \citenamefont {Southworth}}]{picon_hetero-site-specific_2016}%
  \BibitemOpen
  \bibfield  {author} {\bibinfo {author} {\bibfnamefont {A.}~\bibnamefont
  {Pic{\'o}n}}, \bibinfo {author} {\bibfnamefont {C.~S.}\ \bibnamefont
  {Lehmann}}, \bibinfo {author} {\bibfnamefont {C.}~\bibnamefont {Bostedt}},
  \bibinfo {author} {\bibfnamefont {A.}~\bibnamefont {Rudenko}}, \bibinfo
  {author} {\bibfnamefont {A.}~\bibnamefont {Marinelli}}, \bibinfo {author}
  {\bibfnamefont {T.}~\bibnamefont {Osipov}}, \bibinfo {author} {\bibfnamefont
  {D.}~\bibnamefont {Rolles}}, \bibinfo {author} {\bibfnamefont
  {N.}~\bibnamefont {Berrah}}, \bibinfo {author} {\bibfnamefont
  {C.}~\bibnamefont {Bomme}}, \bibinfo {author} {\bibfnamefont
  {M.}~\bibnamefont {Bucher}}, \bibinfo {author} {\bibfnamefont
  {G.}~\bibnamefont {Doumy}}, \bibinfo {author} {\bibfnamefont
  {B.}~\bibnamefont {Erk}}, \bibinfo {author} {\bibfnamefont {K.~R.}\
  \bibnamefont {Ferguson}}, \bibinfo {author} {\bibfnamefont {T.}~\bibnamefont
  {Gorkhover}}, \bibinfo {author} {\bibfnamefont {P.~J.}\ \bibnamefont {Ho}},
  \bibinfo {author} {\bibfnamefont {E.~P.}\ \bibnamefont {Kanter}}, \bibinfo
  {author} {\bibfnamefont {B.}~\bibnamefont {Kr{\"a}ssig}}, \bibinfo {author}
  {\bibfnamefont {J.}~\bibnamefont {Krzywinski}}, \bibinfo {author}
  {\bibfnamefont {A.~A.}\ \bibnamefont {Lutman}}, \bibinfo {author}
  {\bibfnamefont {A.~M.}\ \bibnamefont {March}}, \bibinfo {author}
  {\bibfnamefont {D.}~\bibnamefont {Moonshiram}}, \bibinfo {author}
  {\bibfnamefont {D.}~\bibnamefont {Ray}}, \bibinfo {author} {\bibfnamefont
  {L.}~\bibnamefont {Young}}, \bibinfo {author} {\bibfnamefont {S.~T.}\
  \bibnamefont {Pratt}}, \ and\ \bibinfo {author} {\bibfnamefont {S.~H.}\
  \bibnamefont {Southworth}},\ }\bibfield  {title} {\emph {\bibinfo {title}
  {Hetero-site-specific {X}-ray pump-probe spectroscopy for femtosecond
  intramolecular dynamics}},\ }\href {\doibase 10.1038/ncomms11652} {\bibfield
  {journal} {\bibinfo  {journal} {Nature Communications}\ }\textbf {\bibinfo
  {volume} {7}},\ \bibinfo {pages} {11652} (\bibinfo {year}
  {2016})}\BibitemShut {NoStop}%
\bibitem [{\citenamefont {Jiang}\ \emph {et~al.}(2010)\citenamefont {Jiang},
  \citenamefont {Rudenko}, \citenamefont {Herrwerth}, \citenamefont {Foucar},
  \citenamefont {Kurka}, \citenamefont {K{\"u}hnel}, \citenamefont {Lezius},
  \citenamefont {Kling}, \citenamefont {van Tilborg}, \citenamefont {Belkacem},
  \citenamefont {Ueda}, \citenamefont {D{\"u}sterer}, \citenamefont {Treusch},
  \citenamefont {Schr{\"o}ter}, \citenamefont {Moshammer},\ and\ \citenamefont
  {Ullrich}}]{jiang_ultrafast_2010}%
  \BibitemOpen
  \bibfield  {author} {\bibinfo {author} {\bibfnamefont {Y.~H.}\ \bibnamefont
  {Jiang}}, \bibinfo {author} {\bibfnamefont {A.}~\bibnamefont {Rudenko}},
  \bibinfo {author} {\bibfnamefont {O.}~\bibnamefont {Herrwerth}}, \bibinfo
  {author} {\bibfnamefont {L.}~\bibnamefont {Foucar}}, \bibinfo {author}
  {\bibfnamefont {M.}~\bibnamefont {Kurka}}, \bibinfo {author} {\bibfnamefont
  {K.~U.}\ \bibnamefont {K{\"u}hnel}}, \bibinfo {author} {\bibfnamefont
  {M.}~\bibnamefont {Lezius}}, \bibinfo {author} {\bibfnamefont {M.~F.}\
  \bibnamefont {Kling}}, \bibinfo {author} {\bibfnamefont {J.}~\bibnamefont
  {van Tilborg}}, \bibinfo {author} {\bibfnamefont {A.}~\bibnamefont
  {Belkacem}}, \bibinfo {author} {\bibfnamefont {K.}~\bibnamefont {Ueda}},
  \bibinfo {author} {\bibfnamefont {S.}~\bibnamefont {D{\"u}sterer}}, \bibinfo
  {author} {\bibfnamefont {R.}~\bibnamefont {Treusch}}, \bibinfo {author}
  {\bibfnamefont {C.~D.}\ \bibnamefont {Schr{\"o}ter}}, \bibinfo {author}
  {\bibfnamefont {R.}~\bibnamefont {Moshammer}}, \ and\ \bibinfo {author}
  {\bibfnamefont {J.}~\bibnamefont {Ullrich}},\ }\bibfield  {title} {\emph
  {\bibinfo {title} {Ultrafast {Extreme} {Ultraviolet} {Induced}
  {Isomerization} of {Acetylene} {Cations}}},\ }\href {\doibase
  10.1103/PhysRevLett.105.263002} {\bibfield  {journal} {\bibinfo  {journal}
  {Physical Review Letters}\ }\textbf {\bibinfo {volume} {105}},\ \bibinfo
  {pages} {263002} (\bibinfo {year} {2010})}\BibitemShut {NoStop}%
\bibitem [{\citenamefont {Liekhus-Schmaltz}\ \emph {et~al.}(2015)\citenamefont
  {Liekhus-Schmaltz}, \citenamefont {Tenney}, \citenamefont {Osipov},
  \citenamefont {Sanchez-Gonzalez}, \citenamefont {Berrah}, \citenamefont
  {Boll}, \citenamefont {Bomme}, \citenamefont {Bostedt}, \citenamefont
  {Bozek}, \citenamefont {Carron}, \citenamefont {Coffee}, \citenamefont
  {Devin}, \citenamefont {Erk}, \citenamefont {Ferguson}, \citenamefont
  {Field}, \citenamefont {Foucar}, \citenamefont {Frasinski}, \citenamefont
  {Glownia}, \citenamefont {G{\"u}hr}, \citenamefont {Kamalov}, \citenamefont
  {Krzywinski}, \citenamefont {Li}, \citenamefont {Marangos}, \citenamefont
  {Martinez}, \citenamefont {McFarland}, \citenamefont {Miyabe}, \citenamefont
  {Murphy}, \citenamefont {Natan}, \citenamefont {Rolles}, \citenamefont
  {Rudenko}, \citenamefont {Siano}, \citenamefont {Simpson}, \citenamefont
  {Spector}, \citenamefont {Swiggers}, \citenamefont {Walke}, \citenamefont
  {Wang}, \citenamefont {Weber}, \citenamefont {Bucksbaum},\ and\ \citenamefont
  {Petrovic}}]{liekhus-schmaltz_ultrafast_2015}%
  \BibitemOpen
  \bibfield  {author} {\bibinfo {author} {\bibfnamefont {C.~E.}\ \bibnamefont
  {Liekhus-Schmaltz}}, \bibinfo {author} {\bibfnamefont {I.}~\bibnamefont
  {Tenney}}, \bibinfo {author} {\bibfnamefont {T.}~\bibnamefont {Osipov}},
  \bibinfo {author} {\bibfnamefont {A.}~\bibnamefont {Sanchez-Gonzalez}},
  \bibinfo {author} {\bibfnamefont {N.}~\bibnamefont {Berrah}}, \bibinfo
  {author} {\bibfnamefont {R.}~\bibnamefont {Boll}}, \bibinfo {author}
  {\bibfnamefont {C.}~\bibnamefont {Bomme}}, \bibinfo {author} {\bibfnamefont
  {C.}~\bibnamefont {Bostedt}}, \bibinfo {author} {\bibfnamefont {J.~D.}\
  \bibnamefont {Bozek}}, \bibinfo {author} {\bibfnamefont {S.}~\bibnamefont
  {Carron}}, \bibinfo {author} {\bibfnamefont {R.}~\bibnamefont {Coffee}},
  \bibinfo {author} {\bibfnamefont {J.}~\bibnamefont {Devin}}, \bibinfo
  {author} {\bibfnamefont {B.}~\bibnamefont {Erk}}, \bibinfo {author}
  {\bibfnamefont {K.~R.}\ \bibnamefont {Ferguson}}, \bibinfo {author}
  {\bibfnamefont {R.~W.}\ \bibnamefont {Field}}, \bibinfo {author}
  {\bibfnamefont {L.}~\bibnamefont {Foucar}}, \bibinfo {author} {\bibfnamefont
  {L.~J.}\ \bibnamefont {Frasinski}}, \bibinfo {author} {\bibfnamefont {J.~M.}\
  \bibnamefont {Glownia}}, \bibinfo {author} {\bibfnamefont {M.}~\bibnamefont
  {G{\"u}hr}}, \bibinfo {author} {\bibfnamefont {A.}~\bibnamefont {Kamalov}},
  \bibinfo {author} {\bibfnamefont {J.}~\bibnamefont {Krzywinski}}, \bibinfo
  {author} {\bibfnamefont {H.}~\bibnamefont {Li}}, \bibinfo {author}
  {\bibfnamefont {J.~P.}\ \bibnamefont {Marangos}}, \bibinfo {author}
  {\bibfnamefont {T.~J.}\ \bibnamefont {Martinez}}, \bibinfo {author}
  {\bibfnamefont {B.~K.}\ \bibnamefont {McFarland}}, \bibinfo {author}
  {\bibfnamefont {S.}~\bibnamefont {Miyabe}}, \bibinfo {author} {\bibfnamefont
  {B.}~\bibnamefont {Murphy}}, \bibinfo {author} {\bibfnamefont
  {A.}~\bibnamefont {Natan}}, \bibinfo {author} {\bibfnamefont
  {D.}~\bibnamefont {Rolles}}, \bibinfo {author} {\bibfnamefont
  {A.}~\bibnamefont {Rudenko}}, \bibinfo {author} {\bibfnamefont
  {M.}~\bibnamefont {Siano}}, \bibinfo {author} {\bibfnamefont {E.~R.}\
  \bibnamefont {Simpson}}, \bibinfo {author} {\bibfnamefont {L.}~\bibnamefont
  {Spector}}, \bibinfo {author} {\bibfnamefont {M.}~\bibnamefont {Swiggers}},
  \bibinfo {author} {\bibfnamefont {D.}~\bibnamefont {Walke}}, \bibinfo
  {author} {\bibfnamefont {S.}~\bibnamefont {Wang}}, \bibinfo {author}
  {\bibfnamefont {T.}~\bibnamefont {Weber}}, \bibinfo {author} {\bibfnamefont
  {P.~H.}\ \bibnamefont {Bucksbaum}}, \ and\ \bibinfo {author} {\bibfnamefont
  {V.~S.}\ \bibnamefont {Petrovic}},\ }\bibfield  {title} {\emph {\bibinfo
  {title} {Ultrafast isomerization initiated by {X}-ray core ionization}},\
  }\href {\doibase 10.1038/ncomms9199} {\bibfield  {journal} {\bibinfo
  {journal} {Nature Communications}\ }\textbf {\bibinfo {volume} {6}},\
  \bibinfo {pages} {8199} (\bibinfo {year} {2015})}\BibitemShut {NoStop}%
\bibitem [{\citenamefont {Erk}\ \emph {et~al.}(2014)\citenamefont {Erk},
  \citenamefont {Boll}, \citenamefont {Trippel}, \citenamefont {Anielski},
  \citenamefont {Foucar}, \citenamefont {Rudek}, \citenamefont {Epp},
  \citenamefont {Coffee}, \citenamefont {Carron}, \citenamefont {Schorb},
  \citenamefont {Ferguson}, \citenamefont {Swiggers}, \citenamefont {Bozek},
  \citenamefont {Simon}, \citenamefont {Marchenko}, \citenamefont {K{\"u}pper},
  \citenamefont {Schlichting}, \citenamefont {Ullrich}, \citenamefont
  {Bostedt}, \citenamefont {Rolles},\ and\ \citenamefont
  {Rudenko}}]{erk_imaging_2014}%
  \BibitemOpen
  \bibfield  {author} {\bibinfo {author} {\bibfnamefont {B.}~\bibnamefont
  {Erk}}, \bibinfo {author} {\bibfnamefont {R.}~\bibnamefont {Boll}}, \bibinfo
  {author} {\bibfnamefont {S.}~\bibnamefont {Trippel}}, \bibinfo {author}
  {\bibfnamefont {D.}~\bibnamefont {Anielski}}, \bibinfo {author}
  {\bibfnamefont {L.}~\bibnamefont {Foucar}}, \bibinfo {author} {\bibfnamefont
  {B.}~\bibnamefont {Rudek}}, \bibinfo {author} {\bibfnamefont {S.~W.}\
  \bibnamefont {Epp}}, \bibinfo {author} {\bibfnamefont {R.}~\bibnamefont
  {Coffee}}, \bibinfo {author} {\bibfnamefont {S.}~\bibnamefont {Carron}},
  \bibinfo {author} {\bibfnamefont {S.}~\bibnamefont {Schorb}}, \bibinfo
  {author} {\bibfnamefont {K.~R.}\ \bibnamefont {Ferguson}}, \bibinfo {author}
  {\bibfnamefont {M.}~\bibnamefont {Swiggers}}, \bibinfo {author}
  {\bibfnamefont {J.~D.}\ \bibnamefont {Bozek}}, \bibinfo {author}
  {\bibfnamefont {M.}~\bibnamefont {Simon}}, \bibinfo {author} {\bibfnamefont
  {T.}~\bibnamefont {Marchenko}}, \bibinfo {author} {\bibfnamefont
  {J.}~\bibnamefont {K{\"u}pper}}, \bibinfo {author} {\bibfnamefont
  {I.}~\bibnamefont {Schlichting}}, \bibinfo {author} {\bibfnamefont
  {J.}~\bibnamefont {Ullrich}}, \bibinfo {author} {\bibfnamefont
  {C.}~\bibnamefont {Bostedt}}, \bibinfo {author} {\bibfnamefont
  {D.}~\bibnamefont {Rolles}}, \ and\ \bibinfo {author} {\bibfnamefont
  {A.}~\bibnamefont {Rudenko}},\ }\bibfield  {title} {\emph {\bibinfo {title}
  {Imaging charge transfer in iodomethane upon x-ray photoabsorption}},\ }\href
  {\doibase 10.1126/science.1253607} {\bibfield  {journal} {\bibinfo  {journal}
  {Science}\ }\textbf {\bibinfo {volume} {345}},\ \bibinfo {pages} {288}
  (\bibinfo {year} {2014})}\BibitemShut {NoStop}%
\bibitem [{\citenamefont {Boll}\ \emph {et~al.}(2016)\citenamefont {Boll},
  \citenamefont {Erk}, \citenamefont {Coffee}, \citenamefont {Trippel},
  \citenamefont {Kierspel}, \citenamefont {Bomme}, \citenamefont {Bozek},
  \citenamefont {Burkett}, \citenamefont {Carron}, \citenamefont {Ferguson},
  \citenamefont {Foucar}, \citenamefont {K{\"u}pper}, \citenamefont
  {Marchenko}, \citenamefont {Miron}, \citenamefont {Patanen}, \citenamefont
  {Osipov}, \citenamefont {Schorb}, \citenamefont {Simon}, \citenamefont
  {Swiggers}, \citenamefont {Techert}, \citenamefont {Ueda}, \citenamefont
  {Bostedt}, \citenamefont {Rolles},\ and\ \citenamefont
  {Rudenko}}]{boll_charge_2016}%
  \BibitemOpen
  \bibfield  {author} {\bibinfo {author} {\bibfnamefont {R.}~\bibnamefont
  {Boll}}, \bibinfo {author} {\bibfnamefont {B.}~\bibnamefont {Erk}}, \bibinfo
  {author} {\bibfnamefont {R.}~\bibnamefont {Coffee}}, \bibinfo {author}
  {\bibfnamefont {S.}~\bibnamefont {Trippel}}, \bibinfo {author} {\bibfnamefont
  {T.}~\bibnamefont {Kierspel}}, \bibinfo {author} {\bibfnamefont
  {C.}~\bibnamefont {Bomme}}, \bibinfo {author} {\bibfnamefont {J.~D.}\
  \bibnamefont {Bozek}}, \bibinfo {author} {\bibfnamefont {M.}~\bibnamefont
  {Burkett}}, \bibinfo {author} {\bibfnamefont {S.}~\bibnamefont {Carron}},
  \bibinfo {author} {\bibfnamefont {K.~R.}\ \bibnamefont {Ferguson}}, \bibinfo
  {author} {\bibfnamefont {L.}~\bibnamefont {Foucar}}, \bibinfo {author}
  {\bibfnamefont {J.}~\bibnamefont {K{\"u}pper}}, \bibinfo {author}
  {\bibfnamefont {T.}~\bibnamefont {Marchenko}}, \bibinfo {author}
  {\bibfnamefont {C.}~\bibnamefont {Miron}}, \bibinfo {author} {\bibfnamefont
  {M.}~\bibnamefont {Patanen}}, \bibinfo {author} {\bibfnamefont
  {T.}~\bibnamefont {Osipov}}, \bibinfo {author} {\bibfnamefont
  {S.}~\bibnamefont {Schorb}}, \bibinfo {author} {\bibfnamefont
  {M.}~\bibnamefont {Simon}}, \bibinfo {author} {\bibfnamefont
  {M.}~\bibnamefont {Swiggers}}, \bibinfo {author} {\bibfnamefont
  {S.}~\bibnamefont {Techert}}, \bibinfo {author} {\bibfnamefont
  {K.}~\bibnamefont {Ueda}}, \bibinfo {author} {\bibfnamefont {C.}~\bibnamefont
  {Bostedt}}, \bibinfo {author} {\bibfnamefont {D.}~\bibnamefont {Rolles}}, \
  and\ \bibinfo {author} {\bibfnamefont {A.}~\bibnamefont {Rudenko}},\
  }\bibfield  {title} {\emph {\bibinfo {title} {Charge transfer in dissociating
  iodomethane and fluoromethane molecules ionized by intense femtosecond
  {X}-ray pulses}},\ }\href {\doibase 10.1063/1.4944344} {\bibfield  {journal}
  {\bibinfo  {journal} {Structural Dynamics}\ }\textbf {\bibinfo {volume}
  {3}},\ \bibinfo {pages} {043207} (\bibinfo {year} {2016})}\BibitemShut
  {NoStop}%
\bibitem [{\citenamefont {Schnorr}\ \emph {et~al.}(2013)\citenamefont
  {Schnorr}, \citenamefont {Senftleben}, \citenamefont {Kurka}, \citenamefont
  {Rudenko}, \citenamefont {Foucar}, \citenamefont {Schmid}, \citenamefont
  {Broska}, \citenamefont {Pfeifer}, \citenamefont {Meyer}, \citenamefont
  {Anielski}, \citenamefont {Boll}, \citenamefont {Rolles}, \citenamefont
  {K{\"u}bel}, \citenamefont {Kling}, \citenamefont {Jiang}, \citenamefont
  {Mondal}, \citenamefont {Tachibana}, \citenamefont {Ueda}, \citenamefont
  {Marchenko}, \citenamefont {Simon}, \citenamefont {Brenner}, \citenamefont
  {Treusch}, \citenamefont {Scheit}, \citenamefont {Averbukh}, \citenamefont
  {Ullrich}, \citenamefont {Schr{\"o}ter},\ and\ \citenamefont
  {Moshammer}}]{schnorr_time-resolved_2013}%
  \BibitemOpen
  \bibfield  {author} {\bibinfo {author} {\bibfnamefont {K.}~\bibnamefont
  {Schnorr}}, \bibinfo {author} {\bibfnamefont {A.}~\bibnamefont {Senftleben}},
  \bibinfo {author} {\bibfnamefont {M.}~\bibnamefont {Kurka}}, \bibinfo
  {author} {\bibfnamefont {A.}~\bibnamefont {Rudenko}}, \bibinfo {author}
  {\bibfnamefont {L.}~\bibnamefont {Foucar}}, \bibinfo {author} {\bibfnamefont
  {G.}~\bibnamefont {Schmid}}, \bibinfo {author} {\bibfnamefont
  {A.}~\bibnamefont {Broska}}, \bibinfo {author} {\bibfnamefont
  {T.}~\bibnamefont {Pfeifer}}, \bibinfo {author} {\bibfnamefont
  {K.}~\bibnamefont {Meyer}}, \bibinfo {author} {\bibfnamefont
  {D.}~\bibnamefont {Anielski}}, \bibinfo {author} {\bibfnamefont
  {R.}~\bibnamefont {Boll}}, \bibinfo {author} {\bibfnamefont {D.}~\bibnamefont
  {Rolles}}, \bibinfo {author} {\bibfnamefont {M.}~\bibnamefont {K{\"u}bel}},
  \bibinfo {author} {\bibfnamefont {M.~F.}\ \bibnamefont {Kling}}, \bibinfo
  {author} {\bibfnamefont {Y.~H.}\ \bibnamefont {Jiang}}, \bibinfo {author}
  {\bibfnamefont {S.}~\bibnamefont {Mondal}}, \bibinfo {author} {\bibfnamefont
  {T.}~\bibnamefont {Tachibana}}, \bibinfo {author} {\bibfnamefont
  {K.}~\bibnamefont {Ueda}}, \bibinfo {author} {\bibfnamefont {T.}~\bibnamefont
  {Marchenko}}, \bibinfo {author} {\bibfnamefont {M.}~\bibnamefont {Simon}},
  \bibinfo {author} {\bibfnamefont {G.}~\bibnamefont {Brenner}}, \bibinfo
  {author} {\bibfnamefont {R.}~\bibnamefont {Treusch}}, \bibinfo {author}
  {\bibfnamefont {S.}~\bibnamefont {Scheit}}, \bibinfo {author} {\bibfnamefont
  {V.}~\bibnamefont {Averbukh}}, \bibinfo {author} {\bibfnamefont
  {J.}~\bibnamefont {Ullrich}}, \bibinfo {author} {\bibfnamefont {C.~D.}\
  \bibnamefont {Schr{\"o}ter}}, \ and\ \bibinfo {author} {\bibfnamefont
  {R.}~\bibnamefont {Moshammer}},\ }\bibfield  {title} {\emph {\bibinfo {title}
  {Time-{Resolved} {Measurement} of {Interatomic} {Coulombic} {Decay} in
  {Ne}$_{\textrm{2}}$}},\ }\href {\doibase 10.1103/PhysRevLett.111.093402}
  {\bibfield  {journal} {\bibinfo  {journal} {Physical Review Letters}\
  }\textbf {\bibinfo {volume} {111}},\ \bibinfo {pages} {093402} (\bibinfo
  {year} {2013})}\BibitemShut {NoStop}%
\bibitem [{\citenamefont {L{\'e}gar{\'e}}\ \emph
  {et~al.}(2005{\natexlab{b}})\citenamefont {L{\'e}gar{\'e}}, \citenamefont
  {Lee}, \citenamefont {Litvinyuk}, \citenamefont {Dooley}, \citenamefont
  {Wesolowski}, \citenamefont {Bunker}, \citenamefont {Dombi}, \citenamefont
  {Krausz}, \citenamefont {Bandrauk}, \citenamefont {Villeneuve},\ and\
  \citenamefont {Corkum}}]{legare_laser_2005}%
  \BibitemOpen
  \bibfield  {author} {\bibinfo {author} {\bibfnamefont {F.}~\bibnamefont
  {L{\'e}gar{\'e}}}, \bibinfo {author} {\bibfnamefont {K.}~\bibnamefont {Lee}},
  \bibinfo {author} {\bibfnamefont {I.}~\bibnamefont {Litvinyuk}}, \bibinfo
  {author} {\bibfnamefont {P.}~\bibnamefont {Dooley}}, \bibinfo {author}
  {\bibfnamefont {S.}~\bibnamefont {Wesolowski}}, \bibinfo {author}
  {\bibfnamefont {P.}~\bibnamefont {Bunker}}, \bibinfo {author} {\bibfnamefont
  {P.}~\bibnamefont {Dombi}}, \bibinfo {author} {\bibfnamefont
  {F.}~\bibnamefont {Krausz}}, \bibinfo {author} {\bibfnamefont
  {A.}~\bibnamefont {Bandrauk}}, \bibinfo {author} {\bibfnamefont
  {D.}~\bibnamefont {Villeneuve}}, \ and\ \bibinfo {author} {\bibfnamefont
  {P.}~\bibnamefont {Corkum}},\ }\bibfield  {title} {\emph {\bibinfo {title}
  {Laser {Coulomb}-explosion imaging of small molecules}},\ }\href {\doibase
  10.1103/PhysRevA.71.013415} {\bibfield  {journal} {\bibinfo  {journal}
  {Physical Review A}\ }\textbf {\bibinfo {volume} {71}} (\bibinfo {year}
  {2005}{\natexlab{b}}),\ 10.1103/PhysRevA.71.013415},\ \bibinfo {note}
  {00050}\BibitemShut {NoStop}%
\bibitem [{\citenamefont {L{\'e}gar{\'e}}\ \emph {et~al.}(2006)\citenamefont
  {L{\'e}gar{\'e}}, \citenamefont {Lee}, \citenamefont {Bandrauk},
  \citenamefont {Villeneuve},\ and\ \citenamefont
  {Corkum}}]{legare_laser_2006}%
  \BibitemOpen
  \bibfield  {author} {\bibinfo {author} {\bibfnamefont {F.}~\bibnamefont
  {L{\'e}gar{\'e}}}, \bibinfo {author} {\bibfnamefont {K.~F.}\ \bibnamefont
  {Lee}}, \bibinfo {author} {\bibfnamefont {A.~D.}\ \bibnamefont {Bandrauk}},
  \bibinfo {author} {\bibfnamefont {D.~M.}\ \bibnamefont {Villeneuve}}, \ and\
  \bibinfo {author} {\bibfnamefont {P.~B.}\ \bibnamefont {Corkum}},\ }\bibfield
   {title} {\emph {\bibinfo {title} {Laser {Coulomb} explosion imaging for
  probing ultra-fast molecular dynamics}},\ }\href {\doibase
  10.1088/0953-4075/39/13/S23} {\bibfield  {journal} {\bibinfo  {journal}
  {Journal of Physics B: Atomic, Molecular and Optical Physics}\ }\textbf
  {\bibinfo {volume} {39}},\ \bibinfo {pages} {S503} (\bibinfo {year}
  {2006})}\BibitemShut {NoStop}%
\bibitem [{\citenamefont {Eppink}\ and\ \citenamefont
  {Parker}(1998)}]{eppink_methyl_1998}%
  \BibitemOpen
  \bibfield  {author} {\bibinfo {author} {\bibfnamefont {A.}~\bibnamefont
  {Eppink}}\ and\ \bibinfo {author} {\bibfnamefont {D.~H.}\ \bibnamefont
  {Parker}},\ }\bibfield  {title} {\emph {\bibinfo {title} {Methyl iodide
  {A}-band decomposition study by photofragment velocity imaging}},\ }\href
  {https://notendur.hi.is/agust/rannsoknir/papers/CH3X/CH3I/jcp109-4758-98.pdf}
  {\bibfield  {journal} {\bibinfo  {journal} {The Journal of Chemical Physics}\
  }\textbf {\bibinfo {volume} {109}},\ \bibinfo {pages} {4758} (\bibinfo {year}
  {1998})}\BibitemShut {NoStop}%
\bibitem [{\citenamefont {de~Nalda}\ \emph {et~al.}(2008)\citenamefont
  {de~Nalda}, \citenamefont {Dur{\'a}}, \citenamefont {Garc{\'i}a-Vela},
  \citenamefont {Izquierdo}, \citenamefont {Gonz{\'a}lez-V{\'a}zquez},\ and\
  \citenamefont {Ba{\~n}ares}}]{de_nalda_detailed_2008}%
  \BibitemOpen
  \bibfield  {author} {\bibinfo {author} {\bibfnamefont {R.}~\bibnamefont
  {de~Nalda}}, \bibinfo {author} {\bibfnamefont {J.}~\bibnamefont {Dur{\'a}}},
  \bibinfo {author} {\bibfnamefont {A.}~\bibnamefont {Garc{\'i}a-Vela}},
  \bibinfo {author} {\bibfnamefont {J.~G.}\ \bibnamefont {Izquierdo}}, \bibinfo
  {author} {\bibfnamefont {J.}~\bibnamefont {Gonz{\'a}lez-V{\'a}zquez}}, \ and\
  \bibinfo {author} {\bibfnamefont {L.}~\bibnamefont {Ba{\~n}ares}},\
  }\bibfield  {title} {\emph {\bibinfo {title} {A detailed experimental and
  theoretical study of the femtosecond {A}-band photodissociation of
  {CH}$_{\textrm{3}}${I}}},\ }\href {\doibase 10.1063/1.2943198} {\bibfield
  {journal} {\bibinfo  {journal} {The Journal of Chemical Physics}\ }\textbf
  {\bibinfo {volume} {128}},\ \bibinfo {pages} {244309} (\bibinfo {year}
  {2008})}\BibitemShut {NoStop}%
\bibitem [{\citenamefont {Rubio-Lago}\ \emph {et~al.}(2009)\citenamefont
  {Rubio-Lago}, \citenamefont {Garc{\'i}a-Vela}, \citenamefont {Arregui},
  \citenamefont {Amaral},\ and\ \citenamefont
  {Ba{\~n}ares}}]{rubio-lago_photodissociation_2009}%
  \BibitemOpen
  \bibfield  {author} {\bibinfo {author} {\bibfnamefont {L.}~\bibnamefont
  {Rubio-Lago}}, \bibinfo {author} {\bibfnamefont {A.}~\bibnamefont
  {Garc{\'i}a-Vela}}, \bibinfo {author} {\bibfnamefont {A.}~\bibnamefont
  {Arregui}}, \bibinfo {author} {\bibfnamefont {G.~A.}\ \bibnamefont {Amaral}},
  \ and\ \bibinfo {author} {\bibfnamefont {L.}~\bibnamefont {Ba{\~n}ares}},\
  }\bibfield  {title} {\emph {\bibinfo {title} {The photodissociation of
  {CH}$_{\textrm{3}}${I} in the red edge of the {A}-band: {Comparison} between
  slice imaging experiments and multisurface wave packet calculations}},\
  }\href {\doibase 10.1063/1.3257692} {\bibfield  {journal} {\bibinfo
  {journal} {The Journal of Chemical Physics}\ }\textbf {\bibinfo {volume}
  {131}},\ \bibinfo {pages} {174309} (\bibinfo {year} {2009})}\BibitemShut
  {NoStop}%
\bibitem [{\citenamefont {Murdock}\ \emph {et~al.}(2012)\citenamefont
  {Murdock}, \citenamefont {Crow}, \citenamefont {Ritchie},\ and\ \citenamefont
  {Ashfold}}]{murdock_uv_2012}%
  \BibitemOpen
  \bibfield  {author} {\bibinfo {author} {\bibfnamefont {D.}~\bibnamefont
  {Murdock}}, \bibinfo {author} {\bibfnamefont {M.~B.}\ \bibnamefont {Crow}},
  \bibinfo {author} {\bibfnamefont {G.~A.~D.}\ \bibnamefont {Ritchie}}, \ and\
  \bibinfo {author} {\bibfnamefont {M.~N.~R.}\ \bibnamefont {Ashfold}},\
  }\bibfield  {title} {\emph {\bibinfo {title} {{UV} photodissociation dynamics
  of iodobenzene: {Effects} of fluorination}},\ }\href {\doibase
  10.1063/1.3696892} {\bibfield  {journal} {\bibinfo  {journal} {The Journal of
  Chemical Physics}\ }\textbf {\bibinfo {volume} {136}},\ \bibinfo {pages}
  {124313} (\bibinfo {year} {2012})}\BibitemShut {NoStop}%
\bibitem [{\citenamefont {Sage}\ \emph {et~al.}(2011)\citenamefont {Sage},
  \citenamefont {Oliver}, \citenamefont {Murdock}, \citenamefont {Crow},
  \citenamefont {Ritchie}, \citenamefont {Harvey},\ and\ \citenamefont
  {Ashfold}}]{sage_n*_2011}%
  \BibitemOpen
  \bibfield  {author} {\bibinfo {author} {\bibfnamefont {A.~G.}\ \bibnamefont
  {Sage}}, \bibinfo {author} {\bibfnamefont {T.~A.~A.}\ \bibnamefont {Oliver}},
  \bibinfo {author} {\bibfnamefont {D.}~\bibnamefont {Murdock}}, \bibinfo
  {author} {\bibfnamefont {M.~B.}\ \bibnamefont {Crow}}, \bibinfo {author}
  {\bibfnamefont {G.~A.~D.}\ \bibnamefont {Ritchie}}, \bibinfo {author}
  {\bibfnamefont {J.~N.}\ \bibnamefont {Harvey}}, \ and\ \bibinfo {author}
  {\bibfnamefont {M.~N.~R.}\ \bibnamefont {Ashfold}},\ }\bibfield  {title}
  {\emph {\bibinfo {title} {n$\sigma$* and $\pi$$\sigma$* excited states in
  aryl halide photochemistry: a comprehensive study of the {UV}
  photodissociation dynamics of iodobenzene}},\ }\href {\doibase
  10.1039/c0cp02390f} {\bibfield  {journal} {\bibinfo  {journal} {Physical
  Chemistry Chemical Physics}\ }\textbf {\bibinfo {volume} {13}},\ \bibinfo
  {pages} {8075} (\bibinfo {year} {2011})}\BibitemShut {NoStop}%
\bibitem [{\citenamefont {Yeh}(1993)}]{yeh_atomic_1993}%
  \BibitemOpen
  \bibfield  {author} {\bibinfo {author} {\bibfnamefont {J.-J.}\ \bibnamefont
  {Yeh}},\ }\href {https://vuo.elettra.eu/services/elements/WebElements.html}
  {\emph {\bibinfo {title} {Atomic calculation of photoionization
  cross-sections and asymmetry parameters}}}\ (\bibinfo  {publisher} {Gordon \&
  Breach Science, Publishers},\ \bibinfo {address} {Langhorne, PA},\ \bibinfo
  {year} {1993})\BibitemShut {NoStop}%
\bibitem [{\citenamefont {Str{\"u}der}\ \emph {et~al.}(2010)\citenamefont
  {Str{\"u}der}, \citenamefont {Epp}, \citenamefont {Rolles}, \citenamefont
  {Hartmann}, \citenamefont {Holl}, \citenamefont {Lutz}, \citenamefont
  {Soltau}, \citenamefont {Eckart}, \citenamefont {Reich}, \citenamefont
  {Heinzinger}, \citenamefont {Thamm}, \citenamefont {Rudenko}, \citenamefont
  {Krasniqi}, \citenamefont {K{\"u}hnel}, \citenamefont {Bauer}, \citenamefont
  {Schr{\"o}ter}, \citenamefont {Moshammer}, \citenamefont {Techert},
  \citenamefont {Miessner}, \citenamefont {Porro}, \citenamefont {H{\"o}lker},
  \citenamefont {Meidinger}, \citenamefont {Kimmel}, \citenamefont
  {Andritschke}, \citenamefont {Schopper}, \citenamefont {Weidenspointner},
  \citenamefont {Ziegler}, \citenamefont {Pietschner}, \citenamefont
  {Herrmann}, \citenamefont {Pietsch}, \citenamefont {Walenta}, \citenamefont
  {Leitenberger}, \citenamefont {Bostedt}, \citenamefont {M{\"o}ller},
  \citenamefont {Rupp}, \citenamefont {Adolph}, \citenamefont {Graafsma},
  \citenamefont {Hirsemann}, \citenamefont {G{\"a}rtner}, \citenamefont
  {Richter}, \citenamefont {Foucar}, \citenamefont {Shoeman}, \citenamefont
  {Schlichting},\ and\ \citenamefont {Ullrich}}]{struder_large-format_2010}%
  \BibitemOpen
  \bibfield  {author} {\bibinfo {author} {\bibfnamefont {L.}~\bibnamefont
  {Str{\"u}der}}, \bibinfo {author} {\bibfnamefont {S.}~\bibnamefont {Epp}},
  \bibinfo {author} {\bibfnamefont {D.}~\bibnamefont {Rolles}}, \bibinfo
  {author} {\bibfnamefont {R.}~\bibnamefont {Hartmann}}, \bibinfo {author}
  {\bibfnamefont {P.}~\bibnamefont {Holl}}, \bibinfo {author} {\bibfnamefont
  {G.}~\bibnamefont {Lutz}}, \bibinfo {author} {\bibfnamefont {H.}~\bibnamefont
  {Soltau}}, \bibinfo {author} {\bibfnamefont {R.}~\bibnamefont {Eckart}},
  \bibinfo {author} {\bibfnamefont {C.}~\bibnamefont {Reich}}, \bibinfo
  {author} {\bibfnamefont {K.}~\bibnamefont {Heinzinger}}, \bibinfo {author}
  {\bibfnamefont {C.}~\bibnamefont {Thamm}}, \bibinfo {author} {\bibfnamefont
  {A.}~\bibnamefont {Rudenko}}, \bibinfo {author} {\bibfnamefont
  {F.}~\bibnamefont {Krasniqi}}, \bibinfo {author} {\bibfnamefont {K.-U.}\
  \bibnamefont {K{\"u}hnel}}, \bibinfo {author} {\bibfnamefont
  {C.}~\bibnamefont {Bauer}}, \bibinfo {author} {\bibfnamefont {C.-D.}\
  \bibnamefont {Schr{\"o}ter}}, \bibinfo {author} {\bibfnamefont
  {R.}~\bibnamefont {Moshammer}}, \bibinfo {author} {\bibfnamefont
  {S.}~\bibnamefont {Techert}}, \bibinfo {author} {\bibfnamefont
  {D.}~\bibnamefont {Miessner}}, \bibinfo {author} {\bibfnamefont
  {M.}~\bibnamefont {Porro}}, \bibinfo {author} {\bibfnamefont
  {O.}~\bibnamefont {H{\"o}lker}}, \bibinfo {author} {\bibfnamefont
  {N.}~\bibnamefont {Meidinger}}, \bibinfo {author} {\bibfnamefont
  {N.}~\bibnamefont {Kimmel}}, \bibinfo {author} {\bibfnamefont
  {R.}~\bibnamefont {Andritschke}}, \bibinfo {author} {\bibfnamefont
  {F.}~\bibnamefont {Schopper}}, \bibinfo {author} {\bibfnamefont
  {G.}~\bibnamefont {Weidenspointner}}, \bibinfo {author} {\bibfnamefont
  {A.}~\bibnamefont {Ziegler}}, \bibinfo {author} {\bibfnamefont
  {D.}~\bibnamefont {Pietschner}}, \bibinfo {author} {\bibfnamefont
  {S.}~\bibnamefont {Herrmann}}, \bibinfo {author} {\bibfnamefont
  {U.}~\bibnamefont {Pietsch}}, \bibinfo {author} {\bibfnamefont
  {A.}~\bibnamefont {Walenta}}, \bibinfo {author} {\bibfnamefont
  {W.}~\bibnamefont {Leitenberger}}, \bibinfo {author} {\bibfnamefont
  {C.}~\bibnamefont {Bostedt}}, \bibinfo {author} {\bibfnamefont
  {T.}~\bibnamefont {M{\"o}ller}}, \bibinfo {author} {\bibfnamefont
  {D.}~\bibnamefont {Rupp}}, \bibinfo {author} {\bibfnamefont {M.}~\bibnamefont
  {Adolph}}, \bibinfo {author} {\bibfnamefont {H.}~\bibnamefont {Graafsma}},
  \bibinfo {author} {\bibfnamefont {H.}~\bibnamefont {Hirsemann}}, \bibinfo
  {author} {\bibfnamefont {K.}~\bibnamefont {G{\"a}rtner}}, \bibinfo {author}
  {\bibfnamefont {R.}~\bibnamefont {Richter}}, \bibinfo {author} {\bibfnamefont
  {L.}~\bibnamefont {Foucar}}, \bibinfo {author} {\bibfnamefont {R.~L.}\
  \bibnamefont {Shoeman}}, \bibinfo {author} {\bibfnamefont {I.}~\bibnamefont
  {Schlichting}}, \ and\ \bibinfo {author} {\bibfnamefont {J.}~\bibnamefont
  {Ullrich}},\ }\bibfield  {title} {\emph {\bibinfo {title} {Large-format,
  high-speed, {X}-ray {pnCCDs} combined with electron and ion imaging
  spectrometers in a multipurpose chamber for experiments at 4th generation
  light sources}},\ }\href {\doibase 10.1016/j.nima.2009.12.053} {\bibfield
  {journal} {\bibinfo  {journal} {Nuclear Instruments and Methods in Physics
  Research A}\ }\textbf {\bibinfo {volume} {614}},\ \bibinfo {pages} {483}
  (\bibinfo {year} {2010})}\BibitemShut {NoStop}%
\bibitem [{\citenamefont {Erk}\ \emph {et~al.}(2017)\citenamefont {Erk},
  \citenamefont {M{\"u}ller}, \citenamefont {C{\'e}dric}, \citenamefont {Boll},
  \citenamefont {Brenner}, \citenamefont {Chapman}, \citenamefont {Correa},
  \citenamefont {Dachraoui}, \citenamefont {D{\"u}sterer}, \citenamefont
  {Dziarzhytski}, \citenamefont {Eisebitt}, \citenamefont {Feldhaus},
  \citenamefont {Graafsma}, \citenamefont {Grunewald}, \citenamefont
  {Gumprecht}, \citenamefont {Hartmann}, \citenamefont {Hauser}, \citenamefont
  {Keitel}, \citenamefont {Kuhlmann}, \citenamefont {M{\"u}ller}, \citenamefont
  {Pl{\"o}njes}, \citenamefont {Ramm}, \citenamefont {Rupp}, \citenamefont
  {Rompotis}, \citenamefont {Sauppe}, \citenamefont {Savelyev}, \citenamefont
  {Schlichting}, \citenamefont {Str{\"u}der}, \citenamefont {Swiderski},
  \citenamefont {Techert}, \citenamefont {Tiedtke}, \citenamefont {Tilp},
  \citenamefont {Treusch}, \citenamefont {Ullrich}, \citenamefont {Moshammer},
  \citenamefont {M{\"o}ller},\ and\ \citenamefont
  {Rolles}}]{erk_campflash_2016}%
  \BibitemOpen
  \bibfield  {author} {\bibinfo {author} {\bibfnamefont {B.}~\bibnamefont
  {Erk}}, \bibinfo {author} {\bibfnamefont {J.~P.}\ \bibnamefont {M{\"u}ller}},
  \bibinfo {author} {\bibfnamefont {B.}~\bibnamefont {C{\'e}dric}}, \bibinfo
  {author} {\bibfnamefont {R.}~\bibnamefont {Boll}}, \bibinfo {author}
  {\bibfnamefont {G.}~\bibnamefont {Brenner}}, \bibinfo {author} {\bibfnamefont
  {H.~N.}\ \bibnamefont {Chapman}}, \bibinfo {author} {\bibfnamefont
  {J.}~\bibnamefont {Correa}}, \bibinfo {author} {\bibfnamefont
  {H.}~\bibnamefont {Dachraoui}}, \bibinfo {author} {\bibfnamefont
  {S.}~\bibnamefont {D{\"u}sterer}}, \bibinfo {author} {\bibfnamefont
  {S.}~\bibnamefont {Dziarzhytski}}, \bibinfo {author} {\bibfnamefont
  {S.}~\bibnamefont {Eisebitt}}, \bibinfo {author} {\bibfnamefont
  {J.}~\bibnamefont {Feldhaus}}, \bibinfo {author} {\bibfnamefont
  {H.}~\bibnamefont {Graafsma}}, \bibinfo {author} {\bibfnamefont
  {S.}~\bibnamefont {Grunewald}}, \bibinfo {author} {\bibfnamefont
  {L.}~\bibnamefont {Gumprecht}}, \bibinfo {author} {\bibfnamefont
  {R.}~\bibnamefont {Hartmann}}, \bibinfo {author} {\bibfnamefont
  {G.}~\bibnamefont {Hauser}}, \bibinfo {author} {\bibfnamefont
  {B.}~\bibnamefont {Keitel}}, \bibinfo {author} {\bibfnamefont
  {M.}~\bibnamefont {Kuhlmann}}, \bibinfo {author} {\bibfnamefont
  {E.}~\bibnamefont {M{\"u}ller}}, \bibinfo {author} {\bibfnamefont
  {E.}~\bibnamefont {Pl{\"o}njes}}, \bibinfo {author} {\bibfnamefont
  {D.}~\bibnamefont {Ramm}}, \bibinfo {author} {\bibfnamefont {D.}~\bibnamefont
  {Rupp}}, \bibinfo {author} {\bibfnamefont {D.}~\bibnamefont {Rompotis}},
  \bibinfo {author} {\bibfnamefont {M.}~\bibnamefont {Sauppe}}, \bibinfo
  {author} {\bibfnamefont {E.}~\bibnamefont {Savelyev}}, \bibinfo {author}
  {\bibfnamefont {I.}~\bibnamefont {Schlichting}}, \bibinfo {author}
  {\bibfnamefont {L.}~\bibnamefont {Str{\"u}der}}, \bibinfo {author}
  {\bibfnamefont {A.}~\bibnamefont {Swiderski}}, \bibinfo {author}
  {\bibfnamefont {S.}~\bibnamefont {Techert}}, \bibinfo {author} {\bibfnamefont
  {K.}~\bibnamefont {Tiedtke}}, \bibinfo {author} {\bibfnamefont
  {T.}~\bibnamefont {Tilp}}, \bibinfo {author} {\bibfnamefont {R.}~\bibnamefont
  {Treusch}}, \bibinfo {author} {\bibfnamefont {J.}~\bibnamefont {Ullrich}},
  \bibinfo {author} {\bibfnamefont {R.}~\bibnamefont {Moshammer}}, \bibinfo
  {author} {\bibfnamefont {T.}~\bibnamefont {M{\"o}ller}}, \ and\ \bibinfo
  {author} {\bibfnamefont {D.}~\bibnamefont {Rolles}},\ }\bibfield  {title}
  {\emph {\bibinfo {title} {{CAMP}@{FLASH} - {An} {End}-{Station} for
  {Imaging}, {Electron}- and {Ion}-{Spectroscopy}, and {Pump}-{Probe}
  {Experiments} at the {FLASH} {Free}-{Electron} {Laser}}},\ }\href@noop {}
  {\bibfield  {journal} {\bibinfo  {journal} {in preparation}\ } (\bibinfo
  {year} {2017})}\BibitemShut {NoStop}%
\bibitem [{\citenamefont {Feldhaus}(2010)}]{feldhaus_flashfirst_2010}%
  \BibitemOpen
  \bibfield  {author} {\bibinfo {author} {\bibfnamefont {J.}~\bibnamefont
  {Feldhaus}},\ }\bibfield  {title} {\emph {\bibinfo {title}
  {{FLASH}{\textemdash}the first soft x-ray free electron laser ({FEL}) user
  facility}},\ }\href {\doibase 10.1088/0953-4075/43/19/194002} {\bibfield
  {journal} {\bibinfo  {journal} {Journal of Physics B}\ }\textbf {\bibinfo
  {volume} {43}},\ \bibinfo {pages} {194002} (\bibinfo {year}
  {2010})}\BibitemShut {NoStop}%
\bibitem [{\citenamefont {Savelyev}\ \emph {et~al.}(2017)\citenamefont
  {Savelyev}, \citenamefont {Boll}, \citenamefont {Bomme}, \citenamefont
  {Schirmel}, \citenamefont {Redlin}, \citenamefont {Erk}, \citenamefont
  {D{\"u}sterer}, \citenamefont {M{\"u}ller}, \citenamefont {H{\"o}ppner},
  \citenamefont {Toleikis}, \citenamefont {M{\"u}ller}, \citenamefont
  {Czwalinna}, \citenamefont {Treusch}, \citenamefont {Kierspel}, \citenamefont
  {Mullins}, \citenamefont {Trippel}, \citenamefont {Wiese}, \citenamefont
  {K{\"u}pper}, \citenamefont {Brau$\beta$e}, \citenamefont {Krecinic},
  \citenamefont {Rouz{\'e}e}, \citenamefont {Rudawski}, \citenamefont
  {Johnsson}, \citenamefont {Amini}, \citenamefont {Lauer}, \citenamefont
  {Burt}, \citenamefont {Brouard}, \citenamefont {Christensen}, \citenamefont
  {Th{\o}gersen}, \citenamefont {Stapelfeldt}, \citenamefont {Berrah},
  \citenamefont {M{\"u}ller}, \citenamefont {Ulmer}, \citenamefont {Techert},
  \citenamefont {Rudenko},\ and\ \citenamefont
  {Rolles}}]{savelyev_jitter-correction_2017}%
  \BibitemOpen
  \bibfield  {author} {\bibinfo {author} {\bibfnamefont {E.}~\bibnamefont
  {Savelyev}}, \bibinfo {author} {\bibfnamefont {R.}~\bibnamefont {Boll}},
  \bibinfo {author} {\bibfnamefont {C.}~\bibnamefont {Bomme}}, \bibinfo
  {author} {\bibfnamefont {N.}~\bibnamefont {Schirmel}}, \bibinfo {author}
  {\bibfnamefont {H.}~\bibnamefont {Redlin}}, \bibinfo {author} {\bibfnamefont
  {B.}~\bibnamefont {Erk}}, \bibinfo {author} {\bibfnamefont {S.}~\bibnamefont
  {D{\"u}sterer}}, \bibinfo {author} {\bibfnamefont {E.}~\bibnamefont
  {M{\"u}ller}}, \bibinfo {author} {\bibfnamefont {H.}~\bibnamefont
  {H{\"o}ppner}}, \bibinfo {author} {\bibfnamefont {S.}~\bibnamefont
  {Toleikis}}, \bibinfo {author} {\bibfnamefont {J.}~\bibnamefont
  {M{\"u}ller}}, \bibinfo {author} {\bibfnamefont {M.~K.}\ \bibnamefont
  {Czwalinna}}, \bibinfo {author} {\bibfnamefont {R.}~\bibnamefont {Treusch}},
  \bibinfo {author} {\bibfnamefont {T.}~\bibnamefont {Kierspel}}, \bibinfo
  {author} {\bibfnamefont {T.}~\bibnamefont {Mullins}}, \bibinfo {author}
  {\bibfnamefont {S.}~\bibnamefont {Trippel}}, \bibinfo {author} {\bibfnamefont
  {J.}~\bibnamefont {Wiese}}, \bibinfo {author} {\bibfnamefont
  {J.}~\bibnamefont {K{\"u}pper}}, \bibinfo {author} {\bibfnamefont
  {F.}~\bibnamefont {Brau$\beta$e}}, \bibinfo {author} {\bibfnamefont
  {F.}~\bibnamefont {Krecinic}}, \bibinfo {author} {\bibfnamefont
  {A.}~\bibnamefont {Rouz{\'e}e}}, \bibinfo {author} {\bibfnamefont
  {P.}~\bibnamefont {Rudawski}}, \bibinfo {author} {\bibfnamefont
  {P.}~\bibnamefont {Johnsson}}, \bibinfo {author} {\bibfnamefont
  {K.}~\bibnamefont {Amini}}, \bibinfo {author} {\bibfnamefont
  {A.}~\bibnamefont {Lauer}}, \bibinfo {author} {\bibfnamefont
  {M.}~\bibnamefont {Burt}}, \bibinfo {author} {\bibfnamefont {M.}~\bibnamefont
  {Brouard}}, \bibinfo {author} {\bibfnamefont {L.}~\bibnamefont
  {Christensen}}, \bibinfo {author} {\bibfnamefont {J.}~\bibnamefont
  {Th{\o}gersen}}, \bibinfo {author} {\bibfnamefont {H.}~\bibnamefont
  {Stapelfeldt}}, \bibinfo {author} {\bibfnamefont {N.}~\bibnamefont {Berrah}},
  \bibinfo {author} {\bibfnamefont {M.}~\bibnamefont {M{\"u}ller}}, \bibinfo
  {author} {\bibfnamefont {A.}~\bibnamefont {Ulmer}}, \bibinfo {author}
  {\bibfnamefont {S.}~\bibnamefont {Techert}}, \bibinfo {author} {\bibfnamefont
  {A.}~\bibnamefont {Rudenko}}, \ and\ \bibinfo {author} {\bibfnamefont
  {D.}~\bibnamefont {Rolles}},\ }\bibfield  {title} {\emph {\bibinfo {title}
  {Jitter-correction for {IR}/{UV}-{XUV} pump-probe experiments at the {FLASH}
  free-electron laser}},\ }\href {\doibase 10.1088/1367-2630/aa652d} {\bibfield
   {journal} {\bibinfo  {journal} {New Journal of Physics}\ }\textbf {\bibinfo
  {volume} {19}},\ \bibinfo {pages} {043009} (\bibinfo {year}
  {2017})}\BibitemShut {NoStop}%
\bibitem [{\citenamefont {Amini}\ \emph {et~al.}(2017)\citenamefont {Amini},
  \citenamefont {Boll}, \citenamefont {Lauer}, \citenamefont {Burt},
  \citenamefont {Lee}, \citenamefont {Christensen}, \citenamefont
  {Brau$\beta$e}, \citenamefont {Mullins}, \citenamefont {Savelyev},
  \citenamefont {Ablikim}, \citenamefont {Berrah}, \citenamefont {Bomme},
  \citenamefont {D{\"u}sterer}, \citenamefont {Erk}, \citenamefont
  {H{\"o}ppner}, \citenamefont {Johnsson}, \citenamefont {Kierspel},
  \citenamefont {Krecinic}, \citenamefont {K{\"u}pper}, \citenamefont
  {M{\"u}ller}, \citenamefont {M{\"u}ller}, \citenamefont {Redlin},
  \citenamefont {Rouz{\'e}e}, \citenamefont {Schirmel}, \citenamefont
  {Th{\o}gersen}, \citenamefont {Techert}, \citenamefont {Toleikis},
  \citenamefont {Treusch}, \citenamefont {Trippel}, \citenamefont {Ulmer},
  \citenamefont {Wiese}, \citenamefont {Vallance}, \citenamefont {Rudenko},
  \citenamefont {Stapelfeldt}, \citenamefont {Brouard},\ and\ \citenamefont
  {Rolles}}]{amini_alignment_2017}%
  \BibitemOpen
  \bibfield  {author} {\bibinfo {author} {\bibfnamefont {K.}~\bibnamefont
  {Amini}}, \bibinfo {author} {\bibfnamefont {R.}~\bibnamefont {Boll}},
  \bibinfo {author} {\bibfnamefont {A.}~\bibnamefont {Lauer}}, \bibinfo
  {author} {\bibfnamefont {M.}~\bibnamefont {Burt}}, \bibinfo {author}
  {\bibfnamefont {J.~W.~L.}\ \bibnamefont {Lee}}, \bibinfo {author}
  {\bibfnamefont {L.}~\bibnamefont {Christensen}}, \bibinfo {author}
  {\bibfnamefont {F.}~\bibnamefont {Brau$\beta$e}}, \bibinfo {author}
  {\bibfnamefont {T.}~\bibnamefont {Mullins}}, \bibinfo {author} {\bibfnamefont
  {E.}~\bibnamefont {Savelyev}}, \bibinfo {author} {\bibfnamefont
  {U.}~\bibnamefont {Ablikim}}, \bibinfo {author} {\bibfnamefont
  {N.}~\bibnamefont {Berrah}}, \bibinfo {author} {\bibfnamefont
  {C.}~\bibnamefont {Bomme}}, \bibinfo {author} {\bibfnamefont
  {S.}~\bibnamefont {D{\"u}sterer}}, \bibinfo {author} {\bibfnamefont
  {B.}~\bibnamefont {Erk}}, \bibinfo {author} {\bibfnamefont {H.}~\bibnamefont
  {H{\"o}ppner}}, \bibinfo {author} {\bibfnamefont {P.}~\bibnamefont
  {Johnsson}}, \bibinfo {author} {\bibfnamefont {T.}~\bibnamefont {Kierspel}},
  \bibinfo {author} {\bibfnamefont {F.}~\bibnamefont {Krecinic}}, \bibinfo
  {author} {\bibfnamefont {J.}~\bibnamefont {K{\"u}pper}}, \bibinfo {author}
  {\bibfnamefont {M.}~\bibnamefont {M{\"u}ller}}, \bibinfo {author}
  {\bibfnamefont {E.}~\bibnamefont {M{\"u}ller}}, \bibinfo {author}
  {\bibfnamefont {H.}~\bibnamefont {Redlin}}, \bibinfo {author} {\bibfnamefont
  {A.}~\bibnamefont {Rouz{\'e}e}}, \bibinfo {author} {\bibfnamefont
  {N.}~\bibnamefont {Schirmel}}, \bibinfo {author} {\bibfnamefont
  {J.}~\bibnamefont {Th{\o}gersen}}, \bibinfo {author} {\bibfnamefont
  {S.}~\bibnamefont {Techert}}, \bibinfo {author} {\bibfnamefont
  {S.}~\bibnamefont {Toleikis}}, \bibinfo {author} {\bibfnamefont
  {R.}~\bibnamefont {Treusch}}, \bibinfo {author} {\bibfnamefont
  {S.}~\bibnamefont {Trippel}}, \bibinfo {author} {\bibfnamefont
  {A.}~\bibnamefont {Ulmer}}, \bibinfo {author} {\bibfnamefont
  {J.}~\bibnamefont {Wiese}}, \bibinfo {author} {\bibfnamefont
  {C.}~\bibnamefont {Vallance}}, \bibinfo {author} {\bibfnamefont
  {A.}~\bibnamefont {Rudenko}}, \bibinfo {author} {\bibfnamefont
  {H.}~\bibnamefont {Stapelfeldt}}, \bibinfo {author} {\bibfnamefont
  {M.}~\bibnamefont {Brouard}}, \ and\ \bibinfo {author} {\bibfnamefont
  {D.}~\bibnamefont {Rolles}},\ }\bibfield  {title} {\emph {\bibinfo {title}
  {Alignment, orientation, and {Coulomb} explosion of difluoroiodobenzene
  studied with the pixel imaging mass spectrometry ({PImMS}) camera}},\ }\href
  {\doibase 10.1063/1.4982220} {\bibfield  {journal} {\bibinfo  {journal} {The
  Journal of Chemical Physics}\ }\textbf {\bibinfo {volume} {147}},\ \bibinfo
  {pages} {013933} (\bibinfo {year} {2017})}\BibitemShut {NoStop}%
\bibitem [{\citenamefont {Chang}\ \emph {et~al.}(2015)\citenamefont {Chang},
  \citenamefont {Horke}, \citenamefont {Trippel},\ and\ \citenamefont
  {K{\"u}pper}}]{chang_spatially-controlled_2015}%
  \BibitemOpen
  \bibfield  {author} {\bibinfo {author} {\bibfnamefont {Y.-P.}\ \bibnamefont
  {Chang}}, \bibinfo {author} {\bibfnamefont {D.~A.}\ \bibnamefont {Horke}},
  \bibinfo {author} {\bibfnamefont {S.}~\bibnamefont {Trippel}}, \ and\
  \bibinfo {author} {\bibfnamefont {J.}~\bibnamefont {K{\"u}pper}},\ }\bibfield
   {title} {\emph {\bibinfo {title} {Spatially-controlled complex molecules and
  their applications}},\ }\href {\doibase 10.1080/0144235X.2015.1077838}
  {\bibfield  {journal} {\bibinfo  {journal} {International Reviews in Physical
  Chemistry}\ }\textbf {\bibinfo {volume} {34}},\ \bibinfo {pages} {557}
  (\bibinfo {year} {2015})}\BibitemShut {NoStop}%
\bibitem [{\citenamefont {Filsinger}\ \emph {et~al.}(2009)\citenamefont
  {Filsinger}, \citenamefont {K{\"u}pper}, \citenamefont {Meijer},
  \citenamefont {Holmegaard}, \citenamefont {Nielsen}, \citenamefont {Nevo},
  \citenamefont {Hansen},\ and\ \citenamefont
  {Stapelfeldt}}]{filsinger_quantum-state_2009}%
  \BibitemOpen
  \bibfield  {author} {\bibinfo {author} {\bibfnamefont {F.}~\bibnamefont
  {Filsinger}}, \bibinfo {author} {\bibfnamefont {J.}~\bibnamefont
  {K{\"u}pper}}, \bibinfo {author} {\bibfnamefont {G.}~\bibnamefont {Meijer}},
  \bibinfo {author} {\bibfnamefont {L.}~\bibnamefont {Holmegaard}}, \bibinfo
  {author} {\bibfnamefont {J.~H.}\ \bibnamefont {Nielsen}}, \bibinfo {author}
  {\bibfnamefont {I.}~\bibnamefont {Nevo}}, \bibinfo {author} {\bibfnamefont
  {J.~L.}\ \bibnamefont {Hansen}}, \ and\ \bibinfo {author} {\bibfnamefont
  {H.}~\bibnamefont {Stapelfeldt}},\ }\bibfield  {title} {\emph {\bibinfo
  {title} {Quantum-state selection, alignment, and orientation of large
  molecules using static electric and laser fields}},\ }\href {\doibase
  10.1063/1.3194287} {\bibfield  {journal} {\bibinfo  {journal} {The Journal of
  Chemical Physics}\ }\textbf {\bibinfo {volume} {131}},\ \bibinfo {pages}
  {064309} (\bibinfo {year} {2009})}\BibitemShut {NoStop}%
\bibitem [{\citenamefont {Stapelfeldt}\ and\ \citenamefont
  {Seideman}(2003)}]{stapelfeldt_colloquium:_2003}%
  \BibitemOpen
  \bibfield  {author} {\bibinfo {author} {\bibfnamefont {H.}~\bibnamefont
  {Stapelfeldt}}\ and\ \bibinfo {author} {\bibfnamefont {T.}~\bibnamefont
  {Seideman}},\ }\bibfield  {title} {\emph {\bibinfo {title} {Colloquium:
  {Aligning} molecules with strong laser pulses}},\ }\href
  {http://rmp.aps.org/abstract/RMP/v75/i2/p543_1} {\bibfield  {journal}
  {\bibinfo  {journal} {Reviews of Modern Physics}\ }\textbf {\bibinfo {volume}
  {75}},\ \bibinfo {pages} {543} (\bibinfo {year} {2003})}\BibitemShut
  {NoStop}%
\bibitem [{\citenamefont {Rolles}\ \emph {et~al.}(2014)\citenamefont {Rolles},
  \citenamefont {Boll}, \citenamefont {Adolph}, \citenamefont {Aquila},
  \citenamefont {Bostedt}, \citenamefont {Bozek}, \citenamefont {Chapman},
  \citenamefont {Coffee}, \citenamefont {Coppola}, \citenamefont {Decleva},
  \citenamefont {Delmas}, \citenamefont {Epp}, \citenamefont {Erk},
  \citenamefont {Filsinger}, \citenamefont {Foucar}, \citenamefont {Gumprecht},
  \citenamefont {H{\"o}mke}, \citenamefont {Gorkhover}, \citenamefont
  {Holmegaard}, \citenamefont {Johnsson}, \citenamefont {Kaiser}, \citenamefont
  {Krasniqi}, \citenamefont {K{\"u}hnel}, \citenamefont {Maurer}, \citenamefont
  {Messerschmidt}, \citenamefont {Moshammer}, \citenamefont {Quevedo},
  \citenamefont {Rajkovic}, \citenamefont {Rouz{\'e}e}, \citenamefont {Rudek},
  \citenamefont {Schlichting}, \citenamefont {Schmidt}, \citenamefont {Schorb},
  \citenamefont {Schr{\"o}ter}, \citenamefont {Schulz}, \citenamefont
  {Stapelfeldt}, \citenamefont {Stener}, \citenamefont {Stern}, \citenamefont
  {Techert}, \citenamefont {Th{\o}gersen}, \citenamefont {Vrakking},
  \citenamefont {Rudenko}, \citenamefont {K{\"u}pper},\ and\ \citenamefont
  {Ullrich}}]{rolles_femtosecond_2014}%
  \BibitemOpen
  \bibfield  {author} {\bibinfo {author} {\bibfnamefont {D.}~\bibnamefont
  {Rolles}}, \bibinfo {author} {\bibfnamefont {R.}~\bibnamefont {Boll}},
  \bibinfo {author} {\bibfnamefont {M.}~\bibnamefont {Adolph}}, \bibinfo
  {author} {\bibfnamefont {A.}~\bibnamefont {Aquila}}, \bibinfo {author}
  {\bibfnamefont {C.}~\bibnamefont {Bostedt}}, \bibinfo {author} {\bibfnamefont
  {J.~D.}\ \bibnamefont {Bozek}}, \bibinfo {author} {\bibfnamefont {H.~N.}\
  \bibnamefont {Chapman}}, \bibinfo {author} {\bibfnamefont {R.}~\bibnamefont
  {Coffee}}, \bibinfo {author} {\bibfnamefont {N.}~\bibnamefont {Coppola}},
  \bibinfo {author} {\bibfnamefont {P.}~\bibnamefont {Decleva}}, \bibinfo
  {author} {\bibfnamefont {T.}~\bibnamefont {Delmas}}, \bibinfo {author}
  {\bibfnamefont {S.~W.}\ \bibnamefont {Epp}}, \bibinfo {author} {\bibfnamefont
  {B.}~\bibnamefont {Erk}}, \bibinfo {author} {\bibfnamefont {F.}~\bibnamefont
  {Filsinger}}, \bibinfo {author} {\bibfnamefont {L.}~\bibnamefont {Foucar}},
  \bibinfo {author} {\bibfnamefont {L.}~\bibnamefont {Gumprecht}}, \bibinfo
  {author} {\bibfnamefont {A.}~\bibnamefont {H{\"o}mke}}, \bibinfo {author}
  {\bibfnamefont {T.}~\bibnamefont {Gorkhover}}, \bibinfo {author}
  {\bibfnamefont {L.}~\bibnamefont {Holmegaard}}, \bibinfo {author}
  {\bibfnamefont {P.}~\bibnamefont {Johnsson}}, \bibinfo {author}
  {\bibfnamefont {C.}~\bibnamefont {Kaiser}}, \bibinfo {author} {\bibfnamefont
  {F.}~\bibnamefont {Krasniqi}}, \bibinfo {author} {\bibfnamefont {K.-U.}\
  \bibnamefont {K{\"u}hnel}}, \bibinfo {author} {\bibfnamefont
  {J.}~\bibnamefont {Maurer}}, \bibinfo {author} {\bibfnamefont
  {M.}~\bibnamefont {Messerschmidt}}, \bibinfo {author} {\bibfnamefont
  {R.}~\bibnamefont {Moshammer}}, \bibinfo {author} {\bibfnamefont
  {W.}~\bibnamefont {Quevedo}}, \bibinfo {author} {\bibfnamefont
  {I.}~\bibnamefont {Rajkovic}}, \bibinfo {author} {\bibfnamefont
  {A.}~\bibnamefont {Rouz{\'e}e}}, \bibinfo {author} {\bibfnamefont
  {B.}~\bibnamefont {Rudek}}, \bibinfo {author} {\bibfnamefont
  {I.}~\bibnamefont {Schlichting}}, \bibinfo {author} {\bibfnamefont
  {C.}~\bibnamefont {Schmidt}}, \bibinfo {author} {\bibfnamefont
  {S.}~\bibnamefont {Schorb}}, \bibinfo {author} {\bibfnamefont {C.~D.}\
  \bibnamefont {Schr{\"o}ter}}, \bibinfo {author} {\bibfnamefont
  {J.}~\bibnamefont {Schulz}}, \bibinfo {author} {\bibfnamefont
  {H.}~\bibnamefont {Stapelfeldt}}, \bibinfo {author} {\bibfnamefont
  {M.}~\bibnamefont {Stener}}, \bibinfo {author} {\bibfnamefont
  {S.}~\bibnamefont {Stern}}, \bibinfo {author} {\bibfnamefont
  {S.}~\bibnamefont {Techert}}, \bibinfo {author} {\bibfnamefont
  {J.}~\bibnamefont {Th{\o}gersen}}, \bibinfo {author} {\bibfnamefont
  {M.~J.~J.}\ \bibnamefont {Vrakking}}, \bibinfo {author} {\bibfnamefont
  {A.}~\bibnamefont {Rudenko}}, \bibinfo {author} {\bibfnamefont
  {J.}~\bibnamefont {K{\"u}pper}}, \ and\ \bibinfo {author} {\bibfnamefont
  {J.}~\bibnamefont {Ullrich}},\ }\bibfield  {title} {\emph {\bibinfo {title}
  {Femtosecond x-ray photoelectron diffraction on gas-phase dibromobenzene
  molecules}},\ }\href {\doibase 10.1088/0953-4075/47/12/124035} {\bibfield
  {journal} {\bibinfo  {journal} {Journal of Physics B}\ }\textbf {\bibinfo
  {volume} {47}},\ \bibinfo {pages} {124035} (\bibinfo {year}
  {2014})}\BibitemShut {NoStop}%
\bibitem [{\citenamefont {John}\ \emph {et~al.}(2012)\citenamefont {John},
  \citenamefont {Brouard}, \citenamefont {Clark}, \citenamefont {Crooks},
  \citenamefont {Halford}, \citenamefont {Hill}, \citenamefont {Lee},
  \citenamefont {Nomerotski}, \citenamefont {Pisarczyk}, \citenamefont
  {Sedgwick}, \citenamefont {Slater}, \citenamefont {Turchetta}, \citenamefont
  {Vallance}, \citenamefont {Wilman}, \citenamefont {Winter},\ and\
  \citenamefont {Yuen}}]{john_pimms_2012}%
  \BibitemOpen
  \bibfield  {author} {\bibinfo {author} {\bibfnamefont {J.~J.}\ \bibnamefont
  {John}}, \bibinfo {author} {\bibfnamefont {M.}~\bibnamefont {Brouard}},
  \bibinfo {author} {\bibfnamefont {A.}~\bibnamefont {Clark}}, \bibinfo
  {author} {\bibfnamefont {J.}~\bibnamefont {Crooks}}, \bibinfo {author}
  {\bibfnamefont {E.}~\bibnamefont {Halford}}, \bibinfo {author} {\bibfnamefont
  {L.}~\bibnamefont {Hill}}, \bibinfo {author} {\bibfnamefont {J.~W.~L.}\
  \bibnamefont {Lee}}, \bibinfo {author} {\bibfnamefont {A.}~\bibnamefont
  {Nomerotski}}, \bibinfo {author} {\bibfnamefont {R.}~\bibnamefont
  {Pisarczyk}}, \bibinfo {author} {\bibfnamefont {I.}~\bibnamefont {Sedgwick}},
  \bibinfo {author} {\bibfnamefont {C.~S.}\ \bibnamefont {Slater}}, \bibinfo
  {author} {\bibfnamefont {R.}~\bibnamefont {Turchetta}}, \bibinfo {author}
  {\bibfnamefont {C.}~\bibnamefont {Vallance}}, \bibinfo {author}
  {\bibfnamefont {E.}~\bibnamefont {Wilman}}, \bibinfo {author} {\bibfnamefont
  {B.}~\bibnamefont {Winter}}, \ and\ \bibinfo {author} {\bibfnamefont {W.~H.}\
  \bibnamefont {Yuen}},\ }\bibfield  {title} {\emph {\bibinfo {title} {{PImMS},
  a fast event-triggered monolithic pixel detector with storage of multiple
  timestamps}},\ }\href {\doibase 10.1088/1748-0221/7/08/C08001} {\bibfield
  {journal} {\bibinfo  {journal} {Journal of Instrumentation}\ }\textbf
  {\bibinfo {volume} {7}},\ \bibinfo {pages} {C08001} (\bibinfo {year}
  {2012})}\BibitemShut {NoStop}%
\bibitem [{\citenamefont {Amini}\ \emph {et~al.}(2015)\citenamefont {Amini},
  \citenamefont {Blake}, \citenamefont {Brouard}, \citenamefont {Burt},
  \citenamefont {Halford}, \citenamefont {Lauer}, \citenamefont {Slater},
  \citenamefont {Lee},\ and\ \citenamefont
  {Vallance}}]{amini_three-dimensional_2015}%
  \BibitemOpen
  \bibfield  {author} {\bibinfo {author} {\bibfnamefont {K.}~\bibnamefont
  {Amini}}, \bibinfo {author} {\bibfnamefont {S.}~\bibnamefont {Blake}},
  \bibinfo {author} {\bibfnamefont {M.}~\bibnamefont {Brouard}}, \bibinfo
  {author} {\bibfnamefont {M.~B.}\ \bibnamefont {Burt}}, \bibinfo {author}
  {\bibfnamefont {E.}~\bibnamefont {Halford}}, \bibinfo {author} {\bibfnamefont
  {A.}~\bibnamefont {Lauer}}, \bibinfo {author} {\bibfnamefont {C.~S.}\
  \bibnamefont {Slater}}, \bibinfo {author} {\bibfnamefont {J.~W.~L.}\
  \bibnamefont {Lee}}, \ and\ \bibinfo {author} {\bibfnamefont
  {C.}~\bibnamefont {Vallance}},\ }\bibfield  {title} {\emph {\bibinfo {title}
  {Three-dimensional imaging of carbonyl sulfide and ethyl iodide
  photodissociation using the pixel imaging mass spectrometry camera}},\ }\href
  {\doibase 10.1063/1.4934544} {\bibfield  {journal} {\bibinfo  {journal}
  {Review of Scientific Instruments}\ }\textbf {\bibinfo {volume} {86}},\
  \bibinfo {pages} {103113} (\bibinfo {year} {2015})}\BibitemShut {NoStop}%
\bibitem [{Note1()}]{Note1}%
  \BibitemOpen
  \bibinfo {note} {The finite width of the molecular axis distribution
  perpendicular to the detector plane leads to an effective smearing out of the
  extracted kinetic energy spectra towards lower kinetic energies. However, for
  the data presented here, no significant improvement was found when inverting
  the ion images, so non-inverted data is shown throughout the
  manuscript.}\BibitemShut {Stop}%
\bibitem [{\citenamefont {Fisher-Levine}\ and\ \citenamefont
  {Nomerotski}(2016)}]{fisher-levine_timepixcam:_2016}%
  \BibitemOpen
  \bibfield  {author} {\bibinfo {author} {\bibfnamefont {M.}~\bibnamefont
  {Fisher-Levine}}\ and\ \bibinfo {author} {\bibfnamefont {A.}~\bibnamefont
  {Nomerotski}},\ }\bibfield  {title} {\emph {\bibinfo {title} {{TimepixCam}: a
  fast optical imager with time-stamping}},\ }\href {\doibase
  10.1088/1748-0221/11/03/C03016} {\bibfield  {journal} {\bibinfo  {journal}
  {Journal of Instrumentation}\ }\textbf {\bibinfo {volume} {11}},\ \bibinfo
  {pages} {C03016} (\bibinfo {year} {2016})}\BibitemShut {NoStop}%
\bibitem [{\citenamefont {Fisher-Levine}\ \emph {et~al.}(2017)\citenamefont
  {Fisher-Levine}, \citenamefont {Boll}, \citenamefont {Ziaee}, \citenamefont
  {Bomme}, \citenamefont {Erk}, \citenamefont {Rompotis}, \citenamefont
  {Marchenko}, \citenamefont {Nomerotski},\ and\ \citenamefont
  {Rolles}}]{fisher-levine_time-resolved_2017}%
  \BibitemOpen
  \bibfield  {author} {\bibinfo {author} {\bibfnamefont {M.}~\bibnamefont
  {Fisher-Levine}}, \bibinfo {author} {\bibfnamefont {R.}~\bibnamefont {Boll}},
  \bibinfo {author} {\bibfnamefont {F.}~\bibnamefont {Ziaee}}, \bibinfo
  {author} {\bibfnamefont {C.}~\bibnamefont {Bomme}}, \bibinfo {author}
  {\bibfnamefont {B.}~\bibnamefont {Erk}}, \bibinfo {author} {\bibfnamefont
  {D.}~\bibnamefont {Rompotis}}, \bibinfo {author} {\bibfnamefont
  {T.}~\bibnamefont {Marchenko}}, \bibinfo {author} {\bibfnamefont
  {A.}~\bibnamefont {Nomerotski}}, \ and\ \bibinfo {author} {\bibfnamefont
  {D.}~\bibnamefont {Rolles}},\ }\bibfield  {title} {\emph {\bibinfo {title}
  {Time-{Resolved} {Ion} {Imaging} {Experiments} at {Free}-{Electron} {Lasers}
  {Using} {TimepixCam}}},\ }\href@noop {} {\bibfield  {journal} {\bibinfo
  {journal} {in preparation}\ } (\bibinfo {year} {2017})}\BibitemShut {NoStop}%
\bibitem [{\citenamefont {Long}\ \emph {et~al.}(2017)\citenamefont {Long},
  \citenamefont {Furch}, \citenamefont {Dur{\'a}}, \citenamefont {Tremsin},
  \citenamefont {Vallerga}, \citenamefont {Schulz}, \citenamefont
  {Rouz{\'e}e},\ and\ \citenamefont {Vrakking}}]{long_ion-ion_2017}%
  \BibitemOpen
  \bibfield  {author} {\bibinfo {author} {\bibfnamefont {J.}~\bibnamefont
  {Long}}, \bibinfo {author} {\bibfnamefont {F.~J.}\ \bibnamefont {Furch}},
  \bibinfo {author} {\bibfnamefont {J.}~\bibnamefont {Dur{\'a}}}, \bibinfo
  {author} {\bibfnamefont {A.~S.}\ \bibnamefont {Tremsin}}, \bibinfo {author}
  {\bibfnamefont {J.}~\bibnamefont {Vallerga}}, \bibinfo {author}
  {\bibfnamefont {C.~P.}\ \bibnamefont {Schulz}}, \bibinfo {author}
  {\bibfnamefont {A.}~\bibnamefont {Rouz{\'e}e}}, \ and\ \bibinfo {author}
  {\bibfnamefont {M.~J.}\ \bibnamefont {Vrakking}},\ }\bibfield  {title} {\emph
  {\bibinfo {title} {Ion-ion coincidence imaging at high event rate using an
  in-vacuum pixel detector}},\ }\href
  {http://aip.scitation.org/doi/abs/10.1063/1.4981126} {\bibfield  {journal}
  {\bibinfo  {journal} {The Journal of Chemical Physics}\ }\textbf {\bibinfo
  {volume} {147}},\ \bibinfo {pages} {013919} (\bibinfo {year}
  {2017})}\BibitemShut {NoStop}%
\bibitem [{\citenamefont {Slater}\ \emph {et~al.}(2014)\citenamefont {Slater},
  \citenamefont {Blake}, \citenamefont {Brouard}, \citenamefont {Lauer},
  \citenamefont {Vallance}, \citenamefont {John}, \citenamefont {Turchetta},
  \citenamefont {Nomerotski}, \citenamefont {Christensen}, \citenamefont
  {Nielsen}, \citenamefont {Johansson},\ and\ \citenamefont
  {Stapelfeldt}}]{slater_covariance_2014}%
  \BibitemOpen
  \bibfield  {author} {\bibinfo {author} {\bibfnamefont {C.~S.}\ \bibnamefont
  {Slater}}, \bibinfo {author} {\bibfnamefont {S.}~\bibnamefont {Blake}},
  \bibinfo {author} {\bibfnamefont {M.}~\bibnamefont {Brouard}}, \bibinfo
  {author} {\bibfnamefont {A.}~\bibnamefont {Lauer}}, \bibinfo {author}
  {\bibfnamefont {C.}~\bibnamefont {Vallance}}, \bibinfo {author}
  {\bibfnamefont {J.~J.}\ \bibnamefont {John}}, \bibinfo {author}
  {\bibfnamefont {R.}~\bibnamefont {Turchetta}}, \bibinfo {author}
  {\bibfnamefont {A.}~\bibnamefont {Nomerotski}}, \bibinfo {author}
  {\bibfnamefont {L.}~\bibnamefont {Christensen}}, \bibinfo {author}
  {\bibfnamefont {J.~H.}\ \bibnamefont {Nielsen}}, \bibinfo {author}
  {\bibfnamefont {M.~P.}\ \bibnamefont {Johansson}}, \ and\ \bibinfo {author}
  {\bibfnamefont {H.}~\bibnamefont {Stapelfeldt}},\ }\bibfield  {title} {\emph
  {\bibinfo {title} {Covariance imaging experiments using a pixel-imaging
  mass-spectrometry camera}},\ }\href {\doibase 10.1103/PhysRevA.89.011401}
  {\bibfield  {journal} {\bibinfo  {journal} {Physical Review A}\ }\textbf
  {\bibinfo {volume} {89}},\ \bibinfo {pages} {011401(R)} (\bibinfo {year}
  {2014})}\BibitemShut {NoStop}%
\bibitem [{\citenamefont {Slater}\ \emph {et~al.}(2015)\citenamefont {Slater},
  \citenamefont {Blake}, \citenamefont {Brouard}, \citenamefont {Lauer},
  \citenamefont {Vallance}, \citenamefont {Bohun}, \citenamefont {Christensen},
  \citenamefont {Nielsen}, \citenamefont {Johansson},\ and\ \citenamefont
  {Stapelfeldt}}]{slater_coulomb-explosion_2015}%
  \BibitemOpen
  \bibfield  {author} {\bibinfo {author} {\bibfnamefont {C.~S.}\ \bibnamefont
  {Slater}}, \bibinfo {author} {\bibfnamefont {S.}~\bibnamefont {Blake}},
  \bibinfo {author} {\bibfnamefont {M.}~\bibnamefont {Brouard}}, \bibinfo
  {author} {\bibfnamefont {A.}~\bibnamefont {Lauer}}, \bibinfo {author}
  {\bibfnamefont {C.}~\bibnamefont {Vallance}}, \bibinfo {author}
  {\bibfnamefont {C.~S.}\ \bibnamefont {Bohun}}, \bibinfo {author}
  {\bibfnamefont {L.}~\bibnamefont {Christensen}}, \bibinfo {author}
  {\bibfnamefont {J.~H.}\ \bibnamefont {Nielsen}}, \bibinfo {author}
  {\bibfnamefont {M.~P.}\ \bibnamefont {Johansson}}, \ and\ \bibinfo {author}
  {\bibfnamefont {H.}~\bibnamefont {Stapelfeldt}},\ }\bibfield  {title} {\emph
  {\bibinfo {title} {Coulomb-explosion imaging using a pixel-imaging
  mass-spectrometry camera}},\ }\href {\doibase 10.1103/PhysRevA.91.053424}
  {\bibfield  {journal} {\bibinfo  {journal} {Physical Review A}\ }\textbf
  {\bibinfo {volume} {91}},\ \bibinfo {pages} {053424} (\bibinfo {year}
  {2015})}\BibitemShut {NoStop}%
\bibitem [{\citenamefont {Christensen}\ \emph {et~al.}(2015)\citenamefont
  {Christensen}, \citenamefont {Nielsen}, \citenamefont {Slater}, \citenamefont
  {Lauer}, \citenamefont {Brouard},\ and\ \citenamefont
  {Stapelfeldt}}]{christensen_using_2015}%
  \BibitemOpen
  \bibfield  {author} {\bibinfo {author} {\bibfnamefont {L.}~\bibnamefont
  {Christensen}}, \bibinfo {author} {\bibfnamefont {J.~H.}\ \bibnamefont
  {Nielsen}}, \bibinfo {author} {\bibfnamefont {C.~S.}\ \bibnamefont {Slater}},
  \bibinfo {author} {\bibfnamefont {A.}~\bibnamefont {Lauer}}, \bibinfo
  {author} {\bibfnamefont {M.}~\bibnamefont {Brouard}}, \ and\ \bibinfo
  {author} {\bibfnamefont {H.}~\bibnamefont {Stapelfeldt}},\ }\bibfield
  {title} {\emph {\bibinfo {title} {Using laser-induced {Coulomb} explosion of
  aligned chiral molecules to determine their absolute configuration}},\ }\href
  {\doibase 10.1103/PhysRevA.92.033411} {\bibfield  {journal} {\bibinfo
  {journal} {Physical Review A}\ }\textbf {\bibinfo {volume} {92}},\ \bibinfo
  {pages} {033411} (\bibinfo {year} {2015})}\BibitemShut {NoStop}%
\bibitem [{\citenamefont {Christensen}\ \emph {et~al.}(2016)\citenamefont
  {Christensen}, \citenamefont {Christiansen}, \citenamefont {Shepperson},\
  and\ \citenamefont {Stapelfeldt}}]{christensen_deconvoluting_2016}%
  \BibitemOpen
  \bibfield  {author} {\bibinfo {author} {\bibfnamefont {L.}~\bibnamefont
  {Christensen}}, \bibinfo {author} {\bibfnamefont {L.}~\bibnamefont
  {Christiansen}}, \bibinfo {author} {\bibfnamefont {B.}~\bibnamefont
  {Shepperson}}, \ and\ \bibinfo {author} {\bibfnamefont {H.}~\bibnamefont
  {Stapelfeldt}},\ }\bibfield  {title} {\emph {\bibinfo {title} {Deconvoluting
  nonaxial recoil in {Coulomb} explosion measurements of molecular axis
  alignment}},\ }\href {\doibase 10.1103/PhysRevA.94.023410} {\bibfield
  {journal} {\bibinfo  {journal} {Physical Review A}\ }\textbf {\bibinfo
  {volume} {94}},\ \bibinfo {pages} {023410} (\bibinfo {year}
  {2016})}\BibitemShut {NoStop}%
\bibitem [{\citenamefont {Pickering}\ \emph {et~al.}(2016)\citenamefont
  {Pickering}, \citenamefont {Amini}, \citenamefont {Brouard}, \citenamefont
  {Burt}, \citenamefont {Bush}, \citenamefont {Christensen}, \citenamefont
  {Lauer}, \citenamefont {Nielsen}, \citenamefont {Slater},\ and\ \citenamefont
  {Stapelfeldt}}]{pickering_communication:_2016}%
  \BibitemOpen
  \bibfield  {author} {\bibinfo {author} {\bibfnamefont {J.~D.}\ \bibnamefont
  {Pickering}}, \bibinfo {author} {\bibfnamefont {K.}~\bibnamefont {Amini}},
  \bibinfo {author} {\bibfnamefont {M.}~\bibnamefont {Brouard}}, \bibinfo
  {author} {\bibfnamefont {M.}~\bibnamefont {Burt}}, \bibinfo {author}
  {\bibfnamefont {I.~J.}\ \bibnamefont {Bush}}, \bibinfo {author}
  {\bibfnamefont {L.}~\bibnamefont {Christensen}}, \bibinfo {author}
  {\bibfnamefont {A.}~\bibnamefont {Lauer}}, \bibinfo {author} {\bibfnamefont
  {J.~H.}\ \bibnamefont {Nielsen}}, \bibinfo {author} {\bibfnamefont {C.~S.}\
  \bibnamefont {Slater}}, \ and\ \bibinfo {author} {\bibfnamefont
  {H.}~\bibnamefont {Stapelfeldt}},\ }\bibfield  {title} {\emph {\bibinfo
  {title} {Communication: {Three}-fold covariance imaging of laser-induced
  {Coulomb} explosions}},\ }\href {\doibase 10.1063/1.4947551} {\bibfield
  {journal} {\bibinfo  {journal} {The Journal of Chemical Physics}\ }\textbf
  {\bibinfo {volume} {144}},\ \bibinfo {pages} {161105} (\bibinfo {year}
  {2016})}\BibitemShut {NoStop}%
\bibitem [{\citenamefont {Hansen}\ \emph {et~al.}(2012)\citenamefont {Hansen},
  \citenamefont {Nielsen}, \citenamefont {Madsen}, \citenamefont {Lindhardt},
  \citenamefont {Johansson}, \citenamefont {Skrydstrup}, \citenamefont
  {Madsen},\ and\ \citenamefont {Stapelfeldt}}]{hansen_control_2012}%
  \BibitemOpen
  \bibfield  {author} {\bibinfo {author} {\bibfnamefont {J.~L.}\ \bibnamefont
  {Hansen}}, \bibinfo {author} {\bibfnamefont {J.~H.}\ \bibnamefont {Nielsen}},
  \bibinfo {author} {\bibfnamefont {C.~B.}\ \bibnamefont {Madsen}}, \bibinfo
  {author} {\bibfnamefont {A.~T.}\ \bibnamefont {Lindhardt}}, \bibinfo {author}
  {\bibfnamefont {M.~P.}\ \bibnamefont {Johansson}}, \bibinfo {author}
  {\bibfnamefont {T.}~\bibnamefont {Skrydstrup}}, \bibinfo {author}
  {\bibfnamefont {L.~B.}\ \bibnamefont {Madsen}}, \ and\ \bibinfo {author}
  {\bibfnamefont {H.}~\bibnamefont {Stapelfeldt}},\ }\bibfield  {title} {\emph
  {\bibinfo {title} {Control and femtosecond time-resolved imaging of torsion
  in a chiral molecule}},\ }\href {\doibase 10.1063/1.4719816} {\bibfield
  {journal} {\bibinfo  {journal} {The Journal of Chemical Physics}\ }\textbf
  {\bibinfo {volume} {136}},\ \bibinfo {pages} {204310} (\bibinfo {year}
  {2012})}\BibitemShut {NoStop}%
\bibitem [{\citenamefont {Ryufuku}\ \emph {et~al.}(1980)\citenamefont
  {Ryufuku}, \citenamefont {Sasaki},\ and\ \citenamefont
  {Watanabe}}]{ryufuku_oscillatory_1980}%
  \BibitemOpen
  \bibfield  {author} {\bibinfo {author} {\bibfnamefont {H.}~\bibnamefont
  {Ryufuku}}, \bibinfo {author} {\bibfnamefont {K.}~\bibnamefont {Sasaki}}, \
  and\ \bibinfo {author} {\bibfnamefont {T.}~\bibnamefont {Watanabe}},\
  }\bibfield  {title} {\emph {\bibinfo {title} {Oscillatory behavior of charge
  transfer cross sections as a function of the charge of projectiles in
  low-energy collisions}},\ }\href
  {http://pra.aps.org/abstract/PRA/v21/i3/p745_1} {\bibfield  {journal}
  {\bibinfo  {journal} {Physical Review A}\ }\textbf {\bibinfo {volume} {21}},\
  \bibinfo {pages} {745} (\bibinfo {year} {1980})}\BibitemShut {NoStop}%
\bibitem [{\citenamefont {Niehaus}(1986)}]{niehaus_classical_1986}%
  \BibitemOpen
  \bibfield  {author} {\bibinfo {author} {\bibfnamefont {A.}~\bibnamefont
  {Niehaus}},\ }\bibfield  {title} {\emph {\bibinfo {title} {A classical model
  for multiple-electron capture in slow collisions of highly charged ions with
  atoms}},\ }\href {http://iopscience.iop.org/0022-3700/19/18/021} {\bibfield
  {journal} {\bibinfo  {journal} {Journal of Physics B}\ }\textbf {\bibinfo
  {volume} {19}},\ \bibinfo {pages} {2925} (\bibinfo {year}
  {1986})}\BibitemShut {NoStop}%
\bibitem [{\citenamefont {Schnorr}\ \emph {et~al.}(2014)\citenamefont
  {Schnorr}, \citenamefont {Senftleben}, \citenamefont {Kurka}, \citenamefont
  {Rudenko}, \citenamefont {Schmid}, \citenamefont {Pfeifer}, \citenamefont
  {Meyer}, \citenamefont {K{\"u}bel}, \citenamefont {Kling}, \citenamefont
  {Jiang}, \citenamefont {Treusch}, \citenamefont {D{\"u}sterer}, \citenamefont
  {Siemer}, \citenamefont {W{\"o}stmann}, \citenamefont {Zacharias},
  \citenamefont {Mitzner}, \citenamefont {Zouros}, \citenamefont {Ullrich},
  \citenamefont {Schr{\"o}ter},\ and\ \citenamefont
  {Moshammer}}]{schnorr_electron_2014}%
  \BibitemOpen
  \bibfield  {author} {\bibinfo {author} {\bibfnamefont {K.}~\bibnamefont
  {Schnorr}}, \bibinfo {author} {\bibfnamefont {A.}~\bibnamefont {Senftleben}},
  \bibinfo {author} {\bibfnamefont {M.}~\bibnamefont {Kurka}}, \bibinfo
  {author} {\bibfnamefont {A.}~\bibnamefont {Rudenko}}, \bibinfo {author}
  {\bibfnamefont {G.}~\bibnamefont {Schmid}}, \bibinfo {author} {\bibfnamefont
  {T.}~\bibnamefont {Pfeifer}}, \bibinfo {author} {\bibfnamefont
  {K.}~\bibnamefont {Meyer}}, \bibinfo {author} {\bibfnamefont
  {M.}~\bibnamefont {K{\"u}bel}}, \bibinfo {author} {\bibfnamefont {M.~F.}\
  \bibnamefont {Kling}}, \bibinfo {author} {\bibfnamefont {Y.~H.}\ \bibnamefont
  {Jiang}}, \bibinfo {author} {\bibfnamefont {R.}~\bibnamefont {Treusch}},
  \bibinfo {author} {\bibfnamefont {S.}~\bibnamefont {D{\"u}sterer}}, \bibinfo
  {author} {\bibfnamefont {B.}~\bibnamefont {Siemer}}, \bibinfo {author}
  {\bibfnamefont {M.}~\bibnamefont {W{\"o}stmann}}, \bibinfo {author}
  {\bibfnamefont {H.}~\bibnamefont {Zacharias}}, \bibinfo {author}
  {\bibfnamefont {R.}~\bibnamefont {Mitzner}}, \bibinfo {author} {\bibfnamefont
  {T.~J.~M.}\ \bibnamefont {Zouros}}, \bibinfo {author} {\bibfnamefont
  {J.}~\bibnamefont {Ullrich}}, \bibinfo {author} {\bibfnamefont {C.~D.}\
  \bibnamefont {Schr{\"o}ter}}, \ and\ \bibinfo {author} {\bibfnamefont
  {R.}~\bibnamefont {Moshammer}},\ }\bibfield  {title} {\emph {\bibinfo {title}
  {Electron {Rearrangement} {Dynamics} in {Dissociating} {I}$_{\textrm{2}}$
  {Molecules} {Accessed} by {Extreme} {Ultraviolet} {Pump}-{Probe}
  {Experiments}}},\ }\href {\doibase 10.1103/PhysRevLett.113.073001} {\bibfield
   {journal} {\bibinfo  {journal} {Physical Review Letters}\ }\textbf {\bibinfo
  {volume} {113}},\ \bibinfo {pages} {073001} (\bibinfo {year}
  {2014})}\BibitemShut {NoStop}%
\bibitem [{\citenamefont {Rudenko}\ \emph {et~al.}(2017)\citenamefont
  {Rudenko}, \citenamefont {Inhester}, \citenamefont {Hanasaki}, \citenamefont
  {Li}, \citenamefont {Robatjazi}, \citenamefont {Erk}, \citenamefont {Boll},
  \citenamefont {Toyota}, \citenamefont {Hao}, \citenamefont {Vendrell},
  \citenamefont {Bomme}, \citenamefont {Savelyev}, \citenamefont {Rudek},
  \citenamefont {Foucar}, \citenamefont {Southworth}, \citenamefont {Lehmann},
  \citenamefont {Kraessig}, \citenamefont {Marchenko}, \citenamefont {Simon},
  \citenamefont {Ueda}, \citenamefont {Ferguson}, \citenamefont {Bucher},
  \citenamefont {Gorkhover}, \citenamefont {Carron}, \citenamefont
  {Alonso-Mori}, \citenamefont {Koglin}, \citenamefont {Correa}, \citenamefont
  {Williams}, \citenamefont {Boutet}, \citenamefont {Young}, \citenamefont
  {Bostedt}, \citenamefont {Son}, \citenamefont {Santra},\ and\ \citenamefont
  {Rolles}}]{rudenko_femtosecond_2017}%
  \BibitemOpen
  \bibfield  {author} {\bibinfo {author} {\bibfnamefont {A.}~\bibnamefont
  {Rudenko}}, \bibinfo {author} {\bibfnamefont {L.}~\bibnamefont {Inhester}},
  \bibinfo {author} {\bibfnamefont {K.}~\bibnamefont {Hanasaki}}, \bibinfo
  {author} {\bibfnamefont {X.}~\bibnamefont {Li}}, \bibinfo {author}
  {\bibfnamefont {S.~J.}\ \bibnamefont {Robatjazi}}, \bibinfo {author}
  {\bibfnamefont {B.}~\bibnamefont {Erk}}, \bibinfo {author} {\bibfnamefont
  {R.}~\bibnamefont {Boll}}, \bibinfo {author} {\bibfnamefont {K.}~\bibnamefont
  {Toyota}}, \bibinfo {author} {\bibfnamefont {Y.}~\bibnamefont {Hao}},
  \bibinfo {author} {\bibfnamefont {O.}~\bibnamefont {Vendrell}}, \bibinfo
  {author} {\bibfnamefont {C.}~\bibnamefont {Bomme}}, \bibinfo {author}
  {\bibfnamefont {E.}~\bibnamefont {Savelyev}}, \bibinfo {author}
  {\bibfnamefont {B.}~\bibnamefont {Rudek}}, \bibinfo {author} {\bibfnamefont
  {L.}~\bibnamefont {Foucar}}, \bibinfo {author} {\bibfnamefont {S.~H.}\
  \bibnamefont {Southworth}}, \bibinfo {author} {\bibfnamefont {C.~S.}\
  \bibnamefont {Lehmann}}, \bibinfo {author} {\bibfnamefont {B.}~\bibnamefont
  {Kraessig}}, \bibinfo {author} {\bibfnamefont {T.}~\bibnamefont {Marchenko}},
  \bibinfo {author} {\bibfnamefont {M.}~\bibnamefont {Simon}}, \bibinfo
  {author} {\bibfnamefont {K.}~\bibnamefont {Ueda}}, \bibinfo {author}
  {\bibfnamefont {K.~R.}\ \bibnamefont {Ferguson}}, \bibinfo {author}
  {\bibfnamefont {M.}~\bibnamefont {Bucher}}, \bibinfo {author} {\bibfnamefont
  {T.}~\bibnamefont {Gorkhover}}, \bibinfo {author} {\bibfnamefont
  {S.}~\bibnamefont {Carron}}, \bibinfo {author} {\bibfnamefont
  {R.}~\bibnamefont {Alonso-Mori}}, \bibinfo {author} {\bibfnamefont {J.~E.}\
  \bibnamefont {Koglin}}, \bibinfo {author} {\bibfnamefont {J.}~\bibnamefont
  {Correa}}, \bibinfo {author} {\bibfnamefont {G.~J.}\ \bibnamefont
  {Williams}}, \bibinfo {author} {\bibfnamefont {S.}~\bibnamefont {Boutet}},
  \bibinfo {author} {\bibfnamefont {L.}~\bibnamefont {Young}}, \bibinfo
  {author} {\bibfnamefont {C.}~\bibnamefont {Bostedt}}, \bibinfo {author}
  {\bibfnamefont {S.-K.}\ \bibnamefont {Son}}, \bibinfo {author} {\bibfnamefont
  {R.}~\bibnamefont {Santra}}, \ and\ \bibinfo {author} {\bibfnamefont
  {D.}~\bibnamefont {Rolles}},\ }\bibfield  {title} {\emph {\bibinfo {title}
  {Femtosecond response of polyatomic molecules to ultra-intense hard
  {X}-rays}},\ }\href {\doibase 10.1038/nature22373} {\bibfield  {journal}
  {\bibinfo  {journal} {Nature}\ }\textbf {\bibinfo {volume} {546}},\ \bibinfo
  {pages} {129} (\bibinfo {year} {2017})}\BibitemShut {NoStop}%
\bibitem [{\citenamefont {Thompson}(2009)}]{thompson_x-ray_2009}%
  \BibitemOpen
  \bibinfo {editor} {\bibfnamefont {A.~C.}\ \bibnamefont {Thompson}},\ ed.,\
  \href {http://xdb.lbl.gov/} {\emph {\bibinfo {title} {X-{Ray} {Data}
  {Booklet}}}}\ (\bibinfo {year} {2009})\BibitemShut {NoStop}%
\bibitem [{\citenamefont {Saito}\ and\ \citenamefont
  {Suzuki}(1992)}]{saito_multiple_1992}%
  \BibitemOpen
  \bibfield  {author} {\bibinfo {author} {\bibfnamefont {N.}~\bibnamefont
  {Saito}}\ and\ \bibinfo {author} {\bibfnamefont {I.~H.}\ \bibnamefont
  {Suzuki}},\ }\bibfield  {title} {\emph {\bibinfo {title} {Multiple
  photoionization in {Ne}, {Ar}, {Kr} and {Xe} from 44 to 1300 {eV}}},\ }\href
  {http://www.sciencedirect.com/science/article/pii/0168117692850382}
  {\bibfield  {journal} {\bibinfo  {journal} {International Journal of Mass
  Spectrometry and Ion Processes}\ }\textbf {\bibinfo {volume} {115}},\
  \bibinfo {pages} {157} (\bibinfo {year} {1992})}\BibitemShut {NoStop}%
\bibitem [{\citenamefont {Burt}\ \emph {et~al.}(2017)\citenamefont {Burt},
  \citenamefont {Boll}, \citenamefont {Lee}, \citenamefont {Amini},
  \citenamefont {Koeckert}, \citenamefont {Vallance}, \citenamefont
  {Gentleman}, \citenamefont {Mackenzie}, \citenamefont {Bari}, \citenamefont
  {Bomme}, \citenamefont {D\"usterer}, \citenamefont {Erk}, \citenamefont
  {Manschwetus}, \citenamefont {M\"uller}, \citenamefont {Rompotis},
  \citenamefont {Savelyev}, \citenamefont {Schirmel}, \citenamefont {Techert},
  \citenamefont {Teusch}, \citenamefont {K\"upper}, \citenamefont {Trippel},
  \citenamefont {Wiese}, \citenamefont {Stapelfeldt}, \citenamefont
  {de~Miranda}, \citenamefont {Rudenko}, \citenamefont {Ziaee}, \citenamefont
  {Brouard},\ and\ \citenamefont {Rolles}}]{burt_coulomb_2017}%
  \BibitemOpen
  \bibfield  {author} {\bibinfo {author} {\bibfnamefont {M.}~\bibnamefont
  {Burt}}, \bibinfo {author} {\bibfnamefont {R.}~\bibnamefont {Boll}}, \bibinfo
  {author} {\bibfnamefont {J.}~\bibnamefont {Lee}}, \bibinfo {author}
  {\bibfnamefont {K.}~\bibnamefont {Amini}}, \bibinfo {author} {\bibfnamefont
  {H.}~\bibnamefont {Koeckert}}, \bibinfo {author} {\bibfnamefont
  {C.}~\bibnamefont {Vallance}}, \bibinfo {author} {\bibfnamefont {A.~S.}\
  \bibnamefont {Gentleman}}, \bibinfo {author} {\bibfnamefont {S.~R.}\
  \bibnamefont {Mackenzie}}, \bibinfo {author} {\bibfnamefont {S.}~\bibnamefont
  {Bari}}, \bibinfo {author} {\bibfnamefont {C.}~\bibnamefont {Bomme}},
  \bibinfo {author} {\bibfnamefont {S.}~\bibnamefont {D\"usterer}}, \bibinfo
  {author} {\bibfnamefont {B.}~\bibnamefont {Erk}}, \bibinfo {author}
  {\bibfnamefont {B.}~\bibnamefont {Manschwetus}}, \bibinfo {author}
  {\bibfnamefont {E.}~\bibnamefont {M\"uller}}, \bibinfo {author}
  {\bibfnamefont {D.}~\bibnamefont {Rompotis}}, \bibinfo {author}
  {\bibfnamefont {E.}~\bibnamefont {Savelyev}}, \bibinfo {author}
  {\bibfnamefont {N.}~\bibnamefont {Schirmel}}, \bibinfo {author}
  {\bibfnamefont {S.}~\bibnamefont {Techert}}, \bibinfo {author} {\bibfnamefont
  {R.}~\bibnamefont {Teusch}}, \bibinfo {author} {\bibfnamefont
  {J.}~\bibnamefont {K\"upper}}, \bibinfo {author} {\bibfnamefont
  {S.}~\bibnamefont {Trippel}}, \bibinfo {author} {\bibfnamefont
  {J.}~\bibnamefont {Wiese}}, \bibinfo {author} {\bibfnamefont
  {H.}~\bibnamefont {Stapelfeldt}}, \bibinfo {author} {\bibfnamefont {B.~C.}\
  \bibnamefont {de~Miranda}}, \bibinfo {author} {\bibfnamefont
  {A.}~\bibnamefont {Rudenko}}, \bibinfo {author} {\bibfnamefont
  {F.}~\bibnamefont {Ziaee}}, \bibinfo {author} {\bibfnamefont
  {M.}~\bibnamefont {Brouard}}, \ and\ \bibinfo {author} {\bibfnamefont
  {D.}~\bibnamefont {Rolles}},\ }\bibfield  {title} {\emph {\bibinfo {title}
  {Coulomb explosion imaging of concurrent {CH}$_{\textrm{2}}${BrI}
  photodissociation dynamics}},\ }\href
  {https://journals.aps.org/pra/abstract/10.1103/PhysRevA.96.043415} {\bibfield
   {journal} {\bibinfo  {journal} {Physical Review A}\ }\textbf {\bibinfo
  {volume} {96}},\ \bibinfo {pages} {04341} (\bibinfo {year}
  {2017})}\BibitemShut {NoStop}%
\bibitem [{\citenamefont {Amini}(2017)}]{amini_studies_2017}%
  \BibitemOpen
  \bibfield  {author} {\bibinfo {author} {\bibfnamefont {K.}~\bibnamefont
  {Amini}},\ }\emph {\bibinfo {title} {Studies of {Ultrafast} {Molecular}
  {Photofragmentation} and {Dynamics} {Using} {Fast} {Imaging} {Sensors}}},\
  \href@noop {} {Ph.D. thesis} (\bibinfo {year} {2017})\BibitemShut {NoStop}%
\bibitem [{\citenamefont {Ablikim}\ \emph {et~al.}(2017)\citenamefont
  {Ablikim}, \citenamefont {Bomme}, \citenamefont {Savelyev}, \citenamefont
  {Xiong}, \citenamefont {Kushawaha}, \citenamefont {Boll}, \citenamefont
  {Amini}, \citenamefont {Osipov}, \citenamefont {Kilcoyne}, \citenamefont
  {Rudenko}, \citenamefont {Berrah},\ and\ \citenamefont
  {Rolles}}]{ablikim_isomer-dependent_2017}%
  \BibitemOpen
  \bibfield  {author} {\bibinfo {author} {\bibfnamefont {U.}~\bibnamefont
  {Ablikim}}, \bibinfo {author} {\bibfnamefont {C.}~\bibnamefont {Bomme}},
  \bibinfo {author} {\bibfnamefont {E.}~\bibnamefont {Savelyev}}, \bibinfo
  {author} {\bibfnamefont {H.}~\bibnamefont {Xiong}}, \bibinfo {author}
  {\bibfnamefont {R.}~\bibnamefont {Kushawaha}}, \bibinfo {author}
  {\bibfnamefont {R.}~\bibnamefont {Boll}}, \bibinfo {author} {\bibfnamefont
  {K.}~\bibnamefont {Amini}}, \bibinfo {author} {\bibfnamefont
  {T.}~\bibnamefont {Osipov}}, \bibinfo {author} {\bibfnamefont
  {D.}~\bibnamefont {Kilcoyne}}, \bibinfo {author} {\bibfnamefont
  {A.}~\bibnamefont {Rudenko}}, \bibinfo {author} {\bibfnamefont
  {N.}~\bibnamefont {Berrah}}, \ and\ \bibinfo {author} {\bibfnamefont
  {D.}~\bibnamefont {Rolles}},\ }\bibfield  {title} {\emph {\bibinfo {title}
  {Isomer-dependent fragmentation dynamics of inner-shell photoionized
  difluoroiodobenzene}},\ }\href {\doibase 10.1039/C7CP01379E} {\bibfield
  {journal} {\bibinfo  {journal} {Physical Chemistry Chemical Physics}\
  }\textbf {\bibinfo {volume} {19}},\ \bibinfo {pages} {13419} (\bibinfo {year}
  {2017})}\BibitemShut {NoStop}%
\bibitem [{\citenamefont {Kuleff}\ \emph {et~al.}(2016)\citenamefont {Kuleff},
  \citenamefont {Kryzhevoi}, \citenamefont {Pernpointner},\ and\ \citenamefont
  {Cederbaum}}]{kuleff_core_2016}%
  \BibitemOpen
  \bibfield  {author} {\bibinfo {author} {\bibfnamefont {A.~I.}\ \bibnamefont
  {Kuleff}}, \bibinfo {author} {\bibfnamefont {N.~V.}\ \bibnamefont
  {Kryzhevoi}}, \bibinfo {author} {\bibfnamefont {M.}~\bibnamefont
  {Pernpointner}}, \ and\ \bibinfo {author} {\bibfnamefont {L.~S.}\
  \bibnamefont {Cederbaum}},\ }\bibfield  {title} {\emph {\bibinfo {title}
  {Core {Ionization} {Initiates} {Subfemtosecond} {Charge} {Migration} in the
  {Valence} {Shell} of {Molecules}}},\ }\href {\doibase
  10.1103/PhysRevLett.117.093002} {\bibfield  {journal} {\bibinfo  {journal}
  {Physical Review Letters}\ }\textbf {\bibinfo {volume} {117}},\ \bibinfo
  {pages} {093002} (\bibinfo {year} {2016})}\BibitemShut {NoStop}%
\bibitem [{\citenamefont {Erk}\ \emph {et~al.}(2013{\natexlab{a}})\citenamefont
  {Erk}, \citenamefont {Rolles}, \citenamefont {Foucar}, \citenamefont {Rudek},
  \citenamefont {Epp}, \citenamefont {Cryle}, \citenamefont {Bostedt},
  \citenamefont {Schorb}, \citenamefont {Bozek}, \citenamefont {Rouzee},
  \citenamefont {Hundertmark}, \citenamefont {Marchenko}, \citenamefont
  {Simon}, \citenamefont {Filsinger}, \citenamefont {Christensen},
  \citenamefont {De}, \citenamefont {Trippel}, \citenamefont {K{\"u}pper},
  \citenamefont {Stapelfeldt}, \citenamefont {Wada}, \citenamefont {Ueda},
  \citenamefont {Swiggers}, \citenamefont {Messerschmidt}, \citenamefont
  {Schr{\"o}ter}, \citenamefont {Moshammer}, \citenamefont {Schlichting},
  \citenamefont {Ullrich},\ and\ \citenamefont {Rudenko}}]{erk_ultrafast_2013}%
  \BibitemOpen
  \bibfield  {author} {\bibinfo {author} {\bibfnamefont {B.}~\bibnamefont
  {Erk}}, \bibinfo {author} {\bibfnamefont {D.}~\bibnamefont {Rolles}},
  \bibinfo {author} {\bibfnamefont {L.}~\bibnamefont {Foucar}}, \bibinfo
  {author} {\bibfnamefont {B.}~\bibnamefont {Rudek}}, \bibinfo {author}
  {\bibfnamefont {S.~W.}\ \bibnamefont {Epp}}, \bibinfo {author} {\bibfnamefont
  {M.}~\bibnamefont {Cryle}}, \bibinfo {author} {\bibfnamefont
  {C.}~\bibnamefont {Bostedt}}, \bibinfo {author} {\bibfnamefont
  {S.}~\bibnamefont {Schorb}}, \bibinfo {author} {\bibfnamefont
  {J.}~\bibnamefont {Bozek}}, \bibinfo {author} {\bibfnamefont
  {A.}~\bibnamefont {Rouzee}}, \bibinfo {author} {\bibfnamefont
  {A.}~\bibnamefont {Hundertmark}}, \bibinfo {author} {\bibfnamefont
  {T.}~\bibnamefont {Marchenko}}, \bibinfo {author} {\bibfnamefont
  {M.}~\bibnamefont {Simon}}, \bibinfo {author} {\bibfnamefont
  {F.}~\bibnamefont {Filsinger}}, \bibinfo {author} {\bibfnamefont
  {L.}~\bibnamefont {Christensen}}, \bibinfo {author} {\bibfnamefont
  {S.}~\bibnamefont {De}}, \bibinfo {author} {\bibfnamefont {S.}~\bibnamefont
  {Trippel}}, \bibinfo {author} {\bibfnamefont {J.}~\bibnamefont {K{\"u}pper}},
  \bibinfo {author} {\bibfnamefont {H.}~\bibnamefont {Stapelfeldt}}, \bibinfo
  {author} {\bibfnamefont {S.}~\bibnamefont {Wada}}, \bibinfo {author}
  {\bibfnamefont {K.}~\bibnamefont {Ueda}}, \bibinfo {author} {\bibfnamefont
  {M.}~\bibnamefont {Swiggers}}, \bibinfo {author} {\bibfnamefont
  {M.}~\bibnamefont {Messerschmidt}}, \bibinfo {author} {\bibfnamefont {C.~D.}\
  \bibnamefont {Schr{\"o}ter}}, \bibinfo {author} {\bibfnamefont
  {R.}~\bibnamefont {Moshammer}}, \bibinfo {author} {\bibfnamefont
  {I.}~\bibnamefont {Schlichting}}, \bibinfo {author} {\bibfnamefont
  {J.}~\bibnamefont {Ullrich}}, \ and\ \bibinfo {author} {\bibfnamefont
  {A.}~\bibnamefont {Rudenko}},\ }\bibfield  {title} {\emph {\bibinfo {title}
  {Ultrafast {Charge} {Rearrangement} and {Nuclear} {Dynamics} upon
  {Inner}-{Shell} {Multiple} {Ionization} of {Small} {Polyatomic}
  {Molecules}}},\ }\href {\doibase 10.1103/PhysRevLett.110.053003} {\bibfield
  {journal} {\bibinfo  {journal} {Physical Review Letters}\ }\textbf {\bibinfo
  {volume} {110}},\ \bibinfo {pages} {053003} (\bibinfo {year}
  {2013}{\natexlab{a}})}\BibitemShut {NoStop}%
\bibitem [{\citenamefont {Erk}\ \emph {et~al.}(2013{\natexlab{b}})\citenamefont
  {Erk}, \citenamefont {Rolles}, \citenamefont {Foucar}, \citenamefont {Rudek},
  \citenamefont {Epp}, \citenamefont {Cryle}, \citenamefont {Bostedt},
  \citenamefont {Schorb}, \citenamefont {Bozek}, \citenamefont {Rouzee},
  \citenamefont {Hundertmark}, \citenamefont {Marchenko}, \citenamefont
  {Simon}, \citenamefont {Filsinger}, \citenamefont {Christensen},
  \citenamefont {De}, \citenamefont {Trippel}, \citenamefont {K{\"u}pper},
  \citenamefont {Stapelfeldt}, \citenamefont {Wada}, \citenamefont {Ueda},
  \citenamefont {Swiggers}, \citenamefont {Messerschmidt}, \citenamefont
  {Schr{\"o}ter}, \citenamefont {Moshammer}, \citenamefont {Schlichting},
  \citenamefont {Ullrich},\ and\ \citenamefont
  {Rudenko}}]{erk_inner-shell_2013}%
  \BibitemOpen
  \bibfield  {author} {\bibinfo {author} {\bibfnamefont {B.}~\bibnamefont
  {Erk}}, \bibinfo {author} {\bibfnamefont {D.}~\bibnamefont {Rolles}},
  \bibinfo {author} {\bibfnamefont {L.}~\bibnamefont {Foucar}}, \bibinfo
  {author} {\bibfnamefont {B.}~\bibnamefont {Rudek}}, \bibinfo {author}
  {\bibfnamefont {S.~W.}\ \bibnamefont {Epp}}, \bibinfo {author} {\bibfnamefont
  {M.}~\bibnamefont {Cryle}}, \bibinfo {author} {\bibfnamefont
  {C.}~\bibnamefont {Bostedt}}, \bibinfo {author} {\bibfnamefont
  {S.}~\bibnamefont {Schorb}}, \bibinfo {author} {\bibfnamefont
  {J.}~\bibnamefont {Bozek}}, \bibinfo {author} {\bibfnamefont
  {A.}~\bibnamefont {Rouzee}}, \bibinfo {author} {\bibfnamefont
  {A.}~\bibnamefont {Hundertmark}}, \bibinfo {author} {\bibfnamefont
  {T.}~\bibnamefont {Marchenko}}, \bibinfo {author} {\bibfnamefont
  {M.}~\bibnamefont {Simon}}, \bibinfo {author} {\bibfnamefont
  {F.}~\bibnamefont {Filsinger}}, \bibinfo {author} {\bibfnamefont
  {L.}~\bibnamefont {Christensen}}, \bibinfo {author} {\bibfnamefont
  {S.}~\bibnamefont {De}}, \bibinfo {author} {\bibfnamefont {S.}~\bibnamefont
  {Trippel}}, \bibinfo {author} {\bibfnamefont {J.}~\bibnamefont {K{\"u}pper}},
  \bibinfo {author} {\bibfnamefont {H.}~\bibnamefont {Stapelfeldt}}, \bibinfo
  {author} {\bibfnamefont {S.}~\bibnamefont {Wada}}, \bibinfo {author}
  {\bibfnamefont {K.}~\bibnamefont {Ueda}}, \bibinfo {author} {\bibfnamefont
  {M.}~\bibnamefont {Swiggers}}, \bibinfo {author} {\bibfnamefont
  {M.}~\bibnamefont {Messerschmidt}}, \bibinfo {author} {\bibfnamefont {C.~D.}\
  \bibnamefont {Schr{\"o}ter}}, \bibinfo {author} {\bibfnamefont
  {R.}~\bibnamefont {Moshammer}}, \bibinfo {author} {\bibfnamefont
  {I.}~\bibnamefont {Schlichting}}, \bibinfo {author} {\bibfnamefont
  {J.}~\bibnamefont {Ullrich}}, \ and\ \bibinfo {author} {\bibfnamefont
  {A.}~\bibnamefont {Rudenko}},\ }\bibfield  {title} {\emph {\bibinfo {title}
  {Inner-shell multiple ionization of polyatomic molecules with an intense
  x-ray free-electron laser studied by coincident ion momentum imaging}},\
  }\href {\doibase 10.1088/0953-4075/46/16/164031} {\bibfield  {journal}
  {\bibinfo  {journal} {Journal of Physics B}\ }\textbf {\bibinfo {volume}
  {46}},\ \bibinfo {pages} {164031} (\bibinfo {year}
  {2013}{\natexlab{b}})}\BibitemShut {NoStop}%
\bibitem [{\citenamefont {Brausse}\ \emph {et~al.}(2017)\citenamefont
  {Brausse}, \citenamefont {Goldsztejn}, \citenamefont {Amini}, \citenamefont
  {Boll}, \citenamefont {Bari}, \citenamefont {Bomme}, \citenamefont {Brouard},
  \citenamefont {Burt}, \citenamefont {D\"usterer}, \citenamefont {Erk},
  \citenamefont {Manschwetus}, \citenamefont {M\"uller}, \citenamefont {Lee},
  \citenamefont {Koeckert}, \citenamefont {Vallance}, \citenamefont
  {Gentleman}, \citenamefont {Mackenzie}, \citenamefont {Rompotis},
  \citenamefont {Savelyev}, \citenamefont {Schirmel}, \citenamefont {Techert},
  \citenamefont {Teusch}, \citenamefont {K\"upper}, \citenamefont {Trippel},
  \citenamefont {Wiese}, \citenamefont {de~Miranda}, \citenamefont {Rudenko},
  \citenamefont {Holland}, \citenamefont {Marchenko}, \citenamefont
  {Rouz\'ee},\ and\ \citenamefont {Rolles}}]{brausse_time-resolved_2017}%
  \BibitemOpen
  \bibfield  {author} {\bibinfo {author} {\bibfnamefont {F.}~\bibnamefont
  {Brausse}}, \bibinfo {author} {\bibfnamefont {G.}~\bibnamefont {Goldsztejn}},
  \bibinfo {author} {\bibfnamefont {K.}~\bibnamefont {Amini}}, \bibinfo
  {author} {\bibfnamefont {R.}~\bibnamefont {Boll}}, \bibinfo {author}
  {\bibfnamefont {S.}~\bibnamefont {Bari}}, \bibinfo {author} {\bibfnamefont
  {C.}~\bibnamefont {Bomme}}, \bibinfo {author} {\bibfnamefont
  {M.}~\bibnamefont {Brouard}}, \bibinfo {author} {\bibfnamefont
  {M.}~\bibnamefont {Burt}}, \bibinfo {author} {\bibfnamefont {S.}~\bibnamefont
  {D\"usterer}}, \bibinfo {author} {\bibfnamefont {B.}~\bibnamefont {Erk}},
  \bibinfo {author} {\bibfnamefont {B.}~\bibnamefont {Manschwetus}}, \bibinfo
  {author} {\bibfnamefont {E.}~\bibnamefont {M\"uller}}, \bibinfo {author}
  {\bibfnamefont {J.}~\bibnamefont {Lee}}, \bibinfo {author} {\bibfnamefont
  {H.}~\bibnamefont {Koeckert}}, \bibinfo {author} {\bibfnamefont
  {C.}~\bibnamefont {Vallance}}, \bibinfo {author} {\bibfnamefont {A.~S.}\
  \bibnamefont {Gentleman}}, \bibinfo {author} {\bibfnamefont {S.~R.}\
  \bibnamefont {Mackenzie}}, \bibinfo {author} {\bibfnamefont {D.}~\bibnamefont
  {Rompotis}}, \bibinfo {author} {\bibfnamefont {E.}~\bibnamefont {Savelyev}},
  \bibinfo {author} {\bibfnamefont {N.}~\bibnamefont {Schirmel}}, \bibinfo
  {author} {\bibfnamefont {S.}~\bibnamefont {Techert}}, \bibinfo {author}
  {\bibfnamefont {R.}~\bibnamefont {Teusch}}, \bibinfo {author} {\bibfnamefont
  {J.}~\bibnamefont {K\"upper}}, \bibinfo {author} {\bibfnamefont
  {S.}~\bibnamefont {Trippel}}, \bibinfo {author} {\bibfnamefont
  {J.}~\bibnamefont {Wiese}}, \bibinfo {author} {\bibfnamefont {B.~C.}\
  \bibnamefont {de~Miranda}}, \bibinfo {author} {\bibfnamefont
  {A.}~\bibnamefont {Rudenko}}, \bibinfo {author} {\bibfnamefont {D.~M.~P.}\
  \bibnamefont {Holland}}, \bibinfo {author} {\bibfnamefont {T.}~\bibnamefont
  {Marchenko}}, \bibinfo {author} {\bibfnamefont {A.}~\bibnamefont {Rouz\'ee}},
  \ and\ \bibinfo {author} {\bibfnamefont {D.}~\bibnamefont {Rolles}},\
  }\bibfield  {title} {\emph {\bibinfo {title} {Time-resolved inner-shell
  photoelectron spectroscopy: from a bound molecule to an isolated atom}},\
  }\href@noop {} {\bibfield  {journal} {\bibinfo  {journal} {submitted}\ }
  (\bibinfo {year} {2017})}\BibitemShut {NoStop}%
\end{thebibliography}%
\end{document}